\documentclass[twocolumn,twocolappendix]{aastex631}

\pdfoutput=1

\received{2022 September 15}
\revised{2022 December 12}
\accepted{2022 December 18; Published 2023 MM DD}

\usepackage{txfonts}

\usepackage{amsmath}
\usepackage{fancyhdr}
\usepackage{xcolor}

\begin{document}

\title{\Large Photometric Calibrations of M-dwarf Metallicity with Markov Chain Monte Carlo and Bayesian Inference}

\correspondingauthor{C. Duque-Arribas}
\email{chrduque@ucm.es}

\author[0000-0002-1758-3973]{C. Duque-Arribas}
\affiliation{Departamento de F{\'i}sica de la Tierra y Astrof{\'i}sica \& IPARCOS-UCM (Instituto de F\'{i}sica de Part\'{i}culas y del Cosmos de la UCM), Facultad de Ciencias F{\'i}sicas, Universidad Complutense de Madrid, E-28040 Madrid, Spain}

\author[0000-0002-7779-238X]{D. Montes}
\affiliation{Departamento de F{\'i}sica de la Tierra y Astrof{\'i}sica \& IPARCOS-UCM (Instituto de F\'{i}sica de Part\'{i}culas y del Cosmos de la UCM), Facultad de Ciencias F{\'i}sicas, Universidad Complutense de Madrid, E-28040 Madrid, Spain}

\author[0000-0002-8087-4298]{H. M. Tabernero}
\affiliation{Centro de Astrobiolog{\'i}a (CAB), CSIC-INTA, carretera de Ajalvir km 4, E-28850 Torrej{\'o}n de Ardoz, Madrid, Spain}

\author[0000-0002-7349-1387]{J. A. Caballero}
\affiliation{Centro de Astrobiolog\'{i}a (CAB), CSIC-INTA, camino bajo del Castillo s/n, ESAC campus, E-28691 Villanueva de la Ca\~{n}ada, Madrid, Spain}

\author[0000-0001-7859-3676]{J. Gorgas}
\affiliation{Departamento de F{\'i}sica de la Tierra y Astrof{\'i}sica \& IPARCOS-UCM (Instituto de F\'{i}sica de Part\'{i}culas y del Cosmos de la UCM), Facultad de Ciencias F{\'i}sicas, Universidad Complutense de Madrid, E-28040 Madrid, Spain}

\author[0000-0001-8907-4775]{E. Marfil}
\affiliation{Departamento de F{\'i}sica de la Tierra y Astrof{\'i}sica \& IPARCOS-UCM (Instituto de F\'{i}sica de Part\'{i}culas y del Cosmos de la UCM), Facultad de Ciencias F{\'i}sicas, Universidad Complutense de Madrid, E-28040 Madrid, Spain}
\affiliation{Instituto de Astrofísica de Canarias, E-38200 La Laguna, Tenerife, Spain}
\affiliation{Departamento de Astrofísica, Universidad de La Laguna, E-38206 La Laguna, Tenerife, Spain}



\begin{abstract}
Knowledge of stellar atmospheric parameters ($T_{\rm eff}$, $\log{g}$, [Fe/H]) of M dwarfs can be used to constrain both theoretical stellar models and Galactic chemical evolutionary models, and guide exoplanet searches, but their determination is difficult due to the complexity of the spectra of their cool atmospheres. In our ongoing effort to characterize M dwarfs, and in particular their chemical composition, we carried out multiband photometric calibrations of metallicity for early- and intermediate-type M dwarfs. The third Gaia data release provides high-precision astrometry and three-band photometry. This information, combined with the 2MASS and CatWISE2020 infrared photometric surveys and a sample of 4919 M dwarfs with metallicity values determined with high-resolution spectroscopy by \texttt{The Cannon} and APOGEE spectra, allowed us to study the effect of the metallicity in color--color and color--magnitude diagrams. We divided this sample into two subsamples: we used 1000 stars to train the calibrations with Bayesian statistics and Markov Chain Monte Carlo techniques, and the remaining 3919 stars to check the accuracy of the estimations. We derived several photometric calibrations of metallicity applicable to M dwarfs in the range of $-0.45\leq\text{[Fe/H]}\leq +0.45$\,dex and spectral types down to M5.0\,V that yield uncertainties down to the $0.10$\,dex level. Lastly, we compared our results with other photometric estimations published in the literature for an additional sample of 46 M dwarfs in wide binary systems with FGK-type primary stars, and found a great predictive performance.
\end{abstract}



\defcitealias{Birky2020ApJ...892...31B}{B20}
\defcitealias{Bonfils2005A&A...442..635B}{B05}
\defcitealias{Johnson2009ApJ...699..933J}{JA09}
\defcitealias{Neves2012A&A...538A..25N}{N12}
\defcitealias{Mann2013AJ....145...52M}{M13}
\defcitealias{Davenport2019RNAAS...3...54D}{DD19}
\defcitealias{Rains2021MNRAS.504.5788R}{R21}

\section{Introduction}

\thispagestyle{fancy}
\fancyhead{} 
\fancyhead[R]{IPARCOS-UCM-23-007}
\renewcommand{\headrulewidth}{0pt}

M-type dwarf stars are the coolest, smallest, and most numerous main-sequence stars in the Galaxy, with effective temperatures of $3900\,\text{K} \gtrsim T_{\rm eff} \gtrsim 2300\,\text{K}$, radii of $0.61\,R_\odot \gtrsim R \gtrsim 0.10\,R_\odot$, and masses of $0.62\,M_\odot \gtrsim M \gtrsim 0.08\,M_\odot$ \citep{Delfosse2000A&A...364..217D, Schweitzer2019A&A...625A..68S, Cifuentes2020A&A...642A.115C}. These stars are so faint that, despite dominating in number the solar neighborhood, none of them is visible to the naked eye \citep{Croswell2002S&T...104a..38C}. These stars are also characterized by very active chromospheres and coronae \citep{West2008AJ....135..785W, Jeffers2018A&A...614A..76J, Kiman2021AJ....161..277K}.

M dwarfs are objects of special interest in multiple branches of astrophysics. These stars have main-sequence lifetimes that exceed the currently known age of the universe, due to the slow fusion process in their mainly convective interiors \citep{Adams1997RvMP...69..337A}, and they are the most abundant main-sequence stars in the Milky Way, accounting for more than 75\,\% of them \citep{Henry2006AJ....132.2360H, Winters2015AJ....149....5W, Reyle2021A&A...650A.201R}. Therefore, M dwarfs stand as excellent probes to study the chemical and dynamical evolution of our Galaxy \citep{Bahcall1980ApJS...44...73B, Reid1997PASP..109..559R, Chabrier2003PASP..115..763C, Ferguson2017ApJ...843..141F}.

Despite the considerable progress with the modeling of stellar spectra over the last decades, there are still disagreements between the observational characteristics of M dwarfs and the values predicted by synthetic spectra. For instance, effective temperatures ($T_\text{eff}$) from models can be up to 200--300\,K hotter than observed values, while radii predictions differ from interferometric measurements by up to 25\,\% \citep{Jones2005MNRAS.358..105J, Sarmento2020A&A...636A..85S}. These deviations may be due to effects caused by the level of activity \citep{Lopez-Morales2005ApJ...631.1120L}, differences in metallicity \citep{Berger2006ApJ...644..475B, Lopez-Morales2007ApJ...660..732L}, or the synthetic gap \citep{Passegger2020A&A...642A..22P, Passegger2022A&A...658A.194P}. Additionally, there are several complications in the model atmospheres of late-type stars. For example, stellar convection in M dwarfs challenges some of the physical assumptions for radiative transfer \citep{Bergemann2017ApJ...847...16B, Olander2021A&A...649A.103O} and there is an incompleteness regarding the line lists and atomic parameters used \citep{Shetrone2015ApJS..221...24S}.

Furthermore, the two most successful techniques for detecting exoplanets, namely the radial velocity and transit methods, are favored in M dwarfs \citep{Nutzman2008PASP..120..317N, Engle2011ASPC..451..285E, Shields2016PhR...663....1S,Reiners2018A&A...612A..49R}. The lower masses of these stars facilitate the detection of exoplanets by analyzing their radial velocity curves and, due to their faintness, their habitable zone is located much closer to the star than for solar-type stars \citep{Tarter2007AsBio...7...30T,Kopparapu2014ApJ...787L..29K,Martinez-Rodriguez2019ApJ...887..261M}. This closeness translates into much shorter orbital periods of the exoplanets in the habitable zone, which makes their detection by the transit method much easier. There are several programs focused on finding exoplanets around M dwarfs. On the one hand, there are transit surveys both from the ground \citep[e.g. MEarth;][]{Irwin2015csss...18..767I} and space \citep[TESS;][]{Ricker2014SPIE.9143E..20R}, including several JWST \citep{Gardner2006SSRv..123..485G} Guaranteed Time Observations (GTO) programs focused on exoplanets\footnote{\url{https://www.stsci.edu/jwst/science-execution/approved-programs/cycle-1-gto}}. On the other hand, there are several ground-based radial velocity instruments such as ESPRESSO 
(\citealt{Pepe2010SPIE.7735E..0FP, Pepe2021A&A...645A..96P}), HARPS
(\citealt{Mayor2003Msngr.114...20M}), HARPS-N (\citealt{Consentino2012SPIE.8446E..1VC}), HPF (\citealt{Mahadevan2012SPIE.8446E..1SM}), IRD 
(\citealt{Kotani2018SPIE10702E..11K}), MAROON-X (\citealt{Seifahrt2020SPIE11447E..1FS}), NEID (\citealt{Schwab2016SPIE.9908E..7HS}), or CARMENES
(\citealt{Alonso-Floriano2015A&A...577A.128A, Quirrenbach2020SPIE11447E..3CQ}, Ribas et al. in press). Of them, CARMENES has been the most successful in discovering and characterizing exoplanets around M dwarfs. The instrument consists of a high-resolution, double-channel spectrograph that provides coverage in the visible ($520$--$960$\,nm) and near-infrared ($960$--$1710$\,nm) with spectral resolutions of $R=94\,600$ and $80\,400$, respectively. The main scientific goal of CARMENES is the detection and characterization of Earth-like exoplanets around M dwarfs. Simultaneous observations in two wavelength ranges help to distinguish between a planetary signal and stellar activity. Both spectrographs are designed to perform high-accuracy radial velocity measurements with a long-term stability of $\sim 1$\, m\,s$^{-1}$, which allows the detection of $2\,M_\oplus$ planets orbiting in the habitable zone of M5\,V stars (\citealt{Trifonov2018A&A...609A.117T,Luque2019A&A...628A..39L,Morales2019Sci...365.1441M,Zechmeister2019A&A...627A..49Z,Caballero2022A&A...665A.120C}).

Moreover, it has been observed that the frequency of gas giant planets increases with stellar metalicity in the case of FGK-type stars, which is known as the planet-metallicity correlation \citep{Gonzalez1997MNRAS.285..403G, Fischer2005ApJ...622.1102F, Brewer2016ApJS..225...32B}. It has been proposed that M dwarfs follow the same tendency: the higher the metallicity, the higher the probability of having icy and gaseous giant planets orbiting around them \citep{Johnson2009ApJ...699..933J, Rojas-Ayala2010ApJ...720L.113R, Terrien2012ApJ...747L..38T, Hobson2018RMxAA..54...65H}. Therefore, studying the correlations between stellar properties, such as metallicity, and the presence of exoplanets can be useful in selecting targets for future exoplanet surveys and in understanding planetary formation mechanisms.

Stellar metallicity is defined as the relative abundance of elements heavier than helium, and the iron abundance ratio [Fe/H] is usually used as a proxy for the overall metallicity. The determination of metallicity of M dwarfs is challenging, and measurements of abundances of these cool stars have been limited due to difficulties in the analysis of their spectra, which are much more complex than those of solar-type or hot stars due to their low atmospheric temperatures. Forests of lines caused by several molecular species (TiO, VO, ZrO, FeH, CaH in the optical regime; H$_2$O, CO in the near infrared) dominate the spectrum, making it difficult to determine atmospheric parameters \citep{Allard1997ARA&A..35..137A, deLaverny2012A&A...544A.126D, VanEck2017A&A...601A..10V, Passegger2018A&A...615A...6P, Marfil2021A&A...656A.162M}.

Consequently, methods for determining the metallicity of M dwarfs require observationally expensive data, such as high-resolution spectra, and a complicated subsequent analysis. For this reason, another series of techniques have been used to estimate the metallicity of these objects. Among these techniques are, for instance, the use of spectral features in the $K$ band
in intermediate-resolution spectra \citep{Rojas-Ayala2010ApJ...720L.113R, Rojas-Ayala2012ApJ...748...93R}, studying binary systems in which the metallicity of the FGK-type primary star is extrapolated to the secondary M dwarf \citep{Montes2018MNRAS.479.1332M, Ishikawa2020PASJ...72..102I}, or with different photometric calibrations using several sky surveys, such as Gaia \citep{Gaia2016A&A...595A...1G}, Two Micron All Sky Survey (2MASS; \citealt{Skrutskie2006AJ....131.1163S}), Wide-field Infrared Survey Explorer (WISE; \citealt{Wright2010AJ....140.1868W}) or Sloan Digital Sky Survey (SDSS; \citealt{Alam2015ApJS..219...12A}), among others. Nevertheless, different methods and models can lead to diverse results, finding different values of the metallicity for the same star in the literature (see \citealt{Passegger2022A&A...658A.194P}).

This paper focuses on the different photometric calibrations of metallicity for M dwarfs. Initially, \cite{Stauffer1986ApJS...61..531S} used broadband photometry to identify nearby M dwarfs with metallicities significantly different from that of the Sun. \cite{Bonfils2005A&A...442..635B} proposed an empirical calibration to derive the metallicity of M-dwarf components in wide visual binaries using the $V-K_s$ vs. $M_{K_s}$ color--magnitude diagram, obtaining a precision of $0.20$\,dex and demonstrating that metallicity explains the dispersion in the empirical $V$-band mass-luminosity relation. Later, \cite{Johnson2009ApJ...699..933J}, \cite{Schlaufman2010A&A...519A.105S}, and \cite{Neves2012A&A...538A..25N} updated the photometric calibration based on the same color--magnitude diagram. \cite{Johnson2012AJ....143..111J} and \cite{Mann2013AJ....145...52M} estimated the metallicity of M dwarfs using the $J-K_s$ vs. $V-K_s$ color--color diagram.
\cite{Hejazi2015AJ....149..140H} used SDSS and 2MASS photometry to derive metallicity from the $(g-K_s)$ vs. $(J-K_s)$ color--color diagram, while \cite{Dittmann2016ApJ...818..153D} exploited 2MASS and MEarth passband in a color--magnitude diagram for this purpose.
\cite{Schmidt2016MNRAS.460.2611S} explored the $r'-z'$ vs. $W1-W2$ color--color diagram, combining SDSS and WISE photometry, to derive new calibrations for late K- and early M-dwarf metallicities. They explored the sensitivity of color indices to metallicity, illustrating the importance of the $W1 - W2$ color index as a metallicity indicator. \cite{Davenport2019RNAAS...3...54D} presented \texttt{ingot}, a k-nearest neighbors regressor to estimate [Fe/H] of low-mass stars using Gaia, 2MASS, and WISE photometry (and Gaia astrometry). \cite{Medan2021AJ....161..234M} trained a Gaussian process regressor to calibrate two photometric metallicity relationships: for K- and early M dwarfs ($3500\,\text{K}<T_\text{eff}<5280$\,K), and for intermediate M dwarfs ($2850\text{\,K}<T_\text{eff}<3500$\,K), combining SDSS and WISE photometry. \cite{Rains2021MNRAS.504.5788R} followed the approaches of \cite{Johnson2009ApJ...699..933J}, \cite{Schlaufman2010A&A...519A.105S}, and \cite{Neves2012A&A...538A..25N}, although using $G_\text{BP}-K_s$ instead of $V-K_s$.

The aim of the present paper is to extend these previous studies and perform different photometric calibrations based on color--color and color--magnitude diagrams applying Bayesian statistics and Markov Chain Monte Carlo (MCMC), and compare them with the ``Leave One Out -- Cross Validation'' criterion (LOO-CV -- \citealt{Vehtari201710.1007/s11222-016-9696-4}).

This manuscript is organized as follows. In Sect.~\ref{methodology} we describe the compilation of the photometric and astrometric data from public catalogs, the star samples considered, and the statistical tools employed. Sect.~\ref{results} describes the calibrations based on color--color and color--magnitude diagrams and their comparison using LOO-CV. Additionally, we compile all the information in three-dimensional color--color--magnitude diagrams and compare our results with other photometric estimations found in the literature. Finally, in Sect.~\ref{summary} we discuss our results and future improvements.

\section{Analysis}
\label{methodology}

\subsection{Star Samples}

\begin{figure}
    \centering
    \includegraphics[width=0.45\textwidth]{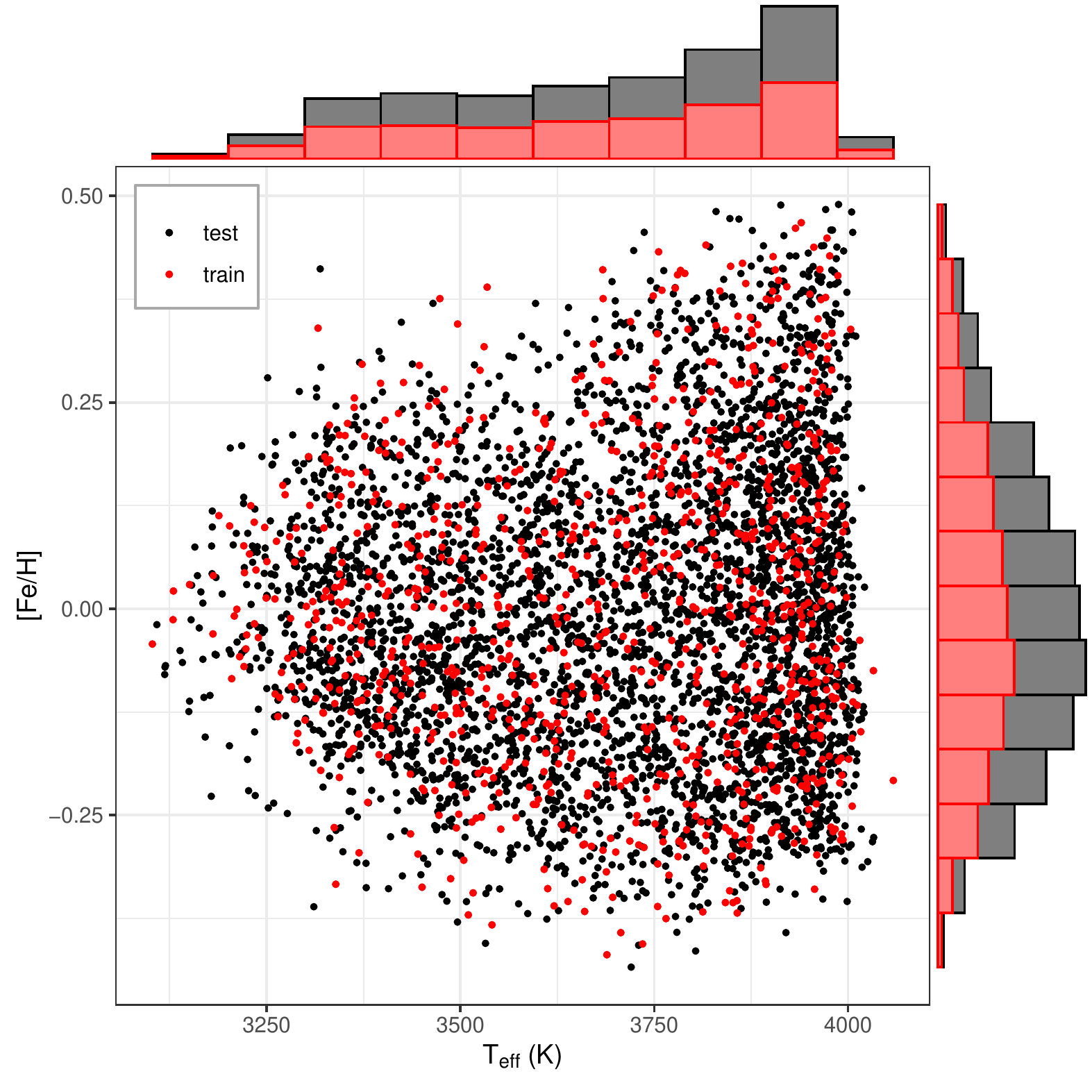}
    \caption{$T_\text{eff}$ vs. [Fe/H] diagram for the training (red) and test (black) subsamples from \citetalias{Birky2020ApJ...892...31B}, showing the histograms for both variables.}
    \label{Teff_FeH_space}
\end{figure}

\citet[hereafter \citetalias{Birky2020ApJ...892...31B}]{Birky2020ApJ...892...31B} presented a sample of 5875 early- and intermediate-type M dwarfs (down to M6\,V) in the Apache Point Observatory Galactic Evolution Experiment (APOGEE; \citealt{Majewski2017AJ....154...94M}, \citealt{Abolfathi2018ApJS..235...42A}) and Gaia DR2 \citep{Gaia2018A&A...616A...1G} surveys. Stellar parameters were inferred for these stars using \texttt{The Cannon} \citep{Ness2015ApJ...808...16N}, a fully empirical model that, beyond the reference labels, employs no line lists or radiative transfer models, transferring labels from high-resolution spectra for which we know parameters to those for which we do not, and circumventing the difficulties of modeling the stellar atmospheres and common issues associated such as incomplete line lists. We used the \citetalias{Birky2020ApJ...892...31B} sample to train our calibrations and check their accuracy (Sect.~\ref{results}). The coverage, distribution, and biases of this star sample are displayed in Fig.~\ref{Teff_FeH_space}. The B20 sample does not cover the $T_\text{eff}$ vs. [Fe/H] space homogeneously, since only early M dwarfs have the highest metallicity values and in the coolest range the sample is biased to solar-metallicity stars.

Moreover, we tested the predictive performance of the calibrations with the sample presented by \citet{Montes2018MNRAS.479.1332M}, who studied 192 binary systems made of late F, G, or early K primaries and late K- or M-dwarf companion candidates. The authors carried out observations with the HERMES spectrograph at the 1.2\,m Mercator telescope \citep{Raskin2011A&A...526A..69R} and obtained high-resolution spectra for the 192 primaries and five secondaries. These spectra were analyzed with the automatic code \textsc{StePar}\footnote{\url{https://github.com/hmtabernero/StePar}} \citep{Tabernero2019A&A...628A.131T}, based on the equivalent width method, to derive precise stellar atmospheric parameters (effective temperature $T_{\rm eff}$, surface gravity $\log{g}$, and metallicity [Fe/H]). Since binaries are assumed to be born at the same time and from the same molecular cloud, the composition and age of the FGK-type primary star can be extrapolated to its secondary M dwarf \citep{Desidera2006A&A...454..581D, Andrews2018MNRAS.473.5393A}. Next we checked our calibrations for these stars and compared them to other photometric estimations found in the literature.

\subsection{Photometry and Data Filtering}
\label{photometry}

\begin{table}
    \centering
    \caption{Data Filtering Criteria Applied to Astrophotometric Data}
    \begin{tabular}{ll} \hline\hline
    \noalign{\smallskip}
    Survey & Filter \\ 
    \noalign{\smallskip}
    \hline
    \noalign{\smallskip}
    Gaia EDR3 & $\texttt{parallax\_over\_error > 10}$ \\
     & $\texttt{ruwe < 1.4}$ \\
     & $\texttt{photo\_g\_mean\_flux\_over\_error > 50}$ \\
     & $\texttt{photo\_bp\_mean\_flux\_over\_error > 20}$ \\
     & $\texttt{photo\_rp\_mean\_flux\_over\_error > 20}$ \\ 
    \noalign{\smallskip}
    \hline
    \noalign{\smallskip}
    2MASS\tablenotemark{a} & $\texttt{Qfl = AAA}$ \\ 
    \hline
    \noalign{\smallskip}
    CatWISE2020\tablenotemark{b} & $\texttt{qph = AA**}$ \\
    \noalign{\smallskip}
    \hline
    \end{tabular}
    \tablenotetext{a}{{\tt Qfl} is the quality flag in 2MASS $JHK_s$ bands.}
    \tablenotetext{b}{{\tt qph} is the quality flag in WISE $W1W2W3W4$ bands.}
    \label{table_filtering}
\end{table}

\begin{figure}
    \centering
    \includegraphics[width=0.44\textwidth]{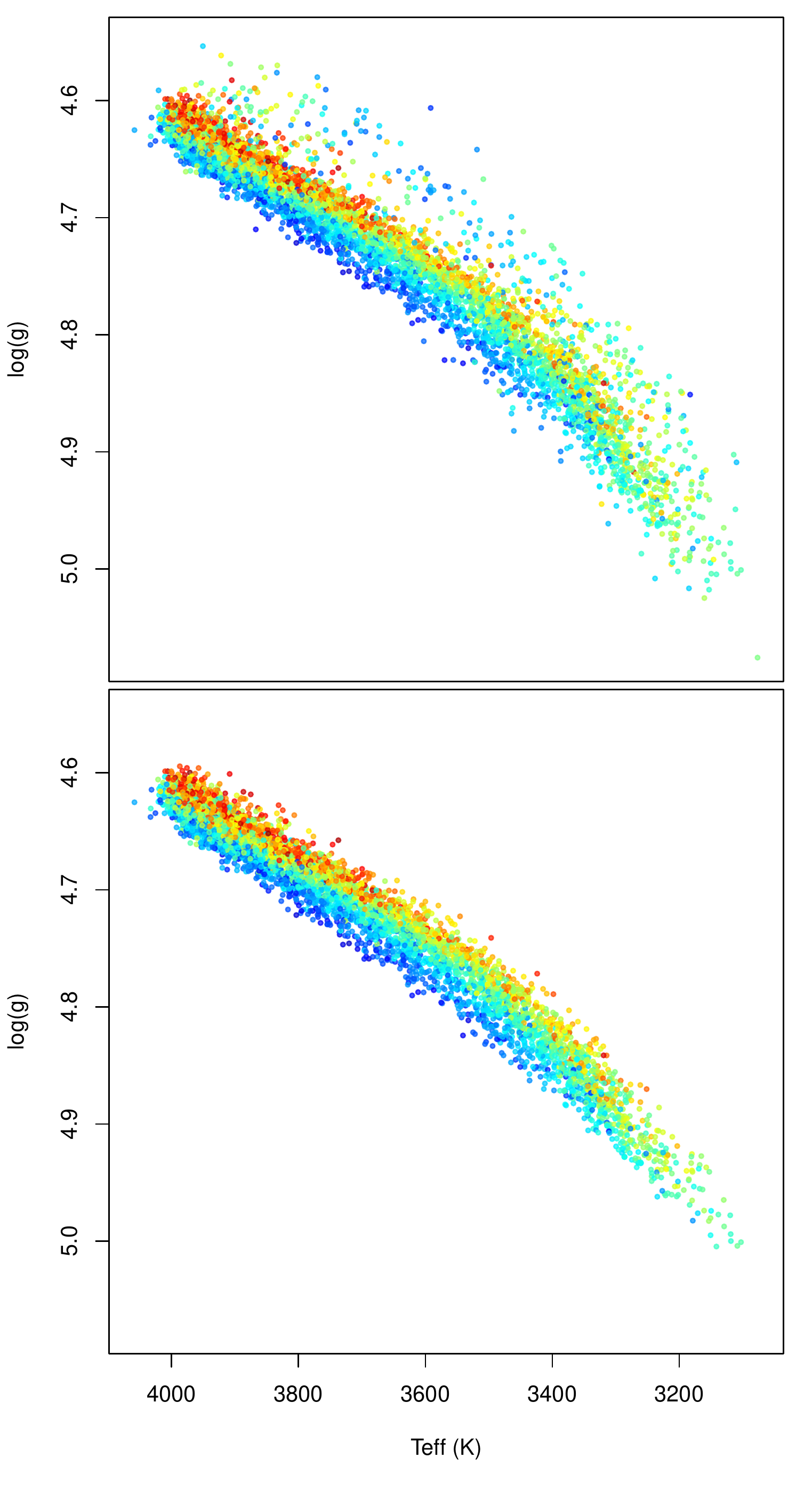}
    
    \includegraphics[width=0.25\textwidth]{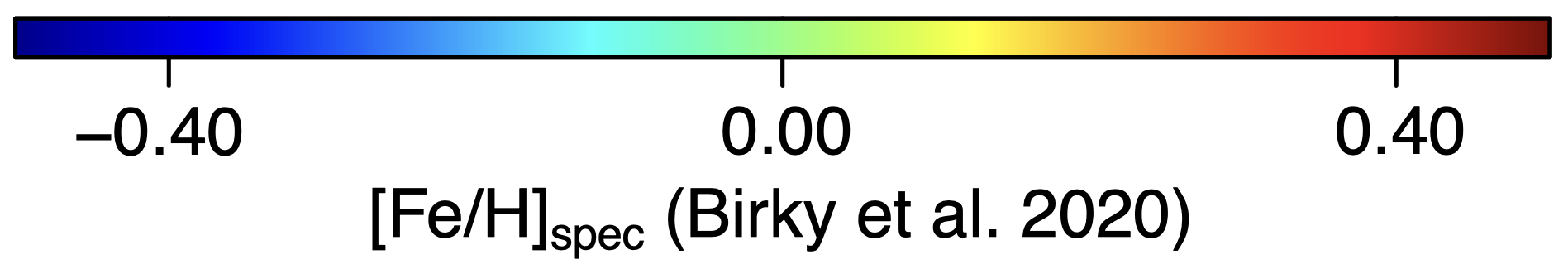}
    \caption{Kiel diagram of the \citetalias{Birky2020ApJ...892...31B} sample, before (upper panel) and after (lower panel) having removed the stars with lower surface gravity, i.e. not in the main sequence.}
    \label{Teff_logg_cleaning}
\end{figure}

The third Gaia data release \citep[Gaia DR3;][]{Gaia2022arXiv220800211G} provides the position and apparent magnitude in the $G$ band (330--1050\,nm) for $1.8$ billion sources. For $1.5$ billion of them, parallax and proper motion data are also available. In addition, photometry in the $G_\text{BP}$ (330--680\,nm) and $G_\text{RP}$ (630--1050\,nm) bands is offered for another $1.5$ billion sources \citep{Gaia2016A&A...595A...1G, Gaia2021A&A...649A...1G, Riello2021A&A...649A...3R}. 
To study M dwarfs, which have the peak of the emission beyond 1000\,nm \citep{Cifuentes2020A&A...642A.115C}, we also used information in the infrared (IR) wavelength range: 2MASS provides magnitudes in the near-IR $J$ ($1229$\,nm), $H$ ($1639$\,nm) and $K_s$ bands ($2152$\,nm), while {\em WISE} offers data in the mid-IR bands $W1$, $W2$, $W3$, and $W4$, centered at 3316\,nm, 4564\,nm, 10\,787\,nm, and 21\,915\,nm, respectively. In particular, we used the data from the updated version 
CatWISE2020 \citep{Marocco2021ApJS..253....8M}, which has enhanced sensitivity and accuracy. Initially, the analysis was performed with the AllWISE version \citep{Cutri2014yCat.2328....0C}, but the uncertainties in the $W1-W2$ color index were a factor of $\sim 2$ larger than those of CatWISE2020.

First, we crossmatched the star samples described above with the Gaia DR3, 2MASS, and CatWISE2020 catalogs. For that, we used the Tool for OPerations on Catalogues And Tables ({\tt TOPCAT}; \citealt{Taylor2005ASPC..347...29T}). In particular, we used the automatic positional crossmatch tool of the Centre de Données astronomiques de Strasbourg, CDS {\tt X-match}, with a search radius of 5\,arcsec and the ``All'' find option. Next, we used the {\tt Aladin} sky atlas \citep{Bonnarel2000A&AS..143...33B} to inspect and correct the possibly mismatched cases.

These data do not have homogeneous quality. Consequently, we applied the data filtering indicated by \citet{Gaia2018A&A...616A..10G}. 
In particular, for color--magnitude diagrams, we made use of the absolute magnitude calculated using the Gaia parallax
and selected only stars that fulfill the 10\,\% relative precision criterion, which corresponds to an uncertainty in $M_G$ lower than $0.22$\,mag. Similarly, we applied filters to the relative flux error on the $G$, $G_\text{BP}$, and $G_\text{RP}$ magnitudes, which led to uncertainties of 0.022\,mag, 0.054\,mag, and 0.054\,mag, respectively. To discard close unresolved or partially resolved binaries, we also applied a conservative filter in the astrometric quality indicator RUWE (renormalized unit weight error) as indicated by \citet{Lindegren2021A&A...649A...2L}, retaining those stars with RUWE values $<1.4$. For sources where the single-star model provides a good fit to the astrometric observations, the RUWE value is expected to be around $1.0$, and value significantly greater ($>1.4$) could indicate that the source is nonsingle or problematic for the astrometric solution.

In addition, we selected the stars with an ``A'' quality flag in the $J$, $H$, $K_s$, $W1$, and $W2$ bands, which corresponds to an approximate signal-to-noise ratio higher than 10. We discarded the $W3$ and $W4$ bands from our analysis, as they tend to present a lower photometric quality. 
All these criteria are compiled in Table~\ref{table_filtering}. Applying these criteria to the 5875 stars presented by \citetalias{Birky2020ApJ...892...31B}, we ended up with a sample of 5453 M dwarfs.

Finally, we removed young objects and/or evolved stars, i.e. stars arriving or leaving the main sequence, since for those cases the age plays an important role in the position of the star in the color--color and color--magnitude diagrams. To do this, we estimated the radii and masses of the stars with the $M_{K_s}$ absolute magnitude using the calibrations given by \citet{Mann2015ApJ...804...64M} [Eq.~5] and \citet{Mann2019ApJ...871...63M} [Eq.~5], respectively. With these two properties, we calculated the surface gravity $\log g$. The pre-main-sequence and the evolved stars are expected to have inflated radii, and thus lower surface gravities. Therefore, we calibrated the surface gravity using the effective temperature and metallicity and removed these lower surface gravity stars, those with a difference between the photometric and fitted surface gravities larger than $0.03$\,dex. Hence we obtained a final sample of 4919 M dwarfs. We show in Fig.~\ref{Teff_logg_cleaning} the Kiel diagram ($T_\text{eff}$ vs. $\log g$) before and after having removed the lower $\log g$ stars. Note the gradient of metallicity present in the main sequence of the Kiel diagram, having decreasing $\log g$ with increasing metallicity for a given effective temperature.

The crossmatch between \citet{Montes2018MNRAS.479.1332M}, Gaia DR3, and CatWISE2020 catalogs resulted in a subsample of 115 M dwarfs among the 192 systems after having eliminated nonphysical pairs \citep{Espada2019} or systems with double-lined spectroscopic binaries. Then, we applied the data filtering mentioned above and constrained to identical values as in \citetalias{Birky2020ApJ...892...31B}, that is, early and mid M dwarfs between M0\,V and M5\,V, and having $1.85$\,mag $\leq G-J\leq$ $3.10$\,mag, $-0.10$\,mag $\leq W1-W2\leq$ $0.24$\,mag, and $-0.5\leq\text{[Fe/H]}\leq 0.5$\,dex, and retrieved a final sample of 46 FGK+M systems to test the calibration.

\subsection{Calibrations, Statistical Analysis, and Model Selection}

\begin{figure*}
    \centering
    \includegraphics[width=0.7\textwidth]{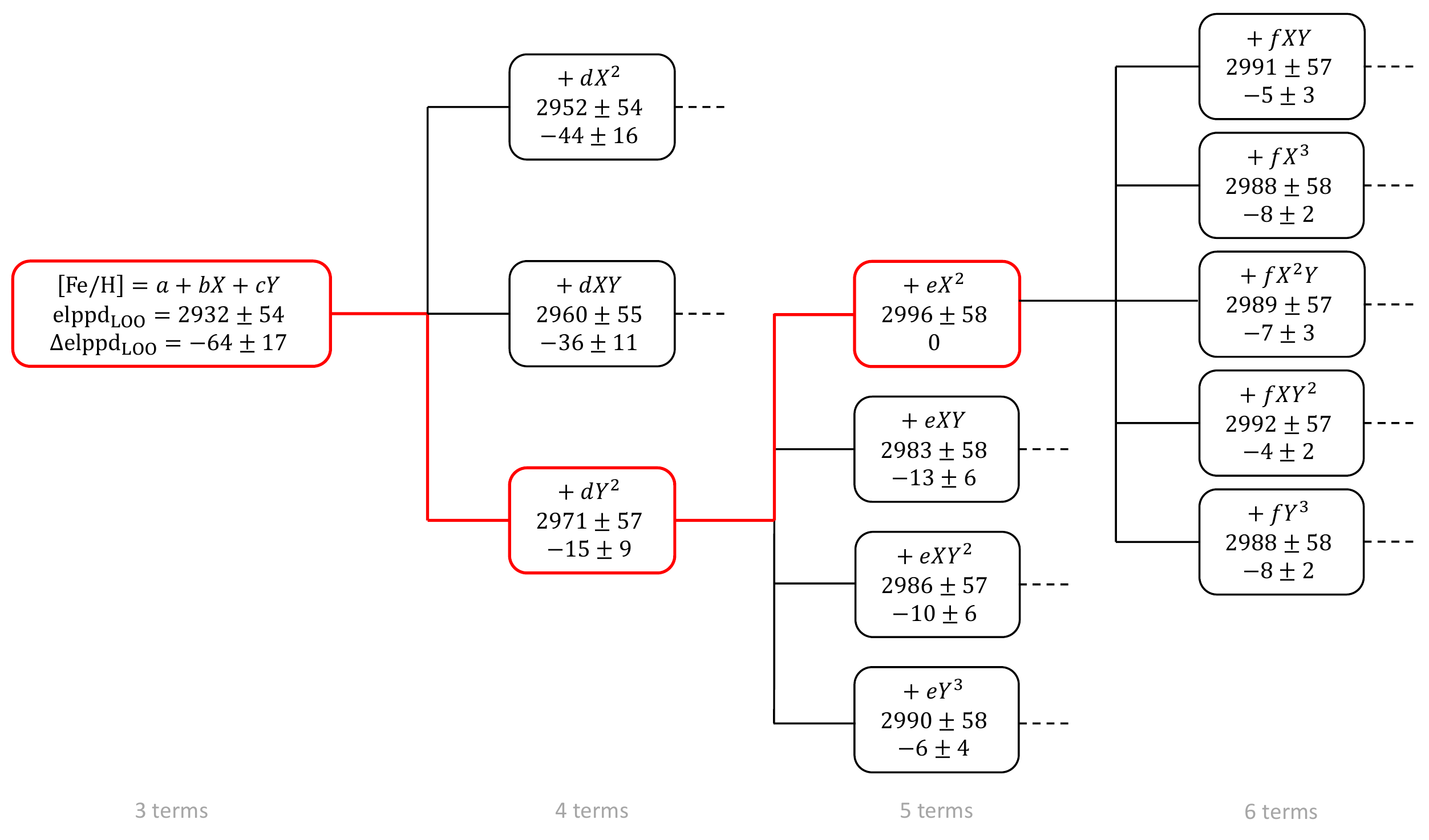}
    \caption{Comparison of different metallicity photometric calibrations with LOO-CV following a stepwise regression (forward selection) procedure, where $X=W1-W2$, $Y=G_\text{BP}-G_\text{RP}$, $\text{elppd}_\text{LOO}$ is the expected log-pointwise predictive density given by Eq.~\ref{loo_eq}, and $\Delta \text{elppd}_\text{LOO}$ is the difference with respect to the model with the largest $\text{elppd}_\text{LOO}$, given by Eq.~\ref{model2b} and marked in red.}
    \label{tree1}
\end{figure*}

We divided the 4919 crossmatched, filtered M dwarfs from \citetalias{Birky2020ApJ...892...31B} into two subsamples:
1000 stars constitute the calibration or training sample, and the remaining 3919 stars are the test sample to check the accuracy of the calibrations.
The $T_\text{eff}$--$\text{[Fe/H]}$ space and their corresponding histograms for both subsamples are shown in Fig.~\ref{Teff_FeH_space}.

The calibrations were derived with MCMC using \texttt{Stan} \citep{Carpenter2017JSS....76....1C} through its R interface, namely \texttt{RStan}. \texttt{Stan} is a C++ library for Bayesian modeling and inference that incorporates, among other components, the Hamiltonian Monte Carlo no-U-turn sampler (HMC$+$NUTS) algorithm. After deriving different calibrations, we compared them with the LOO-CV criterion, which allowed us to choose the calibration that best reproduces the metallicity values by penalizing the more complicated models (i.e. with more free parameters) with respect to the simplest ones. The LOO-CV criterion defines the expected log-pointwise predictive density as:
\begin{equation}
    \text{elppd}_\text{LOO}=\sum_{i=1}^N\log(P(x_i|\pmb{x_{-i}}))
    \label{loo_eq}
\end{equation}
where $P(x_i|\pmb{x_{-i}})$ denotes the probability of predicting $x_i$ using the data without the $i$th observation \citep{Gelman201410.1007/s11222-013-9416-2, Vehtari201710.1007/s11222-016-9696-4}. The value of $\text{elppd}_\text{LOO}$ can be either positive or negative since it uses the probability density, not the probability itself. The model with the largest $\text{elppd}_\text{LOO}$ presents the best predictive accuracy. The computed $\text{elppd}_\text{LOO}$ is defined as the sum of $N$ independent components, so its standard error can be computed as the standard deviation of the $N$ components divided by $\sqrt{N}$. We used the R package \texttt{loo} for implementing the necessary functions and for estimating $\text{elppd}_\text{LOO}$ with the Pareto smoothed importance sampling method \citep{Vehtari2015arXiv150702646V}.

\begin{figure}
       \centering
       \includegraphics[width=0.36\textwidth]{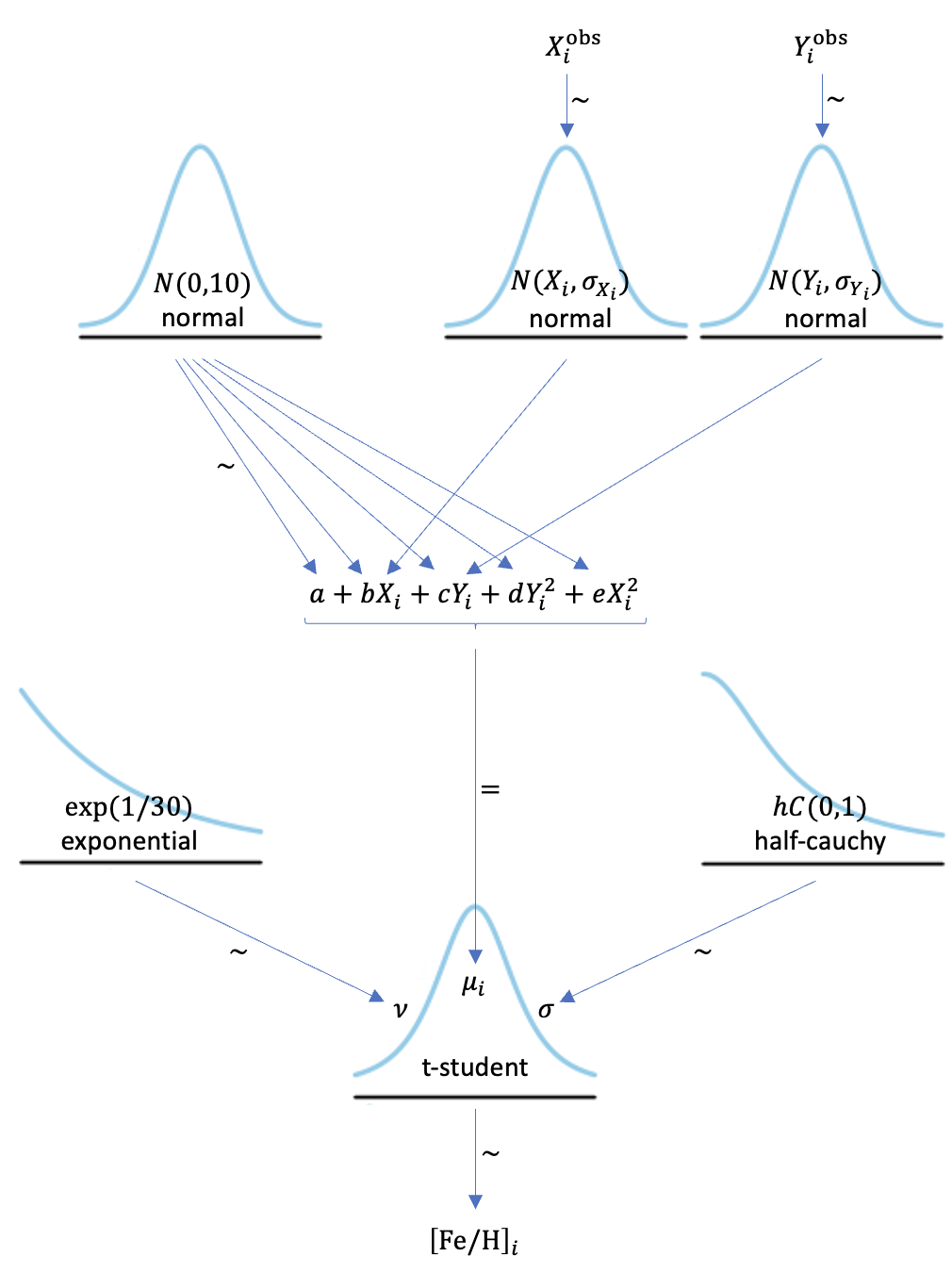}
       \caption{Summary of the distributions used in the Stan model for the robust regression, where $X_i^{\rm obs}$ and $Y_i^{\rm obs}$ indicate the observed values of the predictors, with their observational errors $\sigma_{X_i}$ and $\sigma_{Y_i}$, respectively. Therefore the observed values of the predictors come from a normal distribution (with standard deviations equal to the observed errors) around the `true' values $X_i$ and $Y_i$. In the same way, $\mu$ is the predicted value of the metallicity, which relates to the observed metallicity through a $t$-distribution with parameters $\nu$ (degrees of freedom) and $\sigma$ (scale parameter). For the coefficients we assumed normal priors, while an exponential distribution and a half-Cauchy distribution are used as priors for $\nu$ and $\sigma$, respectively. See \cite{Kruschke2014.book} for more details on the interpretation of this kind of graph.}
       \label{distributions}
\end{figure}

In order to derive the best calibration using the LOO-CV criterion, we tried different calibrations, starting with a linear model and increasingly adding more terms following a stepwise regression procedure (forward selection), shown in Fig.~\ref{tree1} for the $W1-W2$ vs. $G_\text{BP}-G_\text{RP}$ diagram as an example.
For this case, we performed the stepwise regression including up to six terms and found that the model with the best predictive performance (i.e., the largest $\text{elppd}_\text{LOO-CV}$) is given by:
\begin{equation}
   \text{[Fe/H]}=a+bX+cY+dY^2+eX^2
   \label{model2b}
\end{equation}
The distribution of the residuals of the calibrations exhibited extended wings and could not be fitted by a Gaussian distribution. As a result, we used a generalized linear model with a $t$-distribution instead of a Gaussian one to model the corresponding likelihood. Furthermore, this robust regression (with $t$-distribution instead of the Gaussian likelihood) significantly increased the $\text{elppd}_\text{LOO-CV}$. For the robust regression, we used weakly informative priors for the coefficients, that is, $a,b,c,d,e\sim$ \textit{normal}(0,10). The likelihood is given by $\text{[Fe/H]}\sim$ \textit{t-Student}$(\mu,\nu,\sigma)$, where $\mu$ is the expression in Eq.~\ref{model2b}, and the priors for the scale parameter and degrees of freedom are $\sigma \sim$ \textit{half-Cauchy}(0,1) and $\nu\sim$ \textit{exponential}(1/30), respectively.
These are suitable priors for $\sigma$ and $\nu$ since the \textit{half-Cauchy} distribution is a less informative prior than the normal distribution, with heavier tails, and the \textit{exponential}(1/30) distribution captures the behavior of the degrees of freedom in the $t$-distribution, that is, nearly all the variation in the family of $t$-distribution happens when $\nu$ is fairly small and for $\nu>30$ the $t$-distribution is essentially normal. Therefore, since $\nu=30$ is the mean of the \textit{exponential}(1/30) distribution, with this prior we give the same weights to the low and high regimes of $\nu$.
We provide a graphical representation of the model in Fig.~\ref{distributions} (see \citealt{Kruschke2014.book}).

\begin{table*}
\scriptsize
\centering
\caption{Fit Parameters of Color--Color Diagram Calibrations\tablenotemark{a}}
\begin{tabular}{ll ccccc cc cc} 
\hline\hline
\noalign{\smallskip}
$X$ & $Y$ & $a$ & $b$ & $c$ & $d$ & $e$ & $\sigma$ & $\nu$ & $\text{elppd}_\text{LOO}$ & $\Delta\text{elppd}_\text{LOO}$ \\ 
(mag) & (mag) & (dex) & (mag$^{-1}$) & (mag$^{-1}$) & (mag$^{-2}$) & (mag$^{-2}$) & (dex) \\
\noalign{\smallskip}
\hline
\noalign{\smallskip}
$W1-W2$ & $G_\text{BP}-G_\text{RP}$ & $-2.72\pm0.18$ & $-3.96\pm0.13$ & $1.91\pm0.16$ & $-0.253\pm 0.036$ & $0.65\pm 0.90$ & $0.0960\pm0.0049$ & $22.32\pm19.58$ & $2996\pm 58$ & $0$ \\
$W1-W2$ & $G-H$ & $-4.92\pm0.47$ & $-3.54\pm0.11$ & $2.49\pm0.30$ & $-0.266\pm 0.049$ & $2.07\pm 0.79$ & $0.0875\pm0.0050$ & $8.27\pm4.13$ & $2974\pm 61$ & $-21\pm 26$ \\
$W1-W2$ & $G-K_s$ & $-4.39\pm0.43$ & $-3.71\pm0.12$ & $2.03\pm0.26$ & $-0.189\pm 0.040$ & $2.45\pm 0.80$ & $0.0875\pm0.0048$ & $8.62\pm4.81$ & $2965\pm 62$ & $-31\pm25$ \\
$W1-W2$ & $G-J$ & $-3.32\pm0.29$ & $-3.87\pm0.13$ & $2.06\pm0.24$ & $-0.243\pm 0.049$ & $1.11\pm 0.88$ & $0.0953\pm0.0052$ & $15.77\pm13.79$ & $2891\pm 59$ & $-105\pm20$ \\
$W1-W2$ & $G_\text{RP}-K_s$ & $-4.93\pm0.43$ & $-3.52\pm0.12$ & $3.50\pm0.38$ & $-0.536\pm 0.085$ & $3.05\pm 0.79$ & $0.0863\pm0.0048$ & $7.19\pm2.81$ & $2882\pm 64$ & $-113\pm33$ \\
$W1-W2$ & $G_\text{RP}-H$ & $-5.70\pm0.49$ & $-3.20\pm0.11$ & $4.60\pm0.48$ & $-0.83\pm 0.12$ & $2.34\pm 0.80$ & $0.0878\pm0.0049$ & $7.35\pm2.80$ & $2873\pm 62$ & $-123\pm36$ \\
$W1-W2$ & $G_\text{RP}-J$ & $-2.87\pm0.22$ & $-3.70\pm0.13$ & $3.23\pm0.32$ & $-0.70\pm 0.11$ & $1.21\pm 0.90$ & $0.0995\pm0.0052$ & $19.36\pm17.28$ & $2765\pm 59$ & $-231\pm29$ \\
$W1-W2$ & $G-W1$ & $-3.59\pm0.44$ & $-3.43\pm0.14$ & $1.56\pm0.26$ & $-0.131\pm 0.037$ & $2.06\pm 0.87$ & $0.1024\pm0.0055$ & $13.12\pm10.73$ & $2720\pm 58$ & $-276\pm18$ \\
$W1-W2$ & $G-W2$ & $-3.14\pm0.37$ & $-4.14\pm0.16$ & $1.28\pm0.21$ & $-0.089\pm 0.030$ & $2.52\pm 0.93$ & $0.1001\pm0.0059$ & $13.44\pm11.48$ & $2696\pm 60$ & $-300\pm19$ \\
$W1-W2$ & $G_\text{RP}-W1$ & $-3.72\pm0.43$ & $-3.20\pm0.14$ & $2.44\pm0.36$ & $-0.327\pm 0.076$ & $2.46\pm 0.91$ & $0.1061\pm0.0056$ & $11.97\pm8.70$ & $2603\pm 57$ & $-393\pm23$ \\
$W1-W2$ & $G_\text{RP}-W2$ & $-3.00\pm0.35$ & $-4.15\pm0.17$ & $1.80\pm0.29$ & $-0.188\pm 0.059$ & $3.16\pm 0.96$ & $0.1036\pm0.0060$ & $12.48\pm10.09$ & $2568\pm 59$ & $-428\pm24$ \\
\noalign{\smallskip}
 \hline
\end{tabular}
\tablenotetext{a}{The polynomial fits the expression $\text{[Fe/H]}=a+bX+cY+dY^2+eX^2$.}
\label{color_color_coef}
\end{table*}

A Bayesian approach to error measurement can be formulated by treating the true quantities being measured as missing data or latent variables, needing a model of how the measurements are derived from the true values. We can suppose that the `true' values of a predictor $A_i$, for example the magnitude in a given photometric band, are not known, but for each $i$ star, a measurement $A_i^{\rm obs}$ of $A_i$ is available. Then the approach is to assume that the measured values arise from a normal distribution with a mean equal to the true value and some measurement error $\sigma_{A_i}$, that is, $A_i^{\rm obs}\sim normal(A_i,\sigma_{A_i})$. Therefore, the regression is not performed with the measured values $A_i^{\rm obs}$ but with the true ones $A_i$ (see Fig.~\ref{distributions}).

Finally, to ensure the convergence of the MCMC chains we applied the Gelman--Rubin diagnostic or shrink factor $\hat{R}$ \citep{Gelman199210.1214/ss/1177011136, Brooks1998doi:10.1080/10618600.1998.10474787}, which
analyzes the difference between multiple Markov chains. The convergence is assessed by comparing the estimated between-chains and within-chain variances for each model parameter. If $\hat{R}<1.1$ for all parameters, one can be confident that convergence has been reached. In our case, we obtained a shrink factor of $\hat{R}=1.0$ for all parameters.
We typically run three simultaneous chains with 3000 steps in each of them and 500 warm-up iterations.

\section{Results and discussion}
\label{results}

\subsection{Color--Color Diagrams}
\label{results_color_color}

Color--color diagrams allow us to compare apparent magnitudes at different wavelengths. In these diagrams the chemical composition plays a fundamental role, showing a gradient of metallicity that is more noticeable for cool stars \citep{Gaia2018A&A...616A..10G}. In particular, the $W1-W2$ color index constitutes an appropriate metallicity indicator \citep{Schmidt2016MNRAS.460.2611S}.

\begin{figure*}
    \centering
    \begin{tabular}{cc}
        \includegraphics[width=0.35\textwidth]{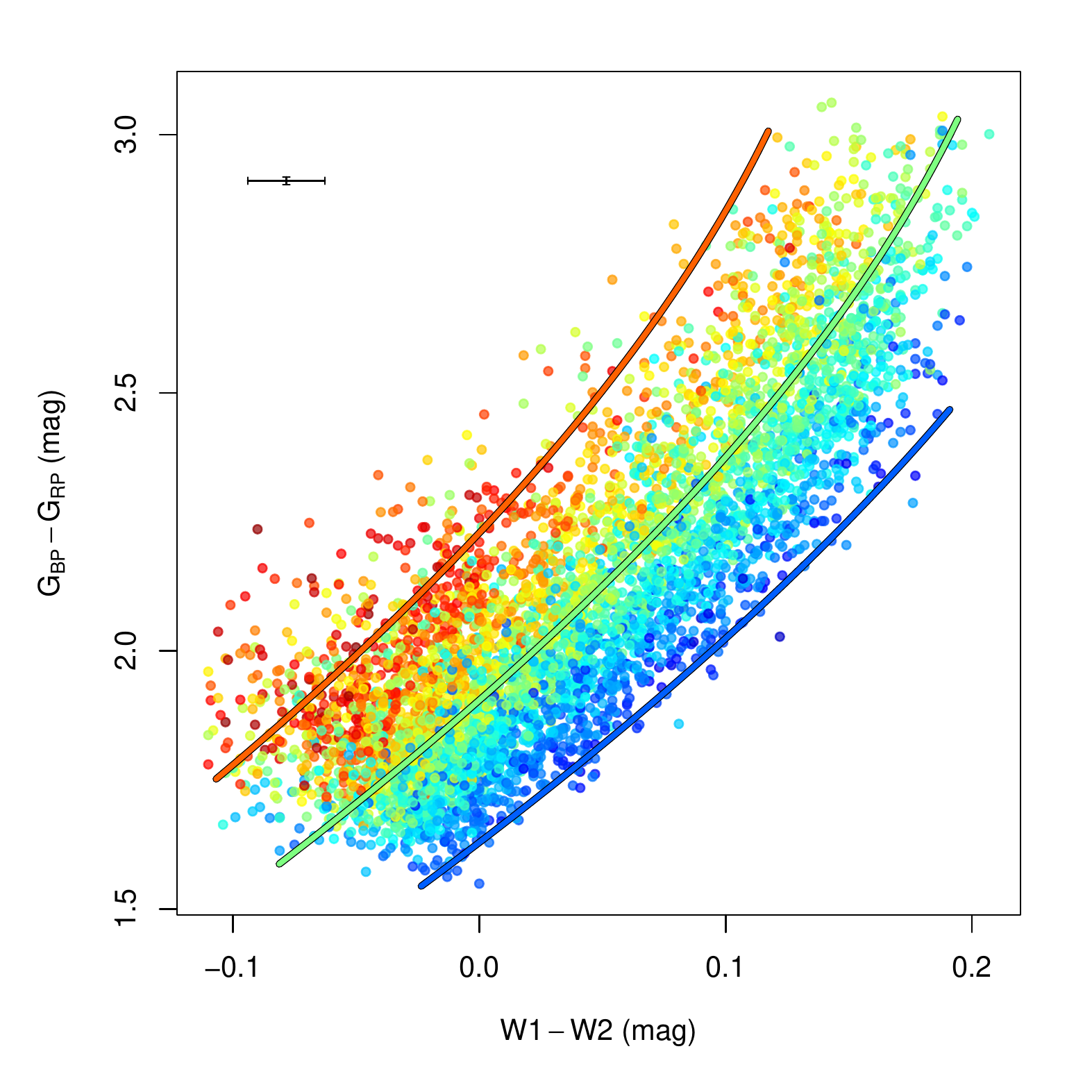} & \includegraphics[width=0.28\textwidth]{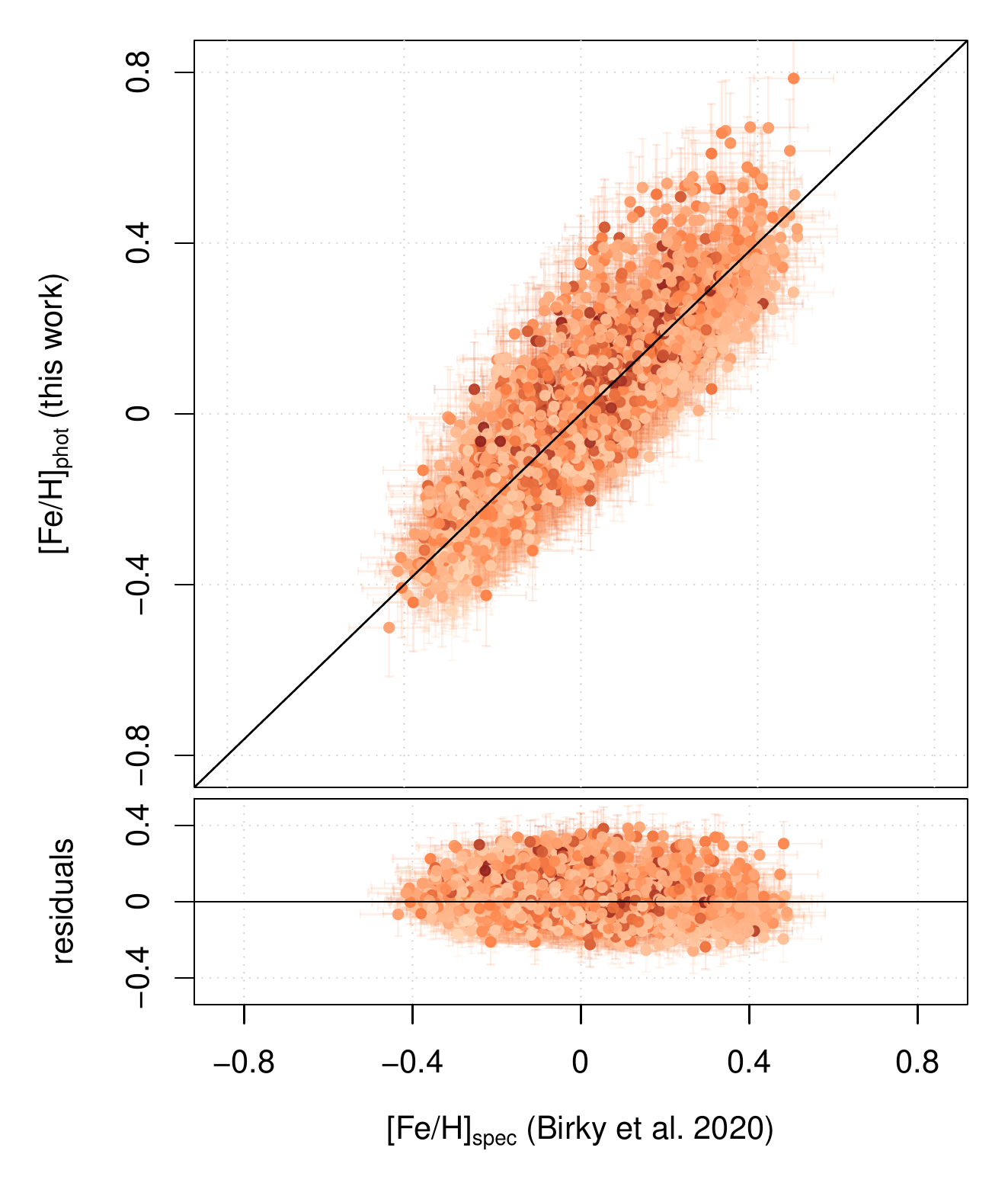} \\
        \includegraphics[width=0.35\textwidth]{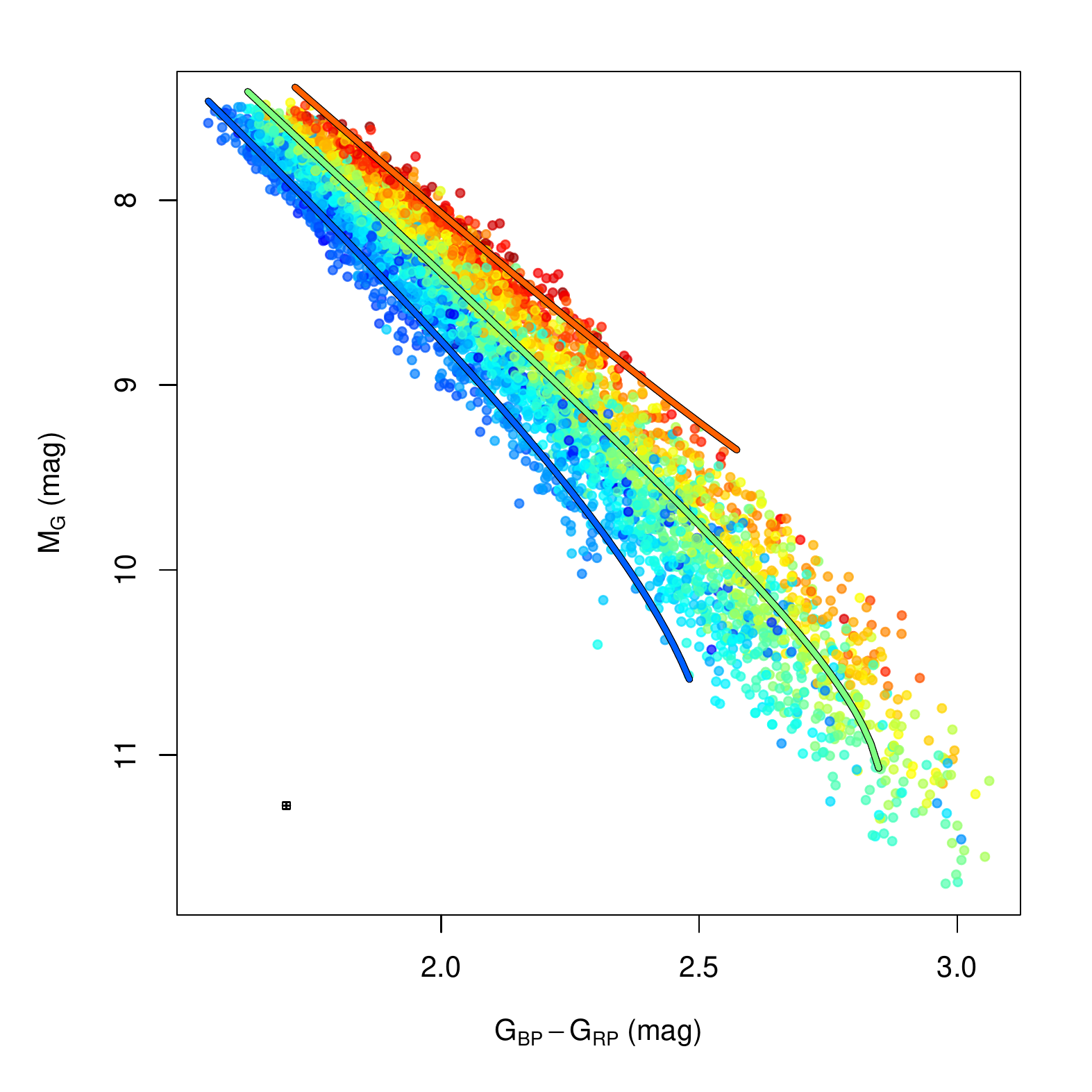} & \includegraphics[width=0.28\textwidth]{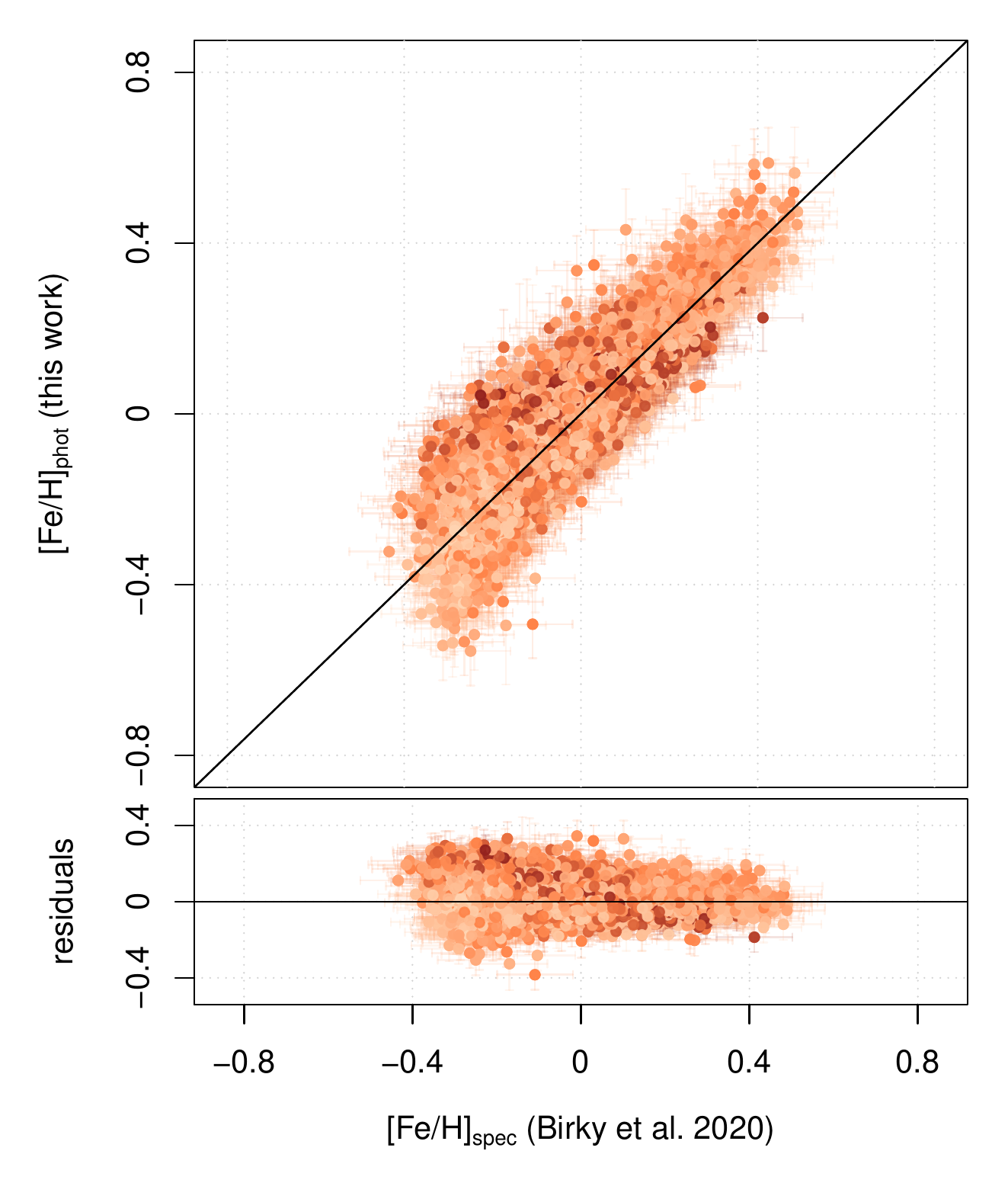} \\
        \includegraphics[width=0.35\textwidth]{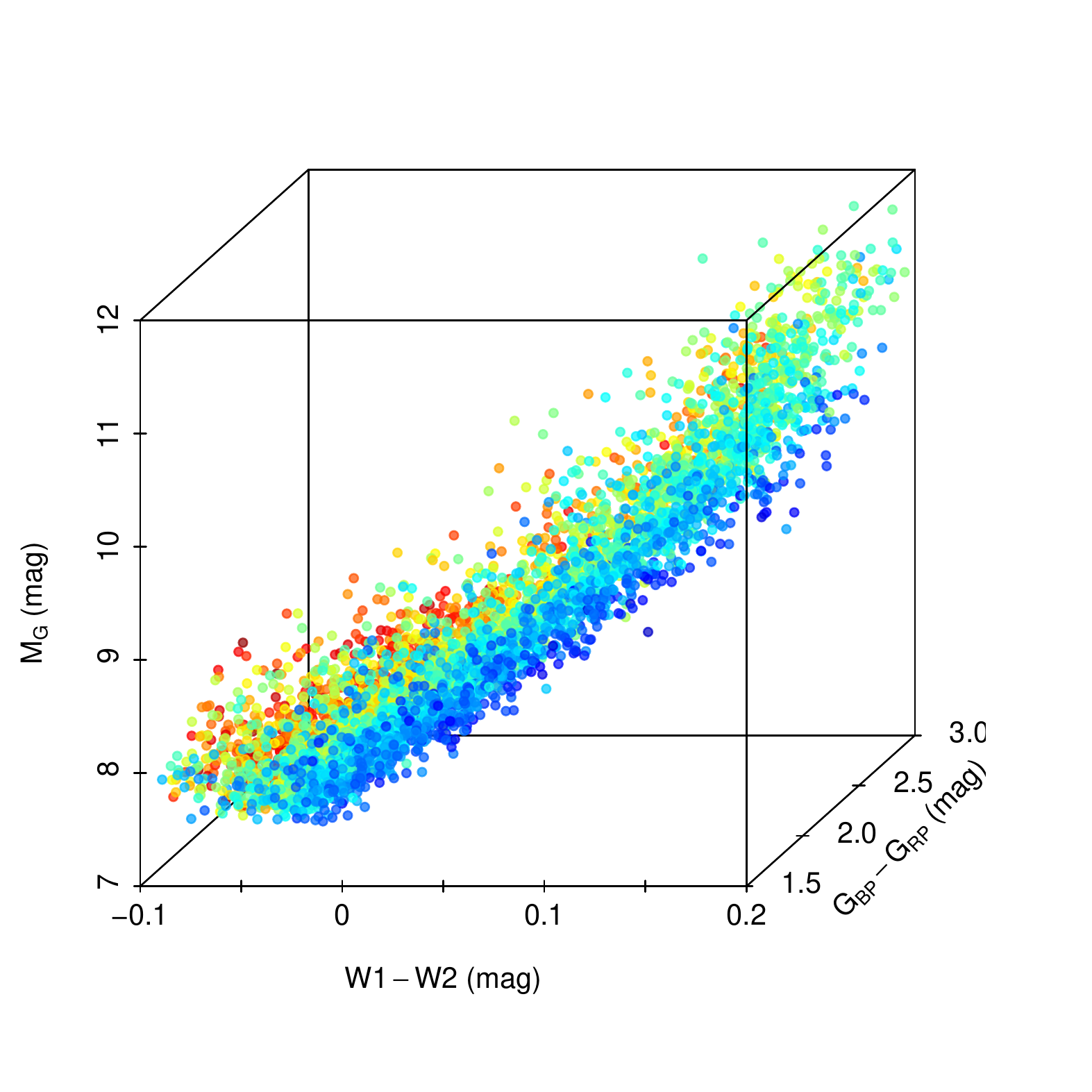} & \includegraphics[width=0.28\textwidth]{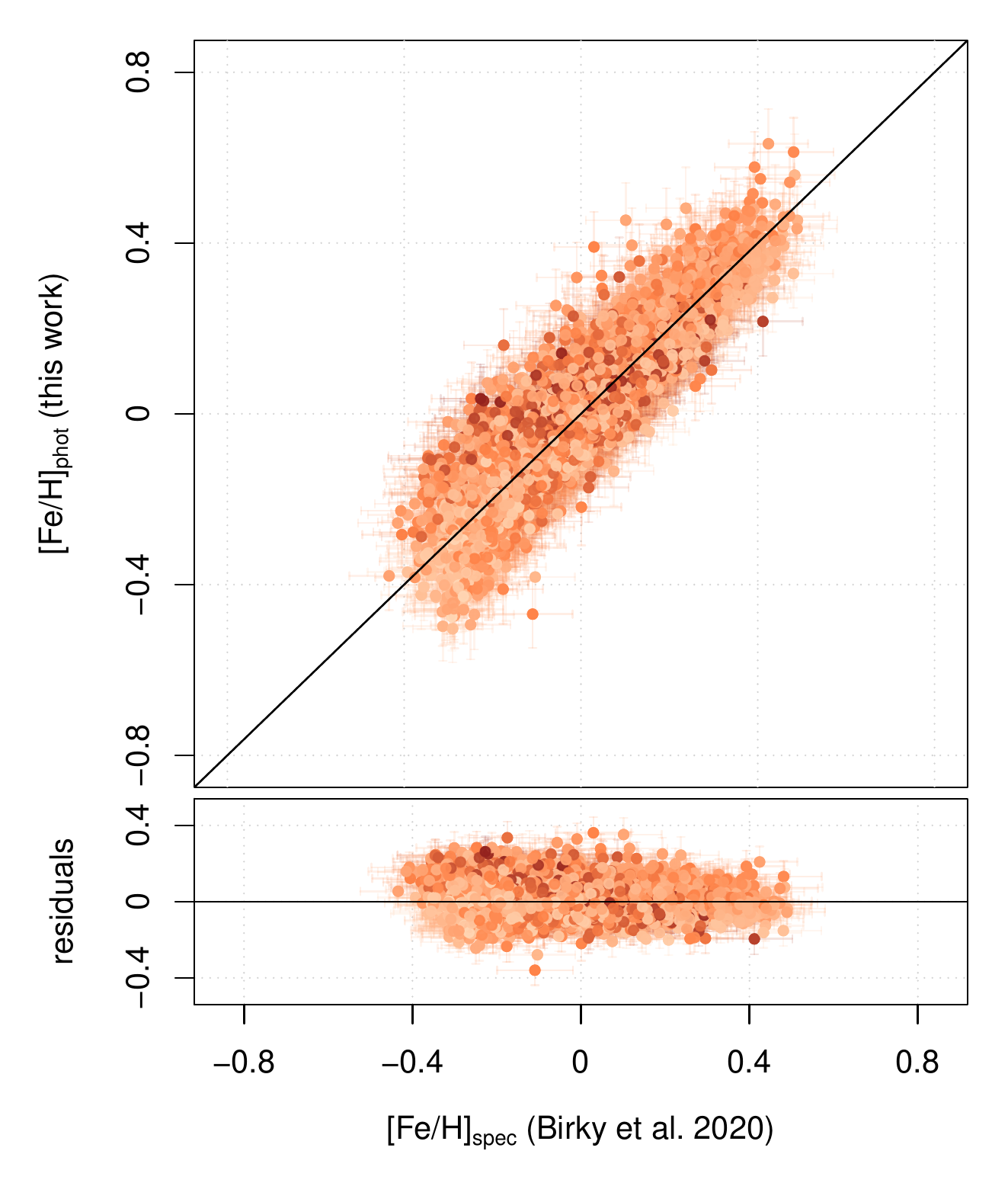} \\
        \includegraphics[width=0.24\textwidth]{metallicity_code.png} & \includegraphics[width=0.24\textwidth]{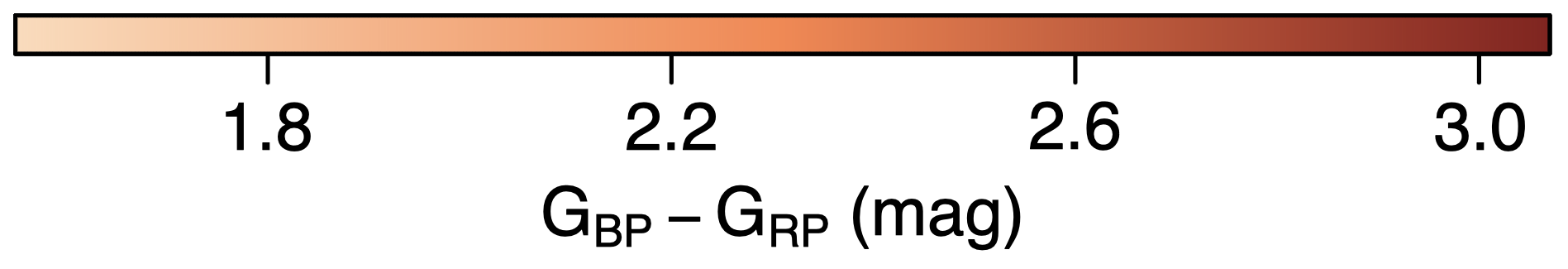}
    \end{tabular}
    \caption{{\em Left panels}: from top to bottom, representative examples of color--color, color--magnitude, and color--color--magnitude diagrams of the stars from \citetalias{Birky2020ApJ...892...31B}, color-coded by [Fe/H]$_\text{spec}$, with the respective calibrations given by this work (red: $\text{[Fe/H]} = 0.3$\,dex, green: $\text{[Fe/H]} = 0.0$\,dex, blue: $\text{[Fe/H]} = -0.3$\,dex). The mean uncertainties of the color indices and absolute magnitudes involved are also shown with a black dot.
    {\em Right panels}: comparison between the spectroscopic metallicity values reported by \citetalias{Birky2020ApJ...892...31B} and the photometric estimations with the corresponding diagrams in the left for the 3919 stars from the test subsample, and respective residuals, color-coded by $G_\text{BP}-G_\text{RP}$ color index (darker symbols: cooler stars; lighter symbols: warmer stars). The solid lines denote the 1:1 relationship and the residuals $=0$.}
    \label{diagrams_comparison}
\end{figure*}

In Table~\ref{color_color_coef} we present the posterior mean and standard deviation of the coefficients of the calibration given by Eq. \ref{model2b} for different color--color combinations, the scale parameter and degrees of freedom of the Student's $t$-distribution, and their comparison with LOO-CV, sorted from the best to the worst combination according to their elppd value. The posterior distributions of the coefficients are Gaussian.
All these calibrations fit the observed metallicities with a residual standard deviation of $0.08$--$0.09$\,dex, of the same order as the [Fe/H] uncertainty provided by \citetalias{Birky2020ApJ...892...31B}, which means we are performing our regression until the observational limit.
An example of color--color diagram is shown in the top left panel of Fig.~\ref{diagrams_comparison}. The remaining color--color diagram combinations are available in Fig.~\ref{color_color_diagrams_appendix}.

In the top right panels of Fig.~\ref{diagrams_comparison} 
we also represent the [Fe/H] values reported by \citetalias{Birky2020ApJ...892...31B} versus the estimated values using the $W1-W2$ vs. $G_\text{BP}-G_\text{RP}$ color--color diagram, and the corresponding residuals, for the 3919 stars from the test subsample, color-coded by $G_\text{BP}-G_\text{RP}$ color index. 
We conclude that most of the estimated metallicities follow the one-to-one relationship and that there is no correlation between them and $T_{\rm eff}$, as expected. Some photometric metallicities are more than $2\sigma$ above or below the spectroscopic value given by \citetalias{Birky2020ApJ...892...31B}, but they are just 112 stars, $2.9\,\%$ of the test sample.

\subsection{Color--Magnitude Diagrams}

Color--magnitude or Hertzsprung--Russell diagrams represent the absolute magnitude or luminosity versus the color index, spectral type, or $T_{\rm eff}$. The position of a star in these diagrams is mainly given by its initial mass, chemical composition, and age, but other effects such as rotation, stellar winds, or magnetic fields also play a role \citep{Gaia2018A&A...616A..10G}. Similar to what is observed in our color--color diagrams, color--magnitude diagrams also present a metallicity gradient.

We proceeded with the analysis as in the previous section. In Table~\ref{color_magnitude_coef} we present the coefficients of the calibration given by Eq.~\ref{model2b} for different color--magnitude combinations and their comparison using the LOO-CV criterion.

In the central panels of Fig.~\ref{diagrams_comparison}, we represent the color--magnitude diagram $G_\text{BP}-G_\text{RP}$ vs. $M_{G}$ of the 4919 stars from \citetalias{Birky2020ApJ...892...31B} (left) and the comparison between the spectroscopic [Fe/H] values and the corresponding estimated values for the 3\,919 stars from the test subsample (right). Again, the rest of color--magnitude diagrams can be found in Figs.~\ref{color_magnitude_diagrams_appendix_G}, \ref{color_magnitude_diagrams_appendix_J}, \ref{color_magnitude_diagrams_appendix_H}, and \ref{color_magnitude_diagrams_appendix_K}.
In the case of the $G_\text{BP}-G_\text{RP}$ vs. $M_G$ color--magnitude diagram, 237 stars ($6.0\,\%$) are more than $2\sigma$ above or below the spectroscopic value.


\begin{table*}
\scriptsize
\centering
\caption{Fit Parameters of Color--Magnitude Diagram Calibrations\tablenotemark{a}}
\begin{tabular}{ll ccccc cccc} 
\hline
\hline
\noalign{\smallskip}
$X$ & $Y$ & $a$ & $b$ & $c$ & $d$ & $e$ & $\sigma$ & $\nu$ & $\text{elppd}_\text{LOO}$ & $\Delta\text{elppd}_\text{LOO}$ \\ 
(mag) & (mag) & (dex) & (mag$^{-1}$) & (mag$^{-1}$) & (mag$^{-2}$) & (mag$^{-2}$) & (dex) \\
\noalign{\smallskip}
\hline
\noalign{\smallskip}

$G_\text{BP}-G_\text{RP}$ & $M_{G}$ & $8.89\pm0.36$ & $5.89\pm0.24$ & $-3.22\pm0.13$ & $0.1432\pm0.0068$ & $-0.933\pm0.054$ & $0.0706\pm0.0033$ & $4.20\pm0.71$ & $3611\pm 61$ & $0$ \\
$G_\text{BP}-G_\text{RP}$ & $M_{K_s}$ & $4.75\pm0.22$ & $2.93\pm0.16$ & $-2.63\pm0.11$ & $0.1790\pm0.0090$ & $-0.416\pm0.036$ & $0.0718\pm0.0034$ & $4.43\pm0.79$ & $3519\pm 62$ & $-92\pm24$ \\
$G_\text{BP}-G_\text{RP}$ & $M_{H}$ & $4.44\pm0.22$ & $3.19\pm0.17$ & $-2.54\pm0.10$ & $0.1660\pm0.0083$ & $-0.468\pm0.037$ & $0.0710\pm0.0035$ & $4.08\pm0.70$ & $3491\pm 62$ & $-120\pm22$ \\
$G_\text{BP}-G_\text{RP}$ & $M_{J}$ & $7.34\pm0.31$ & $3.49\pm0.17$ & $-3.19\pm0.13$ & $0.1953\pm0.0094$ & $-0.525\pm0.039$ & $0.0725\pm0.0035$ & $4.25\pm0.75$ & $3460\pm 62$ & $-151\pm18$ \\
$G-W1$ & $M_{K_s}$ & $2.25\pm0.32$ & $2.39\pm0.23$ & $-2.29\pm0.10$ & $0.1559\pm0.0084$ & $-0.209\pm0.034$ & $0.0761\pm0.0034$ & $4.32\pm0.70$ & $3399\pm 63$ & $-212\pm50$ \\
$G-W1$ & $M_{H}$ & $1.63\pm0.32$ & $2.70\pm0.24$ & $-2.19\pm0.10$ & $0.1430\pm0.0081$ & $-0.250\pm0.036$ & $0.0766\pm0.0035$ & $4.25\pm0.68$ & $3372\pm 63$ & $-239\pm49$ \\
$G-W1$ & $M_{J}$ & $3.91\pm0.34$ & $2.94\pm0.25$ & $-2.72\pm0.12$ & $0.1665\pm0.0089$ & $-0.279\pm0.037$ & $0.0773\pm0.0035$ & $4.38\pm0.74$ & $3367\pm 62$ & $-244\pm47$ \\
$G-K_s$ & $M_{H}$ & $1.16\pm0.35$ & $2.84\pm0.27$ & $-2.072\pm0.098$ & $0.1356\pm0.0079$ & $-0.276\pm0.041$ & $0.0773\pm0.0037$ & $4.66\pm0.85$ & $3351\pm 63$ & $-260\pm46$ \\
$G-K_s$ & $M_{J}$ & $3.24\pm0.38$ & $3.05\pm0.28$ & $-2.55\pm0.12$ & $0.1556\pm0.0089$ & $-0.303\pm0.043$ & $0.0797\pm0.0038$ & $5.26\pm1.11$ & $3323\pm 62$ & $-288\pm44$ \\
$G-H$ & $M_{K_s}$ & $0.65\pm0.40$ & $3.09\pm0.30$ & $-2.009\pm0.099$ & $0.1366\pm0.0082$ & $-0.331\pm0.049$ & $0.0768\pm0.0038$ & $4.30\pm0.78$ & $3311\pm 63$ & $-300\pm45$ \\
$G-W1$ & $M_{G}$ & $2.64\pm0.34$ & $4.73\pm0.34$ & $-2.42\pm0.12$ & $0.1064\pm0.0063$ & $-0.486\pm0.049$ & $0.0807\pm0.0035$ & $4.57\pm0.74$ & $3270\pm 61$ & $-341\pm52$ \\
$G-K_s$ & $M_{K_s}$ & $1.80\pm0.37$ & $2.39\pm0.27$ & $-2.10\pm0.10$ & $0.1427\pm0.0087$ & $-0.215\pm0.042$ & $0.0817\pm0.0038$ & $5.39\pm1.14$ & $3268\pm 62$ & $-343\pm48$ \\
$G-K_s$ & $M_{G}$ & $1.69\pm0.37$ & $4.93\pm0.36$ & $-2.24\pm0.11$ & $0.0984\pm0.0059$ & $-0.530\pm0.054$ & $0.0796\pm0.0039$ & $5.38\pm1.20$ & $3256\pm 63$ & $-355\pm48$ \\
$G-H$ & $M_{J}$ & $1.89\pm0.41$ & $3.66\pm0.31$ & $-2.35\pm0.12$ & $0.1431\pm0.0086$ & $-0.413\pm0.051$ & $0.0785\pm0.0040$ & $4.43\pm0.86$ & $3252\pm 63$ & $-360\pm42$ \\
$G-J$ & $M_{K_s}$ & $3.42\pm0.27$ & $2.84\pm0.25$ & $-2.36\pm0.11$ & $0.1600\pm0.0092$ & $-0.351\pm0.052$ & $0.0825\pm0.0041$ & $5.33\pm1.22$ & $3245\pm 62$ & $-366\pm40$ \\
$G-W2$ & $M_{K_s}$ & $4.43\pm0.29$ & $1.95\pm0.21$ & $-2.72\pm0.12$ & $0.185\pm0.010$ & $-0.143\pm0.031$ & $0.0797\pm0.0037$ & $4.20\pm0.69$ & $3241\pm 63$ & $-370\pm46$ \\
$G_\text{RP}-W1$ & $M_{K_s}$ & $1.49\pm0.33$ & $3.82\pm0.33$ & $-2.16\pm0.10$ & $0.1481\pm0.0085$ & $-0.531\pm0.070$ & $0.0785\pm0.0034$ & $4.19\pm0.63$ & $3223\pm 64$ & $-388\pm59$ \\
$G-W2$ & $M_{H}$ & $3.86\pm0.28$ & $2.30\pm0.22$ & $-2.65\pm0.12$ & $0.1733\pm0.0095$ & $-0.189\pm0.032$ & $0.0794\pm0.0037$ & $3.93\pm0.61$ & $3205\pm 63$ & $-406\pm45$ \\
$G-J$ & $M_{H}$ & $2.97\pm0.26$ & $3.18\pm0.25$ & $-2.27\pm0.11$ & $0.1481\pm0.0087$ & $-0.414\pm0.052$ & $0.0821\pm0.0042$ & $4.91\pm1.08$ & $3203\pm 62$ & $-408\pm39$ \\
$G_\text{RP}-W1$ & $M_{H}$ & $0.91\pm0.32$ & $4.20\pm0.34$ & $-2.057\pm0.099$ & $0.1352\pm0.0080$ & $-0.607\pm0.071$ & $0.0792\pm0.0035$ & $4.15\pm0.63$ & $3193\pm 64$ & $-418\pm58$ \\
$G_\text{RP}-W1$ & $M_{J}$ & $3.04\pm0.36$ & $4.51\pm0.36$ & $-2.55\pm0.12$ & $0.1566\pm0.0089$ & $-0.661\pm0.077$ & $0.0797\pm0.0035$ & $4.25\pm0.65$ & $3186\pm 63$ & $-426\pm56$ \\
$G-H$ & $M_{H}$ & $0.13\pm0.40$ & $3.26\pm0.31$ & $-1.862\pm0.095$ & $0.1212\pm0.0077$ & $-0.358\pm0.051$ & $0.0804\pm0.0040$ & $4.45\pm0.83$ & $3167\pm 62$ & $-445\pm46$ \\
$G-W2$ & $M_{J}$ & $6.75\pm0.35$ & $2.57\pm0.23$ & $-3.32\pm0.15$ & $0.203\pm0.011$ & $-0.221\pm0.034$ & $0.0816\pm0.0039$ & $4.17\pm0.71$ & $3158\pm 62$ & $-453\pm42$ \\
$G-H$ & $M_{G}$ & $-0.02\pm0.41$ & $5.52\pm0.40$ & $-1.97\pm0.10$ & $0.0857\pm0.0055$ & $-0.656\pm0.064$ & $0.0788\pm0.0042$ & $4.45\pm0.87$ & $3154\pm 63$ & $-457\pm46$ \\
$G_\text{RP}-K_s$ & $M_{H}$ & $0.16\pm0.38$ & $4.53\pm0.41$ & $-1.873\pm0.097$ & $0.1228\pm0.0078$ & $-0.699\pm0.090$ & $0.0785\pm0.0039$ & $4.49\pm0.80$ & $3137\pm 65$ & $-474\pm54$ \\
$G_\text{RP}-K_s$ & $M_{J}$ & $2.00\pm0.40$ & $4.80\pm0.42$ & $-2.29\pm0.12$ & $0.1398\pm0.0087$ & $-0.749\pm0.094$ & $0.0810\pm0.0040$ & $5.07\pm1.06$ & $3104\pm 64$ & $-507\pm53$ \\
$G_\text{RP}-K_s$ & $M_{K_s}$ & $0.79\pm0.39$ & $3.87\pm0.40$ & $-1.89\pm0.10$ & $0.1289\pm0.0086$ & $-0.568\pm0.090$ & $0.0834\pm0.0039$ & $5.11\pm0.99$ & $3055\pm 63$ & $-556\pm56$ \\
$G_\text{RP}-W2$ & $M_{K_s}$ & $4.68\pm0.28$ & $2.80\pm0.29$ & $-2.77\pm0.13$ & $0.189\pm0.011$ & $-0.306\pm0.060$ & $0.0824\pm0.0037$ & $3.99\pm0.60$ & $3047\pm 64$ & $-564\pm53$ \\
$G_\text{RP}-H$ & $M_{K_s}$ & $-0.86\pm0.43$ & $5.28\pm0.47$ & $-1.705\pm0.096$ & $0.1156\pm0.0080$ & $-0.94\pm0.12$ & $0.0793\pm0.0041$ & $4.23\pm0.77$ & $3046\pm 64$ & $-565\pm53$ \\
$G-J$ & $M_{J}$ & $5.22\pm0.33$ & $3.27\pm0.27$ & $-2.70\pm0.13$ & $0.1646\pm0.0099$ & $-0.433\pm0.056$ & $0.0905\pm0.0045$ & $6.29\pm1.83$ & $3035\pm 60$ & $-576\pm40$ \\
$G-J$ & $M_{G}$ & $5.12\pm0.33$ & $5.59\pm0.37$ & $-2.57\pm0.13$ & $0.1131\pm0.0069$ & $-0.797\pm0.074$ & $0.0879\pm0.0047$ & $6.03\pm2.06$ & $3024\pm 61$ & $-587\pm40$ \\
$G-W2$ & $M_{G}$ & $6.96\pm0.37$ & $4.76\pm0.34$ & $-3.32\pm0.16$ & $0.1469\pm0.0084$ & $-0.457\pm0.049$ & $0.0859\pm0.0039$ & $4.10\pm0.64$ & $3012\pm 60$ & $-599\pm46$ \\
$G_\text{RP}-W2$ & $M_{H}$ & $4.16\pm0.28$ & $3.27\pm0.30$ & $-2.70\pm0.13$ & $0.177\pm0.010$ & $-0.395\pm0.061$ & $0.0825\pm0.0039$ & $3.80\pm0.57$ & $3005\pm 64$ & $-606\pm52$ \\
$G_\text{RP}-W1$ & $M_{G}$ & $1.49\pm0.34$ & $6.74\pm0.46$ & $-2.13\pm0.11$ & $0.0939\pm0.0061$ & $-1.042\pm0.096$ & $0.0845\pm0.0036$ & $4.24\pm0.63$ & $2985\pm 62$ & $-626\pm63$ \\
$G_\text{RP}-H$ & $M_{J}$ & $0.13\pm0.45$ & $5.92\pm0.50$ & $-1.94\pm0.11$ & $0.1177\pm0.0082$ & $-1.08\pm0.12$ & $0.0815\pm0.0044$ & $4.44\pm0.87$ & $2977\pm 64$ & $-634\pm51$ \\
$G_\text{RP}-W2$ & $M_{J}$ & $7.13\pm0.35$ & $3.59\pm0.32$ & $-3.37\pm0.16$ & $0.207\pm0.012$ & $-0.448\pm0.066$ & $0.0850\pm0.0040$ & $4.04\pm0.65$ & $2947\pm 63$ & $-664\pm49$ \\
$G_\text{RP}-J$ & $M_{K_s}$ & $3.58\pm0.26$ & $4.23\pm0.35$ & $-2.17\pm0.12$ & $0.1468\pm0.0097$ & $-0.94\pm0.13$ & $0.0895\pm0.0046$ & $6.22\pm1.81$ & $2935\pm 62$ & $-676\pm47$ \\
$G_\text{RP}-K_s$ & $M_{G}$ & $0.15\pm0.39$ & $7.11\pm0.51$ & $-1.87\pm0.11$ & $0.0816\pm0.0056$ & $-1.17\pm0.11$ & $0.0806\pm0.0042$ & $4.78\pm0.95$ & $2917\pm 65$ & $-695\pm59$ \\
$G_\text{RP}-H$ & $M_{H}$ & $-1.25\pm0.45$ & $5.36\pm0.50$ & $-1.547\pm0.095$ & $0.1001\pm0.0077$ & $-0.96\pm0.12$ & $0.0841\pm0.0044$ & $4.53\pm0.89$ & $2901\pm 63$ & $-710\pm54$ \\
$G_\text{RP}-J$ & $M_{H}$ & $3.23\pm0.26$ & $4.57\pm0.37$ & $-2.06\pm0.11$ & $0.1338\pm0.0091$ & $-1.05\pm0.13$ & $0.0903\pm0.0048$ & $6.02\pm1.78$ & $2884\pm 62$ & $-727\pm47$ \\
$G_\text{RP}-H$ & $M_{G}$ & $-1.98\pm0.46$ & $8.06\pm0.62$ & $-1.467\pm0.096$ & $0.0628\pm0.0051$ & $-1.51\pm0.15$ & $0.0828\pm0.0047$ & $4.40\pm0.88$ & $2778\pm 65$ & $-833\pm56$ \\
$G_\text{RP}-J$ & $M_{J}$ & $5.20\pm0.36$ & $4.48\pm0.40$ & $-2.38\pm0.14$ & $0.144\pm0.011$ & $-1.03\pm0.14$ & $0.1007\pm0.0051$ & $8.68\pm4.77$ & $2701\pm 59$ & $-910\pm47$ \\
$G_\text{RP}-W2$ & $M_{G}$ & $7.32\pm0.41$ & $6.36\pm0.49$ & $-3.24\pm0.18$ & $0.1430\pm0.0093$ & $-0.877\pm0.098$ & $0.0928\pm0.0041$ & $3.94\pm0.57$ & $2643\pm 61$ & $-968\pm56$ \\
$G_\text{RP}-J$ & $M_{G}$ & $5.03\pm0.39$ & $7.00\pm0.54$ & $-2.08\pm0.14$ & $0.0897\pm0.0074$ & $-1.70\pm0.19$ & $0.1019\pm0.0058$ & $9.45\pm6.34$ & $2530\pm 60$ & $-1082\pm50$ \\

\noalign{\smallskip}
\hline
\end{tabular}
\tablenotetext{a}{The polynomial fits follow the expression $\text{[Fe/H]}=a+bX+cY+dY^2+eX^2$.}
\label{color_magnitude_coef}
\end{table*}

\begin{table*}
\scriptsize
\centering
\caption{Fit Parameters of Color--Color--Magnitude Diagram Calibrations\tablenotemark{a}}
\begin{tabular}{l ccccccc cccc} 
\hline
\hline
\noalign{\smallskip}

$Z$ & $a$ & $b$ & $c$ & $d$ & $e$ & $f$ & $\sigma$ & $\nu$ & $\text{elppd}_\text{LOO-CV}$ & $\Delta\text{elppd}_\text{LOO-CV}$ \\ 
(mag) & (dex) & (mag$^{-1}$) & (mag$^{-1}$) & (mag$^{-1}$) & (mag$^{-2}$) & (mag$^{-2}$) & (dex) \\
\noalign{\smallskip}
\hline
\noalign{\smallskip}

$M_{G}$ & $5.87\pm0.44$ & $-1.40\pm0.12$ & $5.08\pm0.24$ & $-2.45\pm0.14$ & $-0.788\pm0.052$ & $0.1094\pm0.0071$ & $0.0691\pm0.0032$ & $5.43\pm1.19$ & $3961\pm59$ & $0$\\
$M_{K_s}$ & $2.70\pm0.27$ & $-1.39\pm0.12$ & $2.82\pm0.15$ & $-1.99\pm0.11$ & $-0.392\pm0.034$ & $0.1355\pm0.0090$ & $0.0677\pm0.0032$ & $4.82\pm0.93$ & $3864\pm61$ & $-97\pm20$\\
$M_{J}$ & $4.42\pm0.37$ & $-1.54\pm0.12$ & $3.23\pm0.17$ & $-2.34\pm0.14$ & $-0.472\pm0.037$ & $0.1442\pm0.0097$ & $0.0697\pm0.0034$ & $5.37\pm1.19$ & $3859\pm60$ & $-102\pm15$\\
$M_{H}$ & $2.42\pm0.27$ & $-1.42\pm0.13$ & $3.01\pm0.16$ & $-1.90\pm0.11$ & $-0.430\pm0.036$ & $0.1248\pm0.0087$ & $0.0686\pm0.0033$ & $4.86\pm0.95$ & $3839\pm60$ & $-121\pm18$\\

\noalign{\smallskip}
\hline
\end{tabular}
\tablenotetext{a}{The polynomial fits follow the expression $\text{[Fe/H]}=a+bX+cY+dZ+eY^2+fZ^2$, where $X=W1-W2$, $Y=G_\text{BP}-G_\text{RP}$, and $Z$ is the corresponding absolute magnitude.}
\label{color_color_magnitude_coef}
\end{table*}

\subsection{Color--Color--Magnitude Diagrams} 

Some authors, such as \citet{Davenport2019RNAAS...3...54D}, added an absolute magnitude as a third variable in their calibrations in order to improve their estimations, which we refer to as color--color--magnitude diagrams. Thus, we performed a stepwise regression using three variables, finding that the model with the best predictive performance is given by:
\begin{equation}
   \text{[Fe/H]}=a+bX+cY+dZ+eY^2+fZ^2, \label{modelCCM} 
\end{equation}
where, as an example, $X=W1-W2$ and $Y=G_\text{BP}-G_\text{RP}$ are two color indices, and the absolute magnitude $Z=M_G$ is added as a third variable. We display the coefficients of the calibration given by Eq.~\ref{modelCCM} in Table~\ref{color_color_magnitude_coef}. As \cite{Davenport2019RNAAS...3...54D}, we found an improvement by adding $M_{G}$ to the $W1-W2$ vs. $G_\text{BP}-G_\text{RP}$ diagram, having that the residual dispersion $\sigma$ and the $\text{elppd}_\text{LOO}$ improve with the addition of this third independent variable. The three-dimensional color--color--magnitude $W1-W2$ vs. $G_\text{BP}-G_\text{RP}$ vs. $M_{G}$ and the comparison between the spectroscopic and estimated [Fe/H] values for the 3919 test stars are plotted in Fig.~\ref{diagrams_comparison} (bottom panels) and the pairs plot of the coefficients regarding this model is shown in Fig.~\ref{pairsplot}. For this color--color--magnitude diagram, 201 stars ($5.1\,\%$) are more than $2\sigma$ above or below the spectroscopic value.
The statistics of the residuals regarding the three calibrations displayed in Fig.~\ref{diagrams_comparison} are shown in Table~\ref{residual_statistics}.

\begin{table}
    \centering
    \scriptsize
    \caption{Statistics\tablenotemark{a} of the Residuals of the Best Color--Color, Color--Magnitude, and Color--Color--Magnitude calibrations}
    \begin{tabular}{l cccc} 
    \hline
    \hline
    \noalign{\smallskip}
     & $\overline{x}$ & $\tilde{x}$ & $\sigma_x$ & MAD($x$) \\ 
    \noalign{\smallskip}
    \hline
    \noalign{\smallskip}
    
    $W1-W2$ vs. $G_\text{BP}-G_\text{RP}$ & $0.003$ & $-0.004$ & $0.112$ & $0.103$ \\
    $G_\text{BP}-G_\text{RP}$ vs. $M_G$ & $0.012$ & $-0.006$ & $0.095$ & $0.072$ \\
    $W1-W2$ vs. $G_\text{BP}-G_\text{RP}$ vs. $M_G$ & $0.009$ & $-0.006$ & $0.089$ & $0.075$ \\
    
    \noalign{\smallskip}
    \hline
    \end{tabular}
    \tablenotetext{a}{Mean ($\overline{x}$), median ($\tilde{x}$), standard deviation ($\sigma_x$), and median absolute deviation (MAD($x$)).} 
    \label{residual_statistics}
\end{table}

\begin{table}
    \centering
    \footnotesize
    \caption{Statistics\tablenotemark{a} of the Differences between Primary's [Fe/H]$_\text{spec}$ and Secondary's [Fe/H]$_\text{phot}$}
    \begin{tabular}{l cccc cc} 
    \hline
    \hline
    \noalign{\smallskip}
     & $\overline{x}$ & $\tilde{x}$ & $\sigma_x$ & MAD($x$) & $r$ & $r_{\rm S}$ \\ 
    \noalign{\smallskip}
    \hline
    \noalign{\smallskip}
    
    B05 & $0.144$ & $0.111$ & $0.147$ & $0.105$ & $0.707$ & $0.671$ \\
    JA09 & $-0.064$ & $-0.086$ & $0.206$ & $0.144$ & $0.688$ & $0.656$ \\
    N12 & $0.071$ & $0.039$ & $0.149$ & $0.115$ & $0.693$ & $0.659$ \\
    M13 & $0.043$ & $0.063$ & $0.144$ & $0.156$ & $0.649$ & $0.641$ \\
    DD19 & $0.047$ & $0.056$ & $0.176$ & $0.167$ & $0.528$ & $0.518$ \\
    R21 & $-0.007$ & $0.001$ & $0.105$ & $0.089$ & $0.842$ & $0.783$ \\
    This work & $-0.012$ & $-0.004$ & $0.087$ & $0.081$ & $0.897$ & $0.827$ \\
    
    \noalign{\smallskip}
    \hline
    \end{tabular}
    \tablenotetext{a}{Mean ($\overline{x}$), median ($\tilde{x}$), standard deviation ($\sigma_x$), median absolute deviation (MAD($x$)), and Pearson 's ($r$) and Spearman's ($r_{\rm S}$) correlation coefficients.} 
    \label{hist_statistics}
\end{table}

\subsection{Comparison with Previous Photometric Estimations}
   
\begin{figure*}
    \centering
       \begin{tabular}{cccc}
         \includegraphics[width=0.23\textwidth]{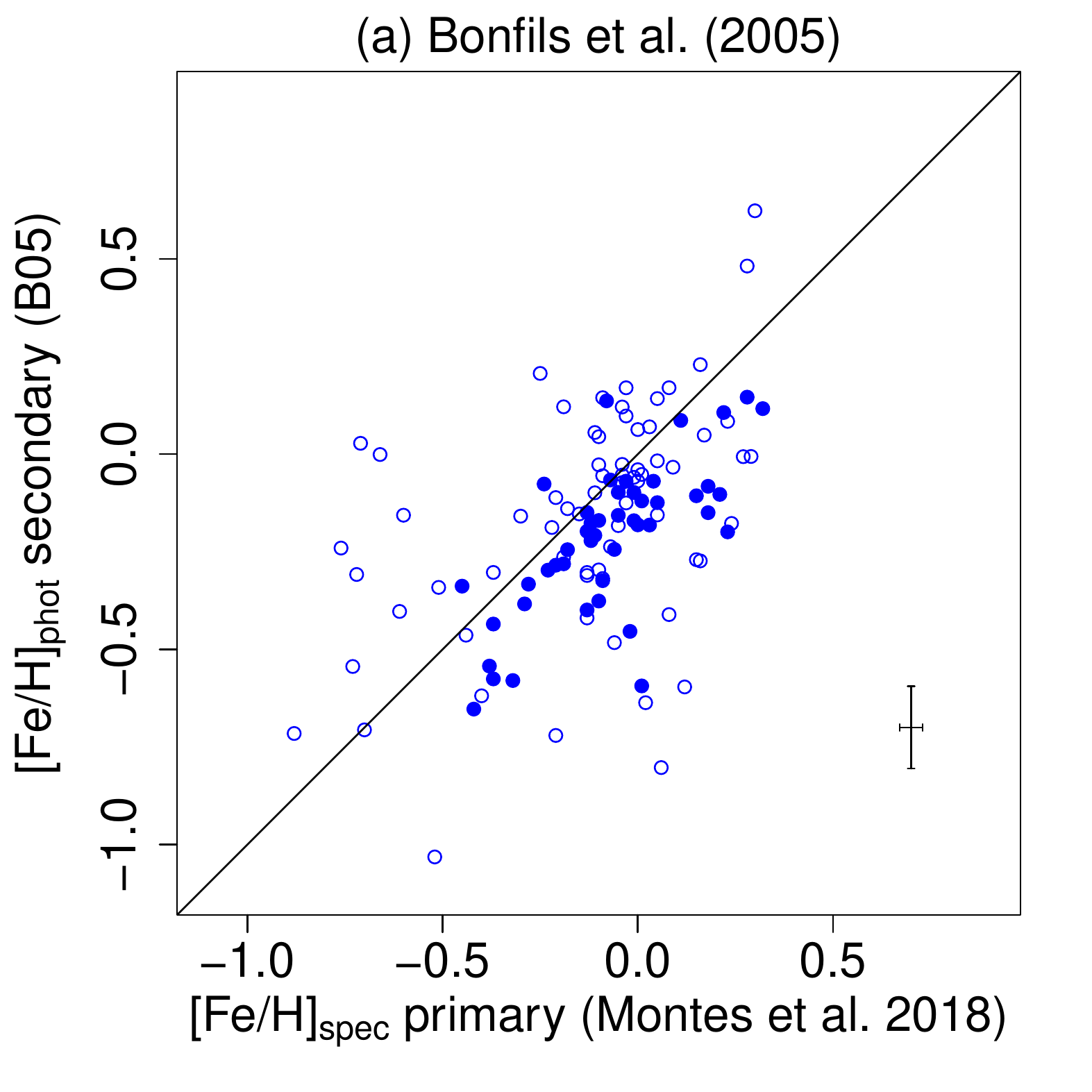} & 
         \includegraphics[width=0.23\textwidth]{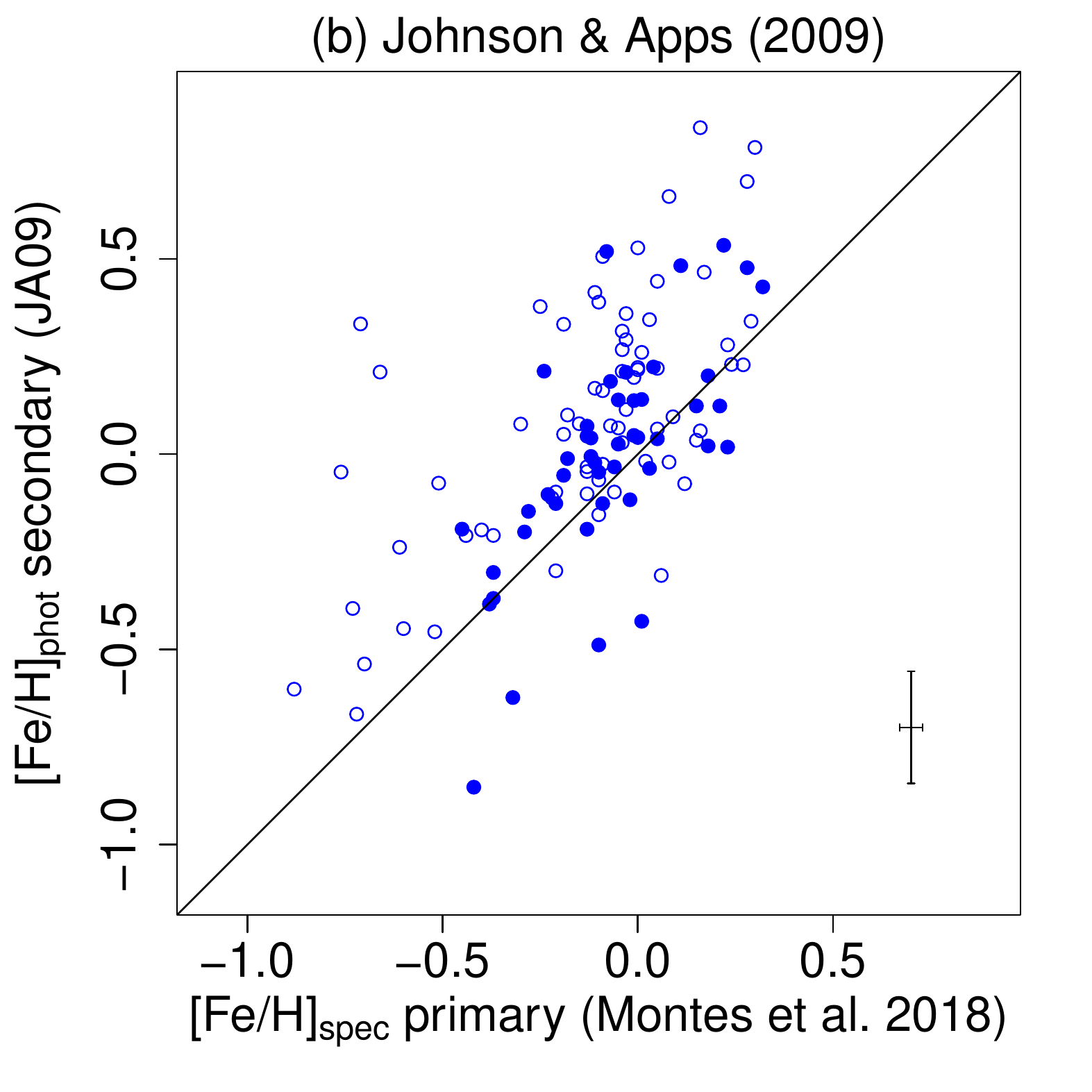} &
         \includegraphics[width=0.23\textwidth]{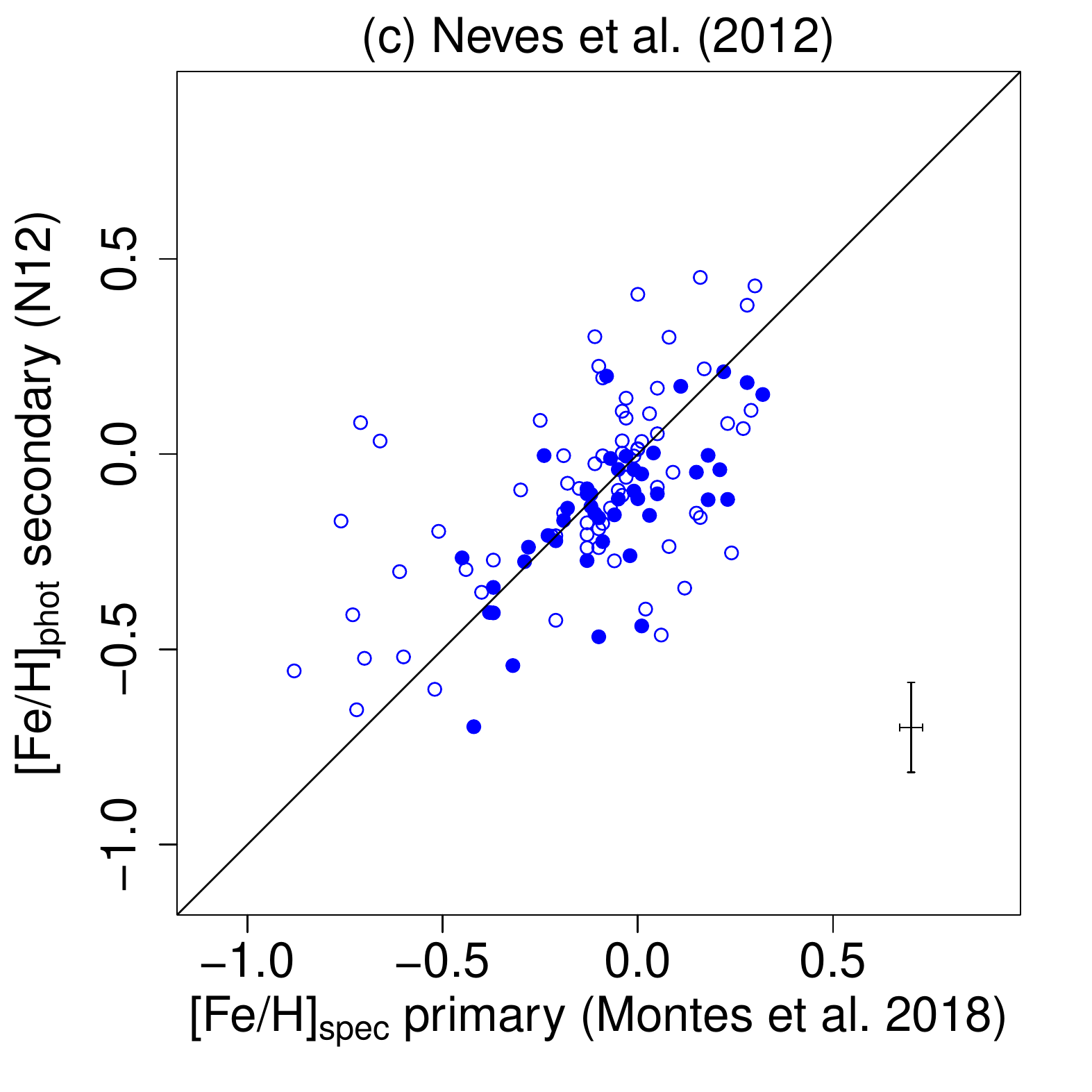} &
         \includegraphics[width=0.23\textwidth]{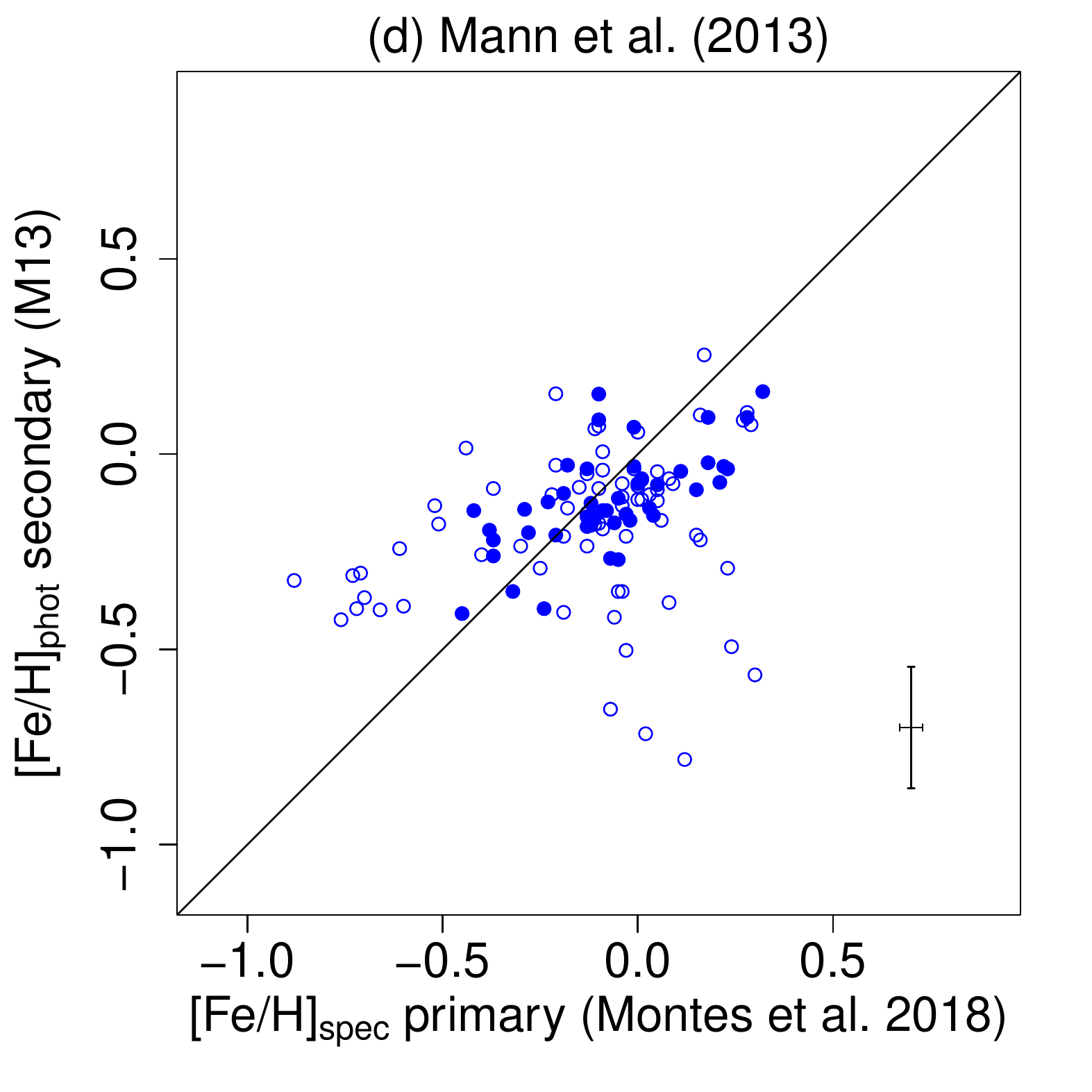}
       \end{tabular}
       \begin{tabular}{ccc}
         \includegraphics[width=0.23\textwidth]{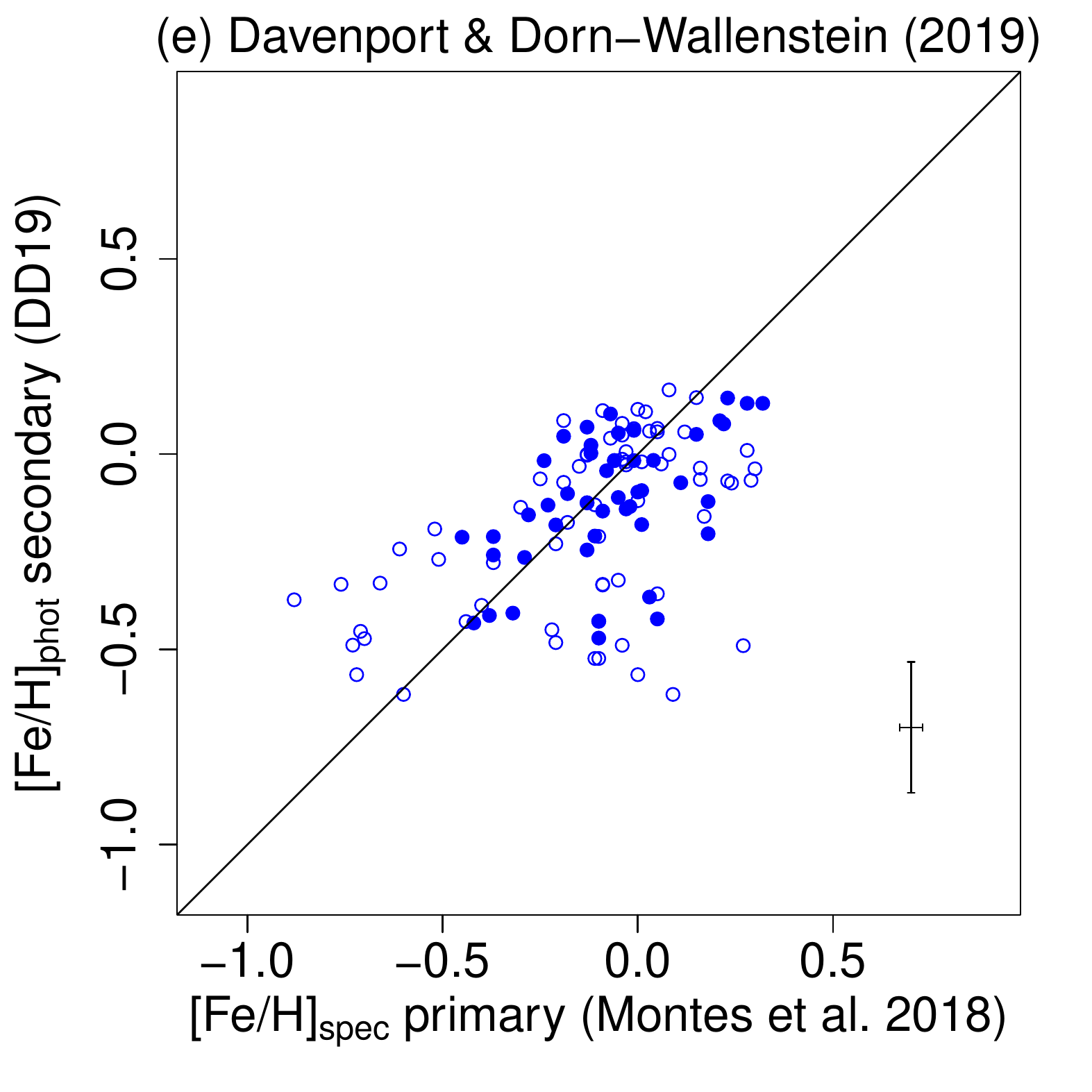} &
         \includegraphics[width=0.23\textwidth]{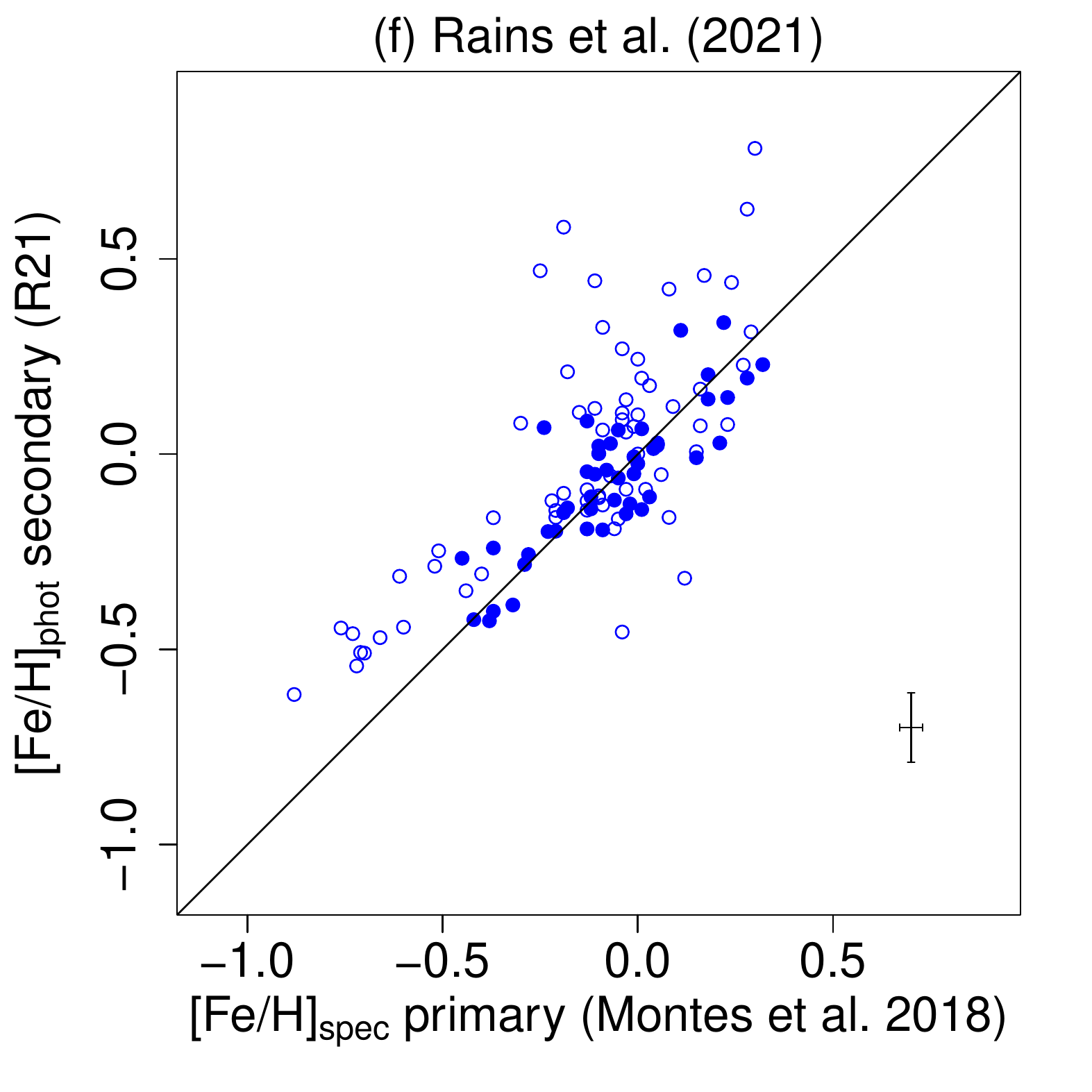} &
         \includegraphics[width=0.23\textwidth]{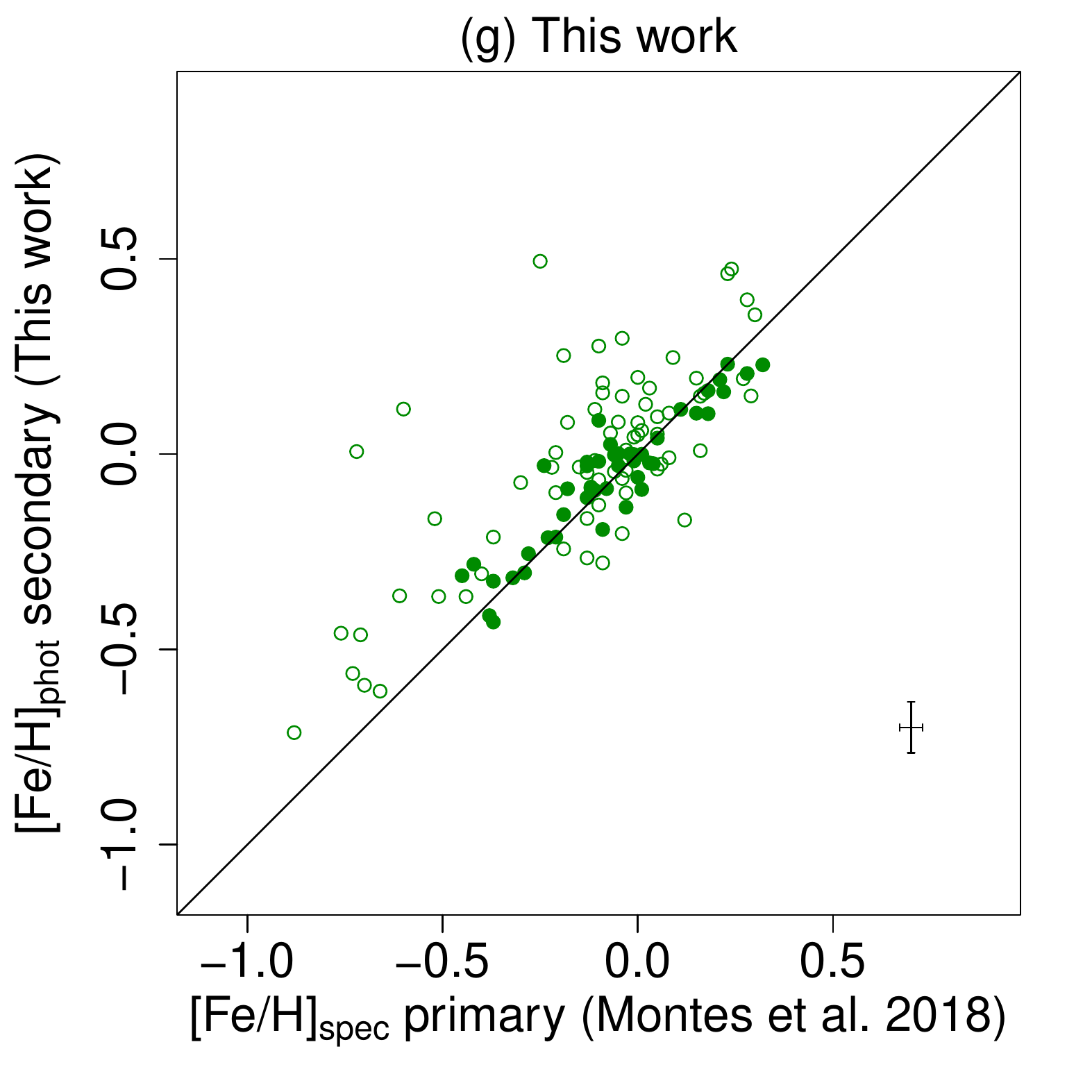}
       \end{tabular}
    \caption{Comparison between the spectroscopic metallicity of the the FGK-type primary stars measured by \citet{Montes2018MNRAS.479.1332M} and the photometric metallicity estimated by \citetalias{Bonfils2005A&A...442..635B}, \citetalias{Johnson2009ApJ...699..933J}, \citetalias{Neves2012A&A...538A..25N}, \citetalias{Mann2013AJ....145...52M}, \citetalias{Davenport2019RNAAS...3...54D}, \citetalias{Rains2021MNRAS.504.5788R} (blue), and by us (green) using the best accurate color--color--magnitude calibration from Table~\ref{color_color_magnitude_coef} of the 46 secondary M dwarfs of Sect.~\ref{methodology}. The open circles represent the 115 stars and the filled circles the 46 stars with good photometric and astrometric data. The solid line denotes the 1:1 relationship. The error bars show the mean spectroscopic uncertainty and the MAD of the differences between the metallicities of each of the literature methods and the spectroscopic values.}
    \label{comparison_binaries_dav}
\end{figure*}

We compared our results with previous photometric estimations from the literature (\citetalias{Bonfils2005A&A...442..635B}: \citealt{Bonfils2005A&A...442..635B} [Eq.~1], \citetalias{Johnson2009ApJ...699..933J}: \citealt{Johnson2009ApJ...699..933J} [Eq.~1], \citetalias{Neves2012A&A...538A..25N}: \citealt{Neves2012A&A...538A..25N} [Eq.~3], \citetalias{Mann2013AJ....145...52M}: \citealt{Mann2013AJ....145...52M} [Eq.~29], \citetalias{Davenport2019RNAAS...3...54D}: \citealt{Davenport2019RNAAS...3...54D} [\texttt{ingot}], \citetalias{Rains2021MNRAS.504.5788R}: \citealt{Rains2021MNRAS.504.5788R} [Eq.~2]).
To do this, we used the stellar sample of FGK+M binary systems presented by \citet{Montes2018MNRAS.479.1332M}. We did not compare with \cite{Schlaufman2010A&A...519A.105S} and \cite{Johnson2012AJ....143..111J} because updated versions of their calibrations are in \citetalias{Neves2012A&A...538A..25N} and \citetalias{Mann2013AJ....145...52M}, respectively. We did not compare either with \cite{Schmidt2016MNRAS.460.2611S}, \cite{Hejazi2015AJ....149..140H} and \cite{Medan2021AJ....161..234M} since less than the $50\,\%$ of the M-dwarf companions presented in \citet{Montes2018MNRAS.479.1332M} have a counterpart in the SDSS catalog, nor with \cite{Dittmann2016ApJ...818..153D} which requires the MEarth photometry. In Fig.~\ref{comparison_binaries_dav} we represent the spectroscopic metallicity values of the primary stars reported by \citet{Montes2018MNRAS.479.1332M} versus the photometrically estimated values for the M-dwarf companions. In our case, we use the best accurate color--color--magnitude calibration. The open circles represent the 115 stars and the filled circles the 46 stars that remain after applying the criteria described in Sect.~\ref{methodology} (i.e. stars with good photometric and astrometric data). Qualitatively, our estimations and the ones from \citetalias{Rains2021MNRAS.504.5788R} for the 46 stars follow the 1:1 relationship with less dispersion than the estimated values from previous studies. In Fig.~\ref{photometric_comparisons} we compared all the photometric estimations for the metallicity of M-dwarf companions. We found that the calibrations by \citetalias{Bonfils2005A&A...442..635B}, \citetalias{Johnson2009ApJ...699..933J} and \citetalias{Neves2012A&A...538A..25N} are highly correlated, since they are based on the same color--magnitude diagram. We also noted that our estimations  show a good consistency with the ones by \citetalias{Rains2021MNRAS.504.5788R}.

\begin{figure*}
       \centering
       \begin{tabular}{cccc}
         \includegraphics[width=0.23\textwidth]{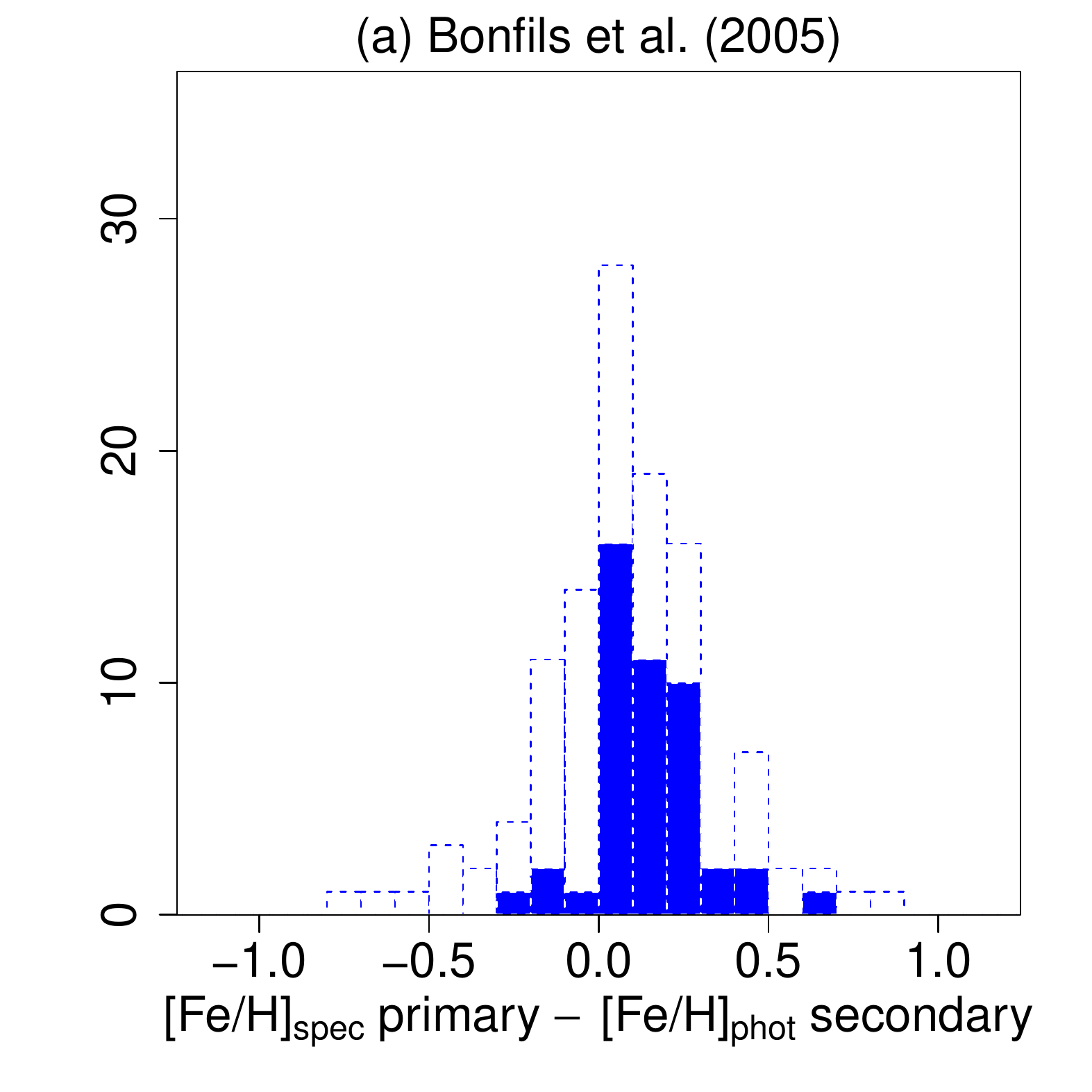} & 
         \includegraphics[width=0.23\textwidth]{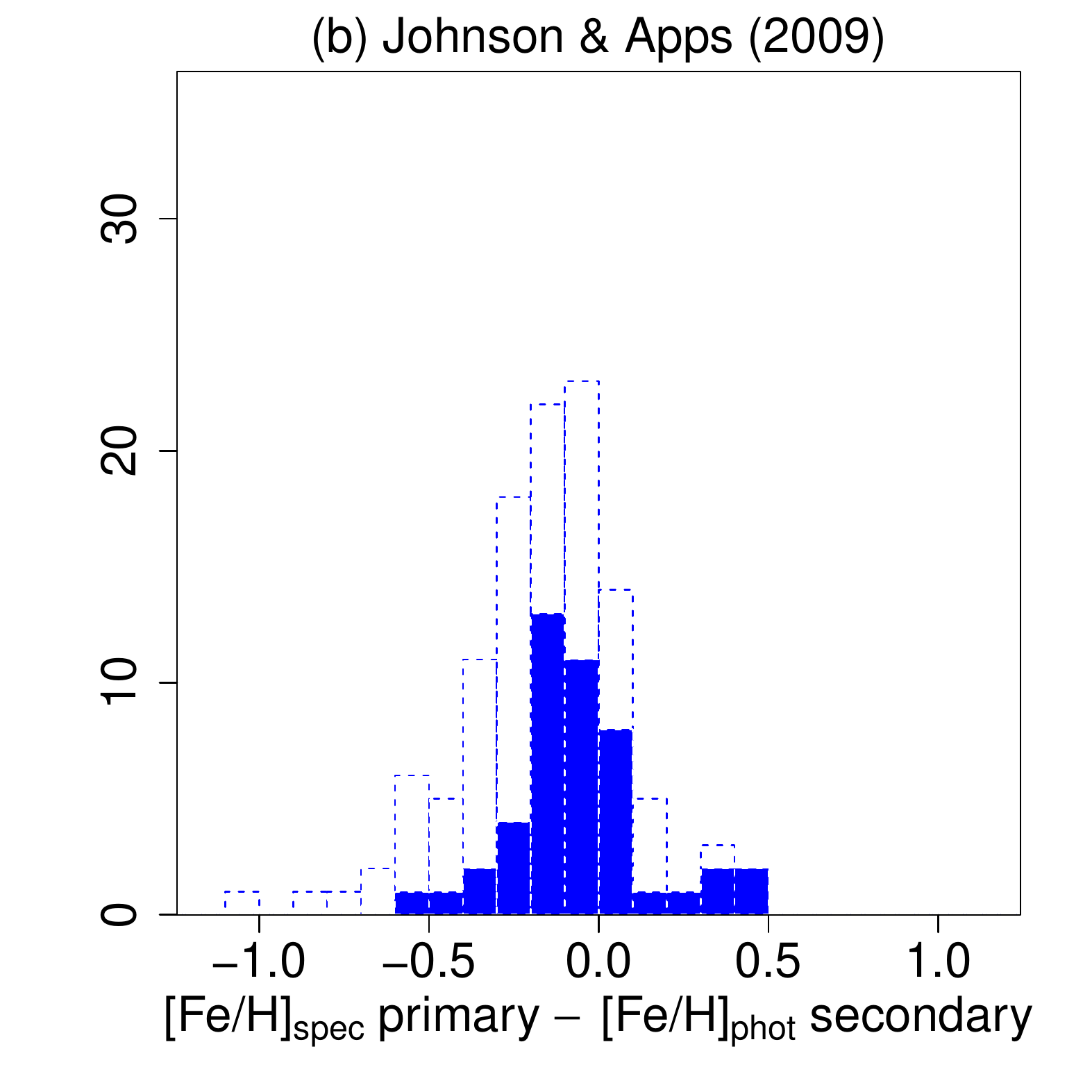} &
         \includegraphics[width=0.23\textwidth]{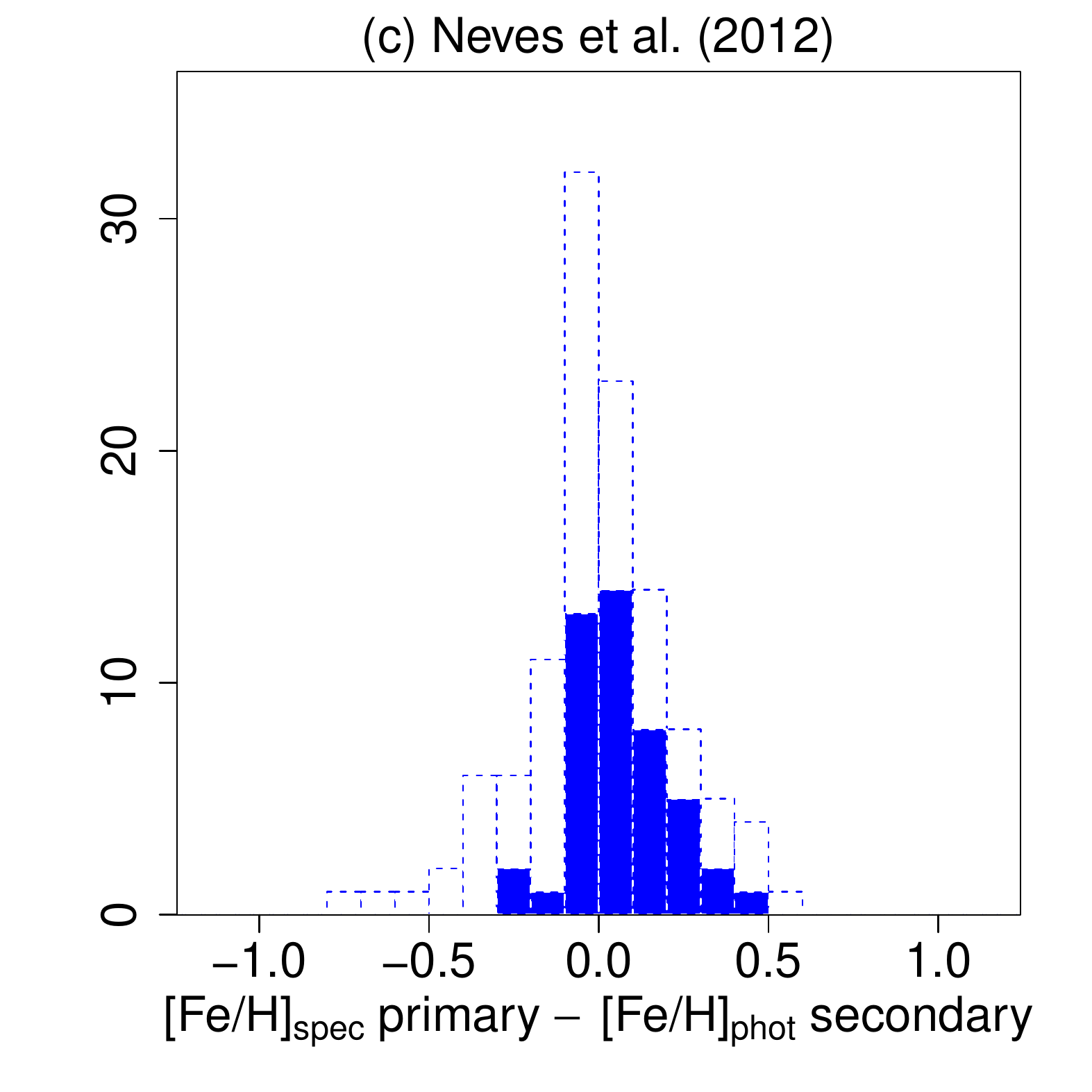} &
         \includegraphics[width=0.23\textwidth]{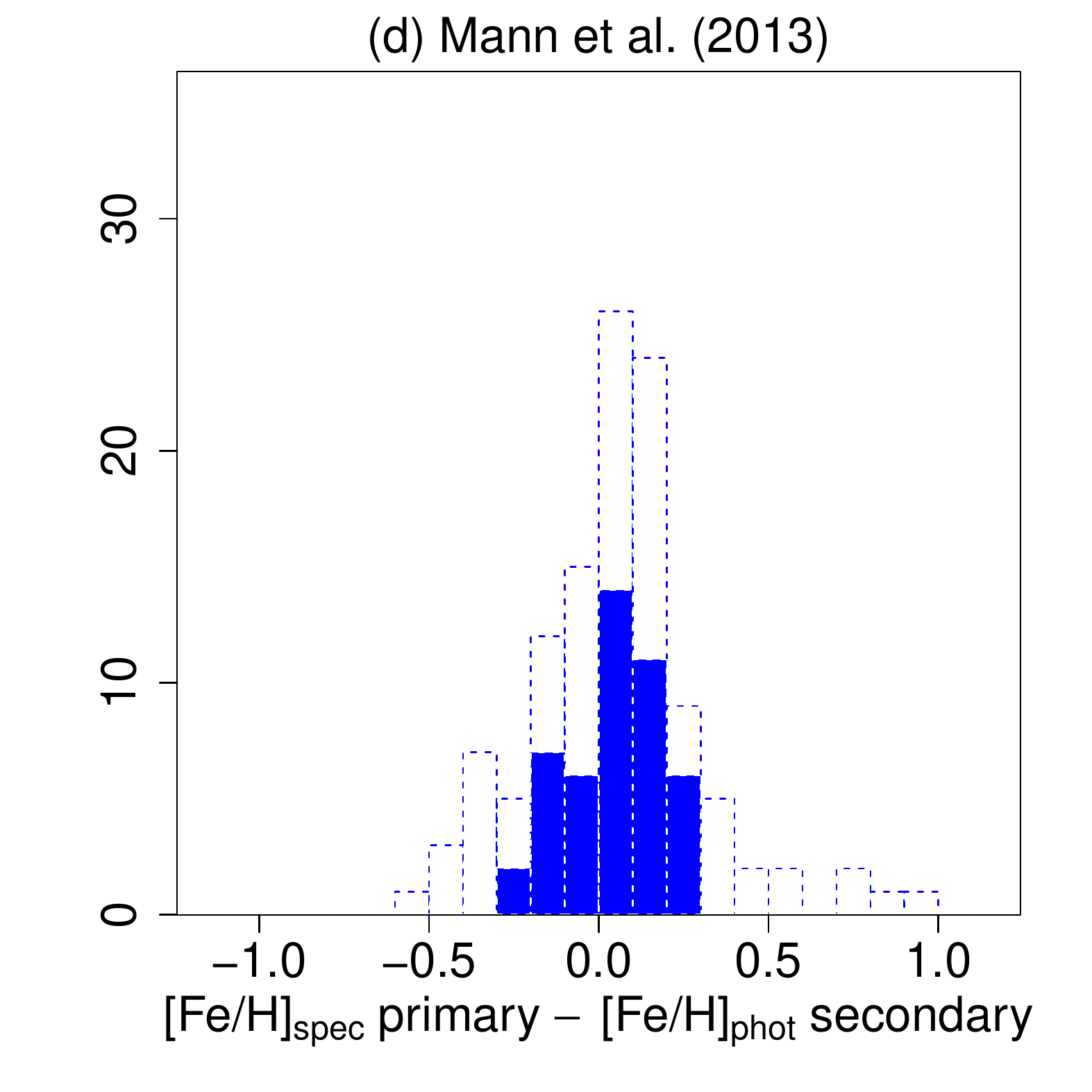}
       \end{tabular}
       \begin{tabular}{ccc}
         \includegraphics[width=0.23\textwidth]{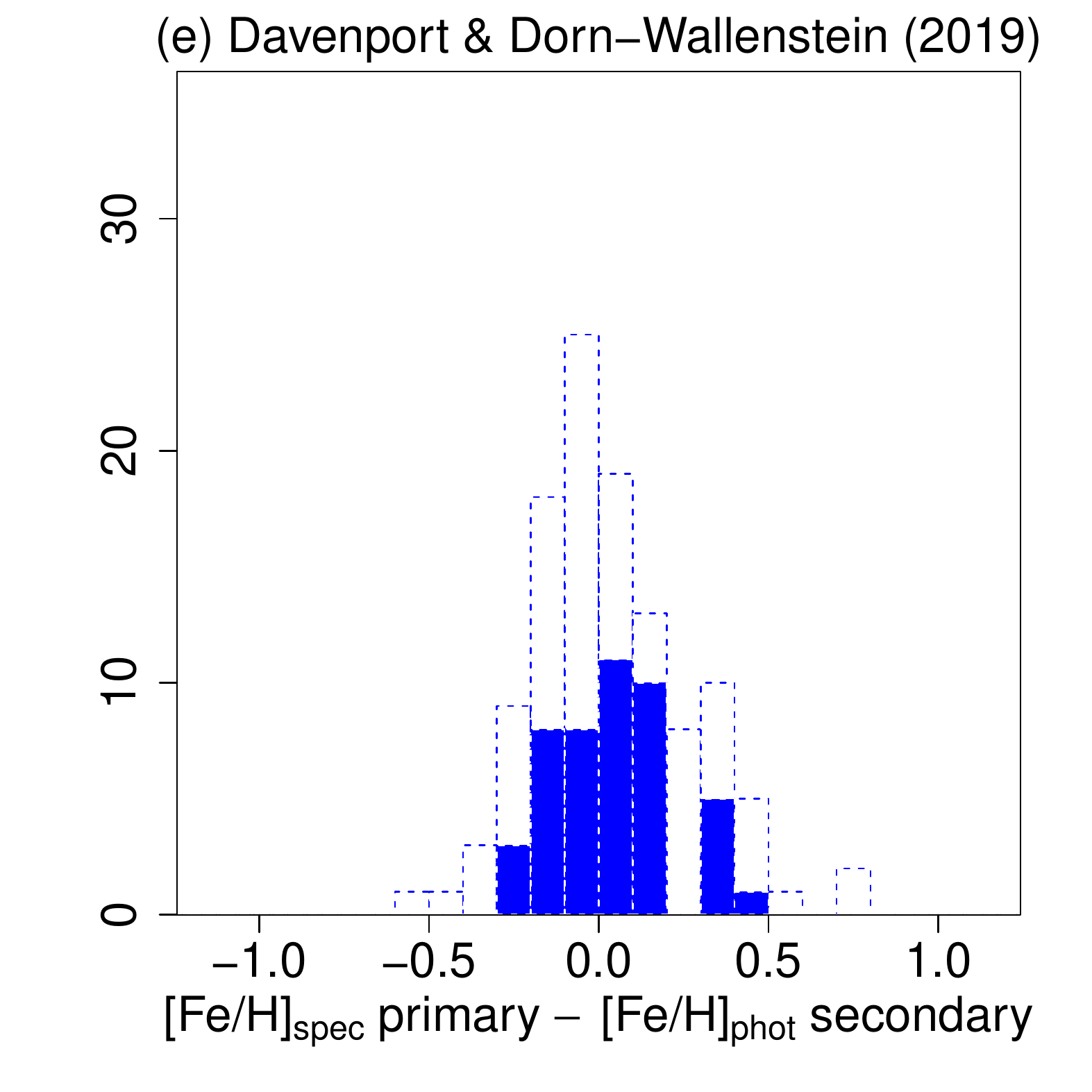} &
         \includegraphics[width=0.23\textwidth]{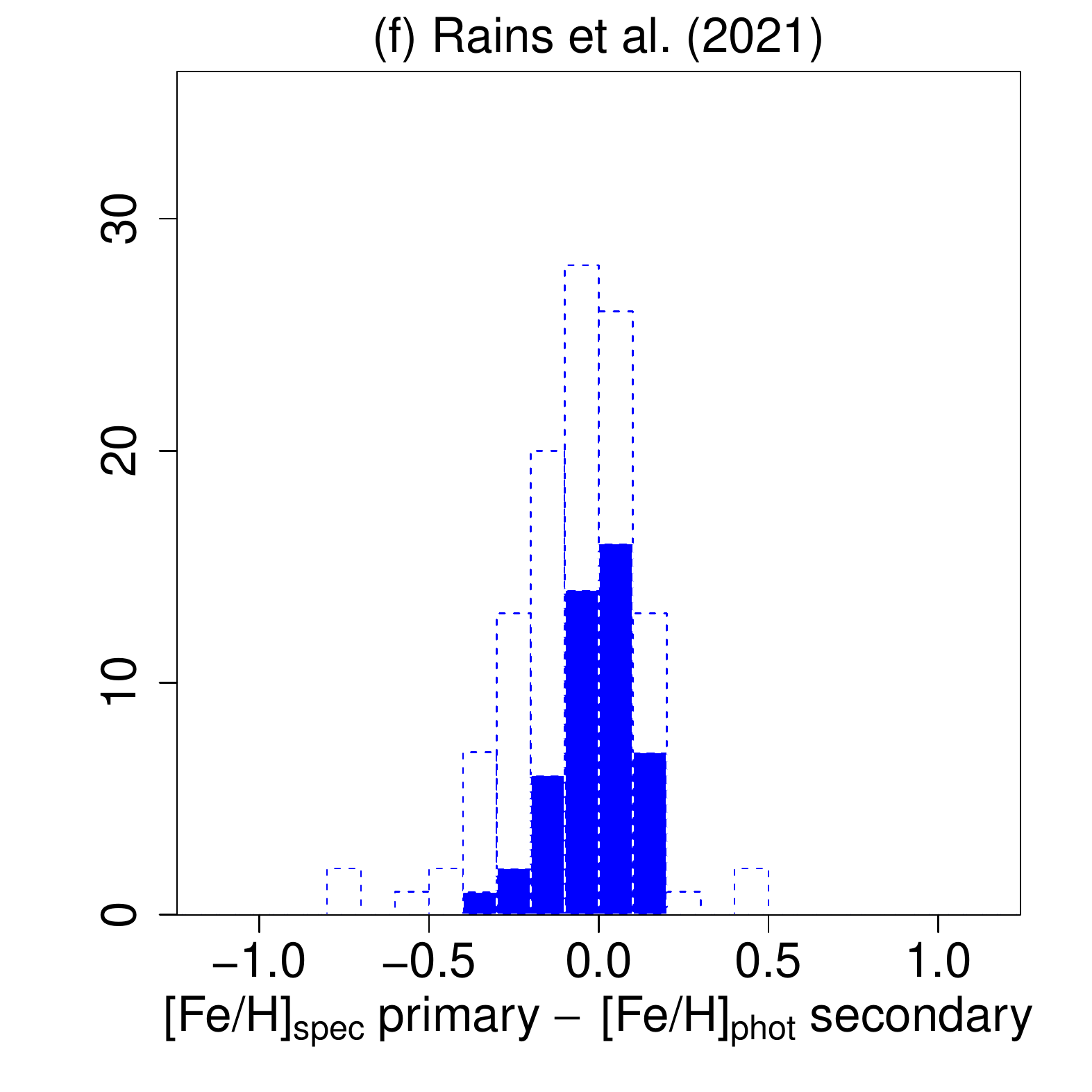} &
         \includegraphics[width=0.23\textwidth]{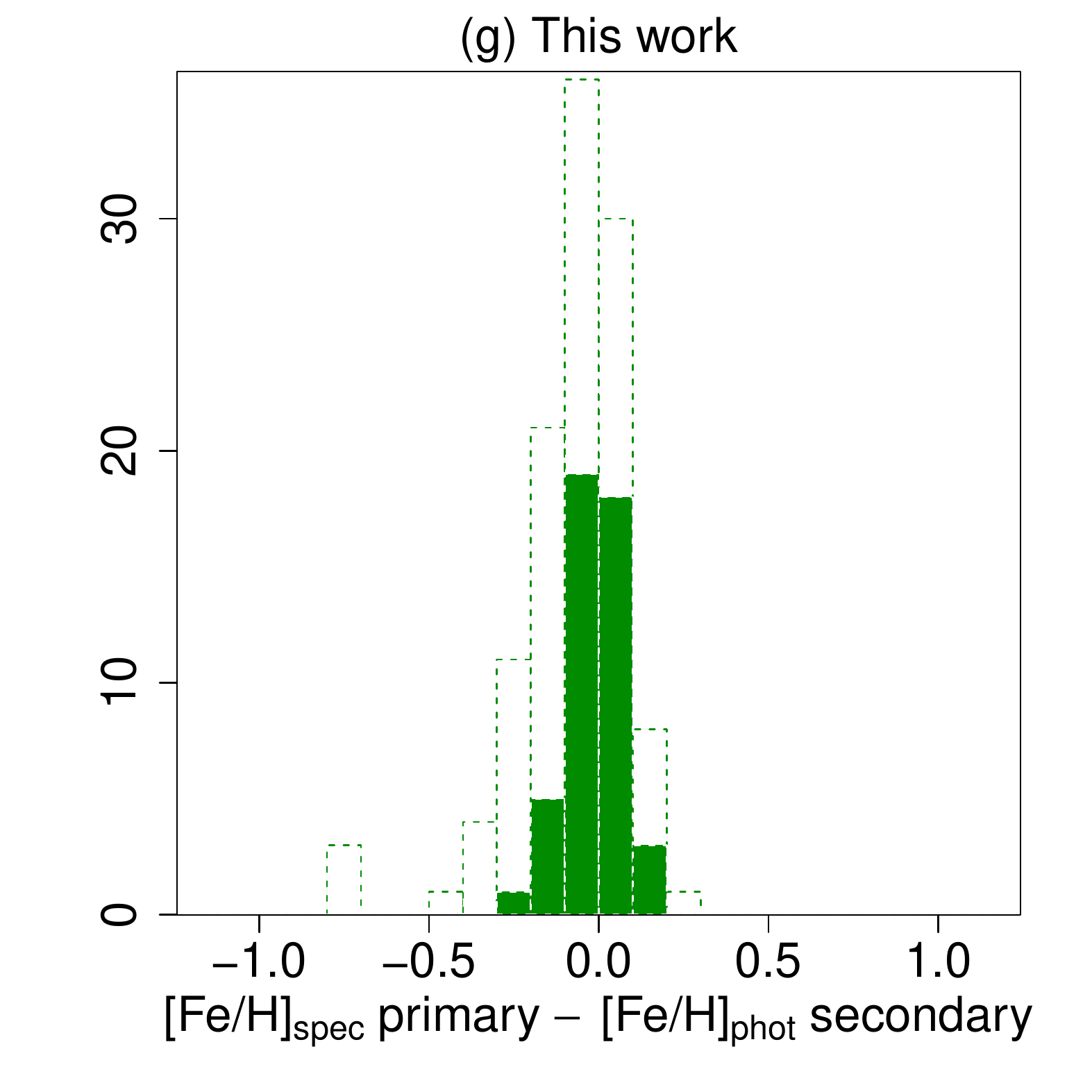}
       \end{tabular}
       \caption{Same as Fig.~\ref{comparison_binaries_dav} but for the distributions of the difference between primary's [Fe/H]$_\text{spec}$ and solid and dashed histograms instead of solid and open circles.}
    \label{hist_comparisons}
\end{figure*}

For a more quantitative comparison, we studied the distributions of the differences between the primary's [Fe/H]$_\text{spec}$ and the secondary's [Fe/H]$_\text{phot}$. The histograms of these distributions are plotted in Fig.~\ref{hist_comparisons}, while some statistics are compiled in Table~\ref{hist_statistics}, along with the Pearson's and Spearman's correlation coefficients for the spectroscopic and photometric metallicity values. Our metallicity estimations and their uncertainties, shown for the 46 M dwarfs in Table~\ref{binaries}, are quantitatively less biased and have less dispersion and a greater correlation than those from previous studies. Our calibration can also be used even when some of the filtering criteria are not met, obtaining a distribution that, despite having outliers, is not biased and reproduces reasonably well the metallicity values.

\section{Summary}
\label{summary}

In the present work we studied the photometric estimations of metallicity for M dwarfs. The precision, accuracy, and homogeneity of both astrometry and photometry from Gaia DR3, complemented by near- and mid-IR  photometry from 2MASS and CatWISE2020, allowed us to study different calibrations based on color--color and color--magnitude diagrams. In order to obtain the best quality for calibrations, we filtered our data and removed multiple stars, lower surface gravity stars or with low photometric or astrometric quality.

Using the sample presented by \citetalias{Birky2020ApJ...892...31B}, we derived several photometric calibrations using MCMC methods with \texttt{Stan}. We studied the metallicity gradient shown in color--color and color--magnitude diagrams in order to estimate the metallicity of early and mid M dwarfs (down to M5.0\,V). We compared the predictive performance of the different calibrations with the LOO-CV criterion, and combined the information in a three-dimensional color--color--magnitude diagram. We obtained an improvement when adding an absolute magnitude as a third variable to the optical-IR color--color diagrams.

Finally, we compared our most accurate calibration with other photometric metallicity estimations found in the literature (\citetalias{Bonfils2005A&A...442..635B}, \citetalias{Johnson2009ApJ...699..933J}, \citetalias{Neves2012A&A...538A..25N}, \citetalias{Mann2013AJ....145...52M}, \citetalias{Davenport2019RNAAS...3...54D}, and \citetalias{Rains2021MNRAS.504.5788R}) for an additional sample of M-dwarf common proper motion companions to FGK-type primary stars with well defined spectroscopic metallicities \citep{Montes2018MNRAS.479.1332M}. Our metallicity estimations are not significantly biased and have a lower dispersion than those of previous photometric studies. Our most accurate calibration is given by:
\begin{equation}
    \text{[Fe/H]}=a+bX+cY+dZ+eY^2+fZ^2,
\end{equation}
where $X=W1-W2$, $Y=G_\text{BP}-G_\text{RP}$, $Z=M_{G}$, and $a$, $b$, $c$, $d$, $e$, and $f$ are the parameters provided in the first row of Table~\ref{color_color_magnitude_coef}.
A code in GitHub\footnote{\url{https://github.com/chrduque/metamorphosis.git}} and a shinyapp\footnote{\url{https://chrduque.shinyapps.io/metamorphosis}} are provided to facilitate the calculation of the metallicity estimations and their uncertainties \citep{Duque-Arribas2022zenodo}.

The Bayesian approach leads to an improvement due to several factors, among which are the objective determination of the terms to be included in the calibrations based on the application of an information criterion such as the LOO-CV, the inclusion of errors in the variables, and the implementation of a robust approach adopting a t-Student likelihood instead of the Gaussian one used in the frequentist analysis. Furthermore, the improvement of our calibration may also be due to the use of the color index $W1-W2$. Some of the previous calibrations did not include this color index, but relied on indices that included instead the visual magnitude $V$, which has been shown to perform poorly in M-dwarf analyses (see \citealt{Cifuentes2020A&A...642A.115C}). 

This work can be extended in several ways.
More passbands from different surveys can be used, mainly in the optical and near-IR, such as SDSS, J-PLUS, and J-PAS \citep{Cenarro2019A&A...622A.176C} from the ground or EUCLID \citep{Laureijs2011arXiv1110.3193L} from space. Furthermore, the calibrations can be extended to late M dwarfs using other star samples, which would allow the obtaining of a complete understanding of the photometric estimations of metallicity for these cool stars.

\section*{Acknowledgments}
We thank the anonymous referee for the instructive report, which improved our manuscript. We acknowledge financial support from the Universidad Complutense de Madrid, from the Agencia Estatal de Investigaci\'on of the Ministerio de Ciencia, Innovaci\'on y Universidades through projects PID2019-109522GB-C5[1:4]/AEI/10.13039/501100011033, and PID2019-107427-GB-31, the Ministerio de Universidades through fellowship FPU15/01476, and the Centre of Excellence ``Mar\'ia de Maeztu'' award to Centro de Astrobiolog\'ia (MDM-2017-0737). We also acknowledge financial support from the European Regional Development Fund (ERDF) and the Gobierno de Canarias through project ProID2021010128. We thank L. M. Sarro for his comments.

%


\software{METaMorPHosis \citep{Duque-Arribas2022zenodo}, Stan \citep{Carpenter2017JSS....76....1C}, TOPCAT \citep{Taylor2005ASPC..347...29T}, \textsc{StePar} \citep{Tabernero2019A&A...628A.131T}}



\newpage

\bibliography{sample631}{}
\bibliographystyle{aasjournal}

\appendix

\section{Figures}

\renewcommand\thefigure{A.1}
\begin{figure*}
    \centering
    \begin{tabular}{ccc}
         \includegraphics[width=0.28\textwidth]{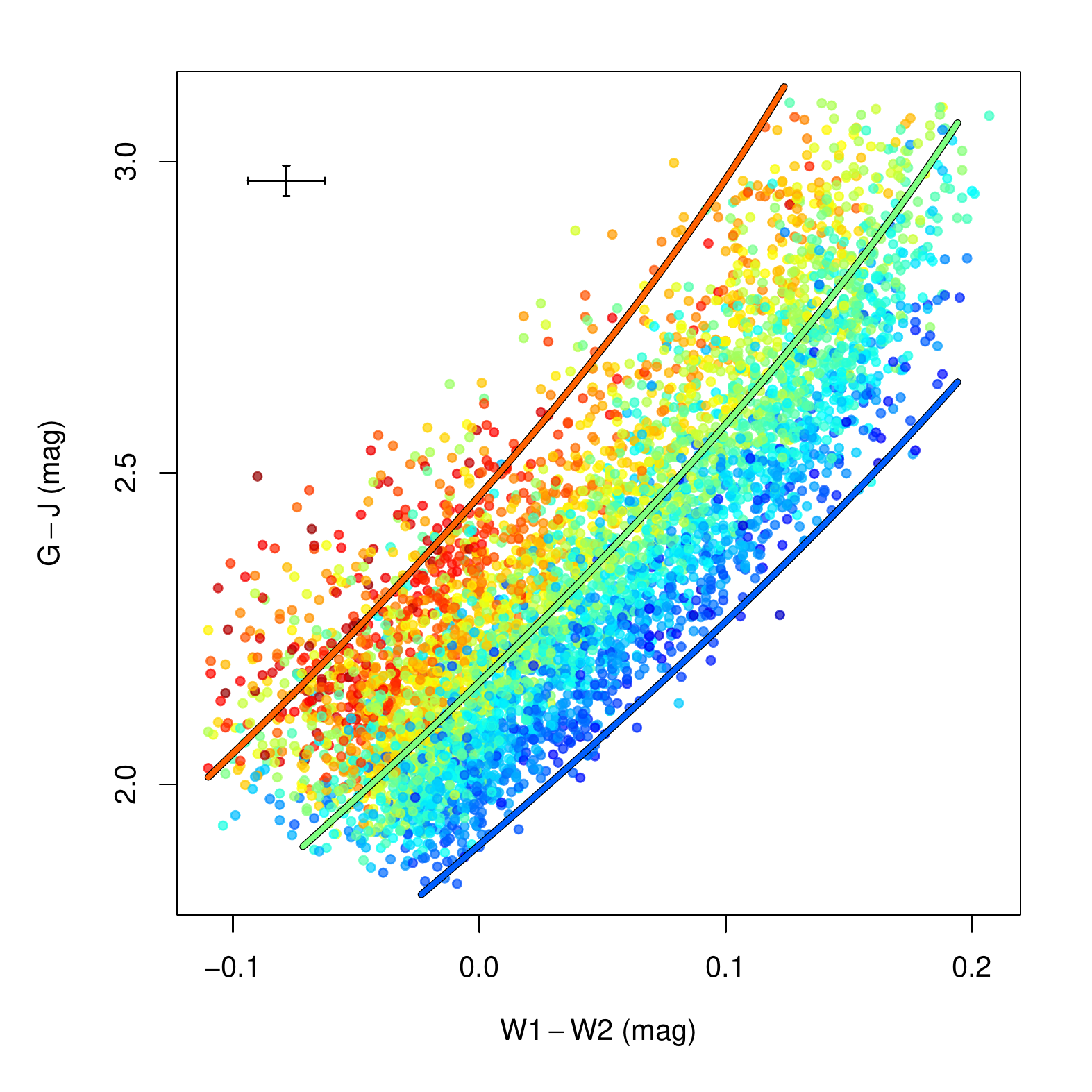} &
         \includegraphics[width=0.28\textwidth]{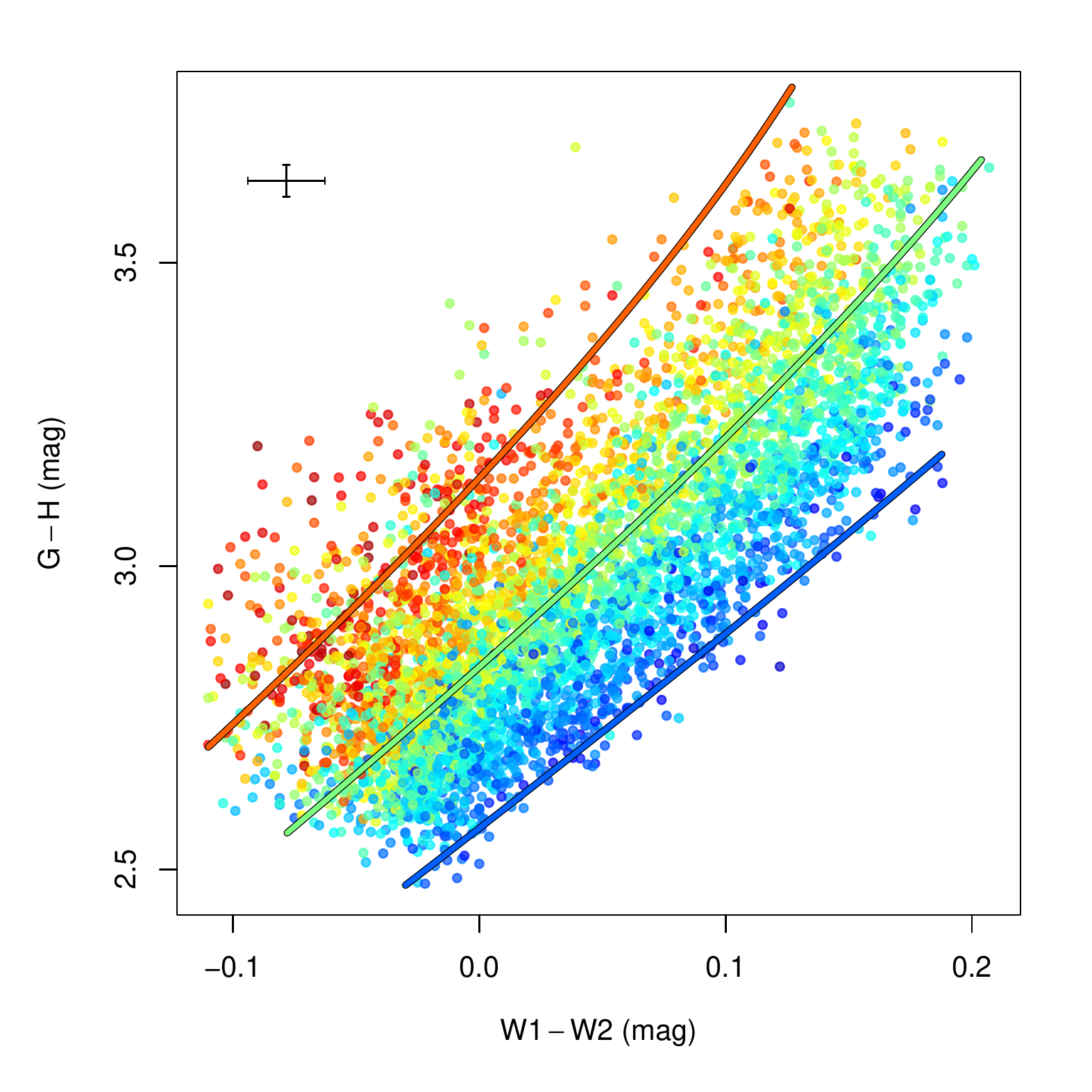} &
         \includegraphics[width=0.28\textwidth]{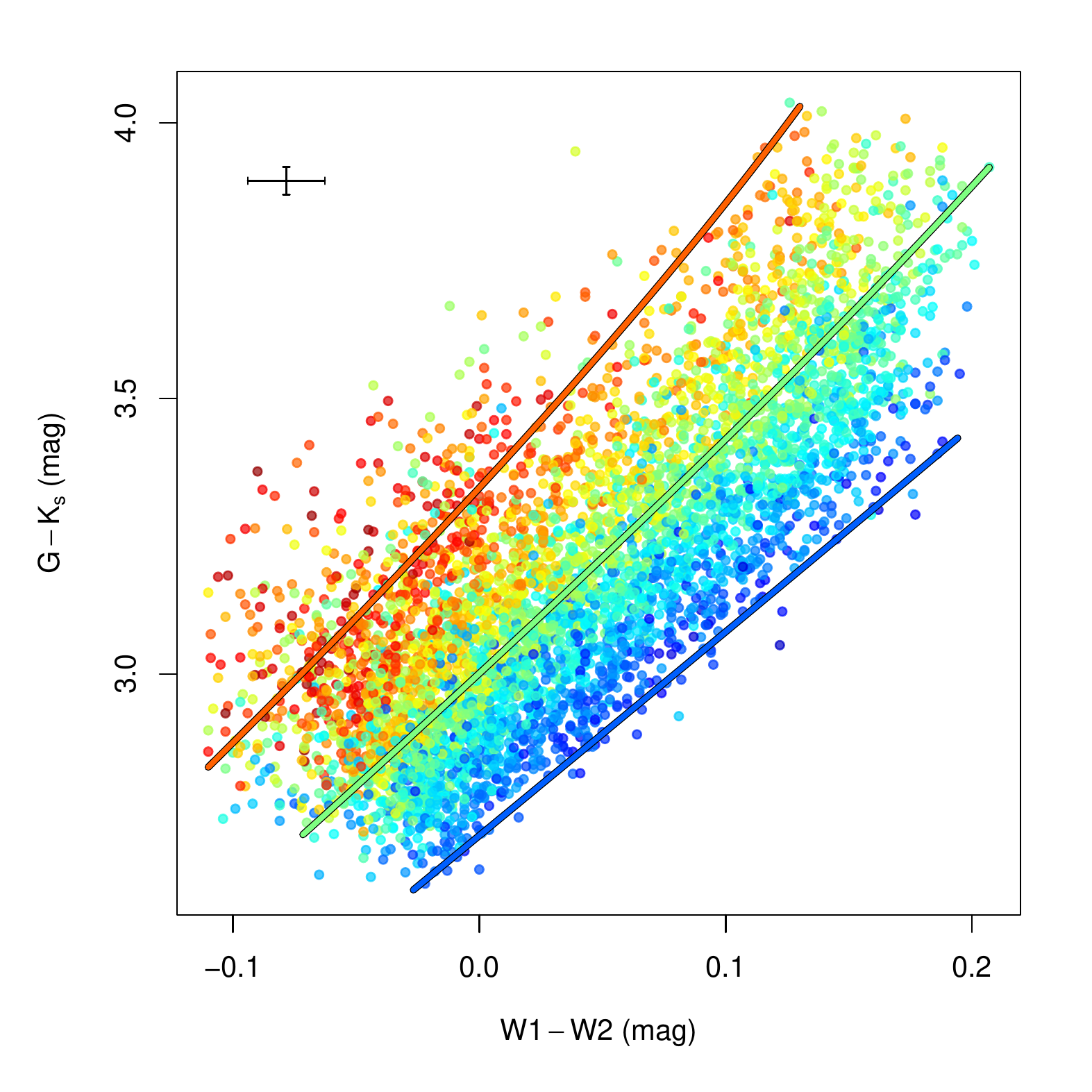} \\
         \includegraphics[width=0.28\textwidth]{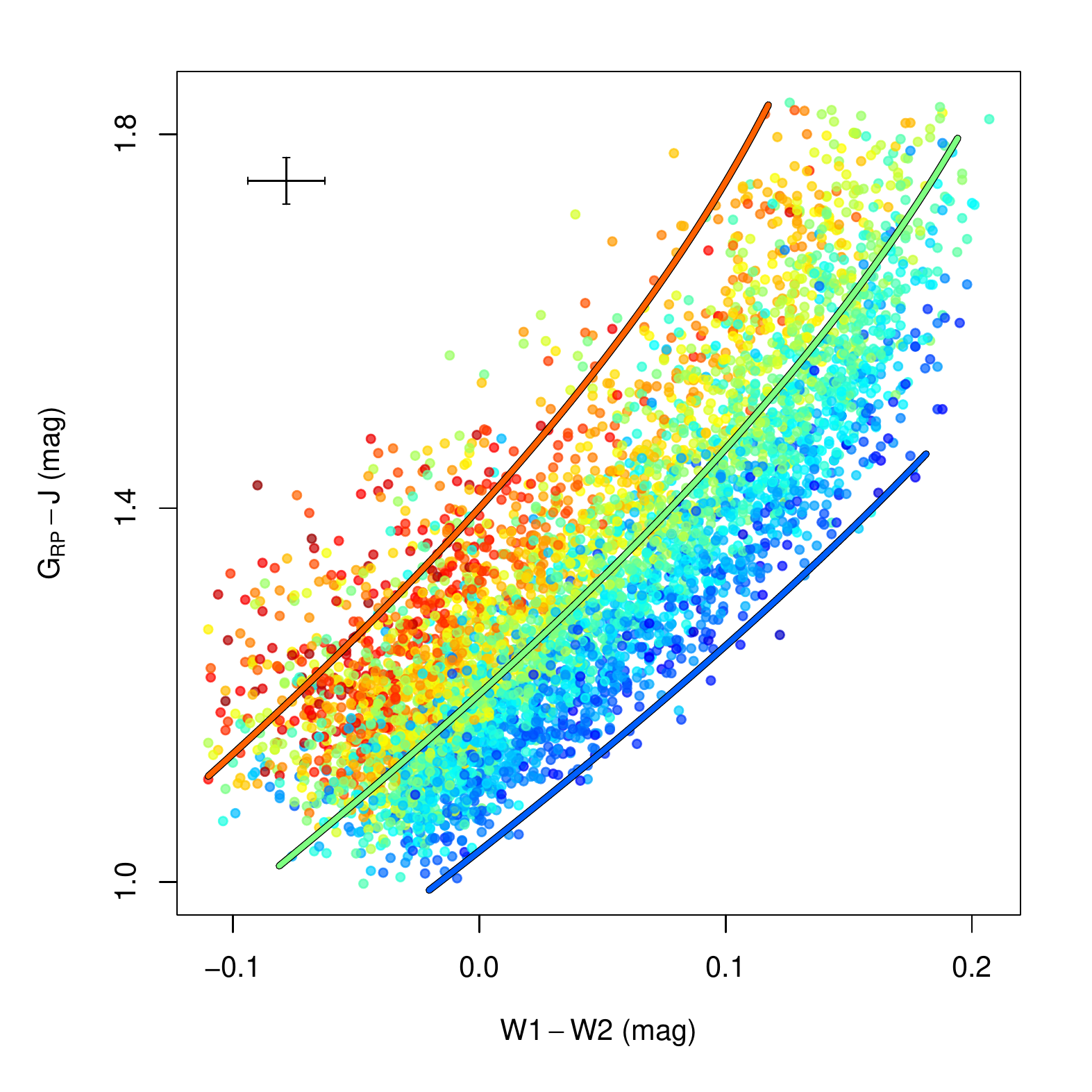} &
         \includegraphics[width=0.28\textwidth]{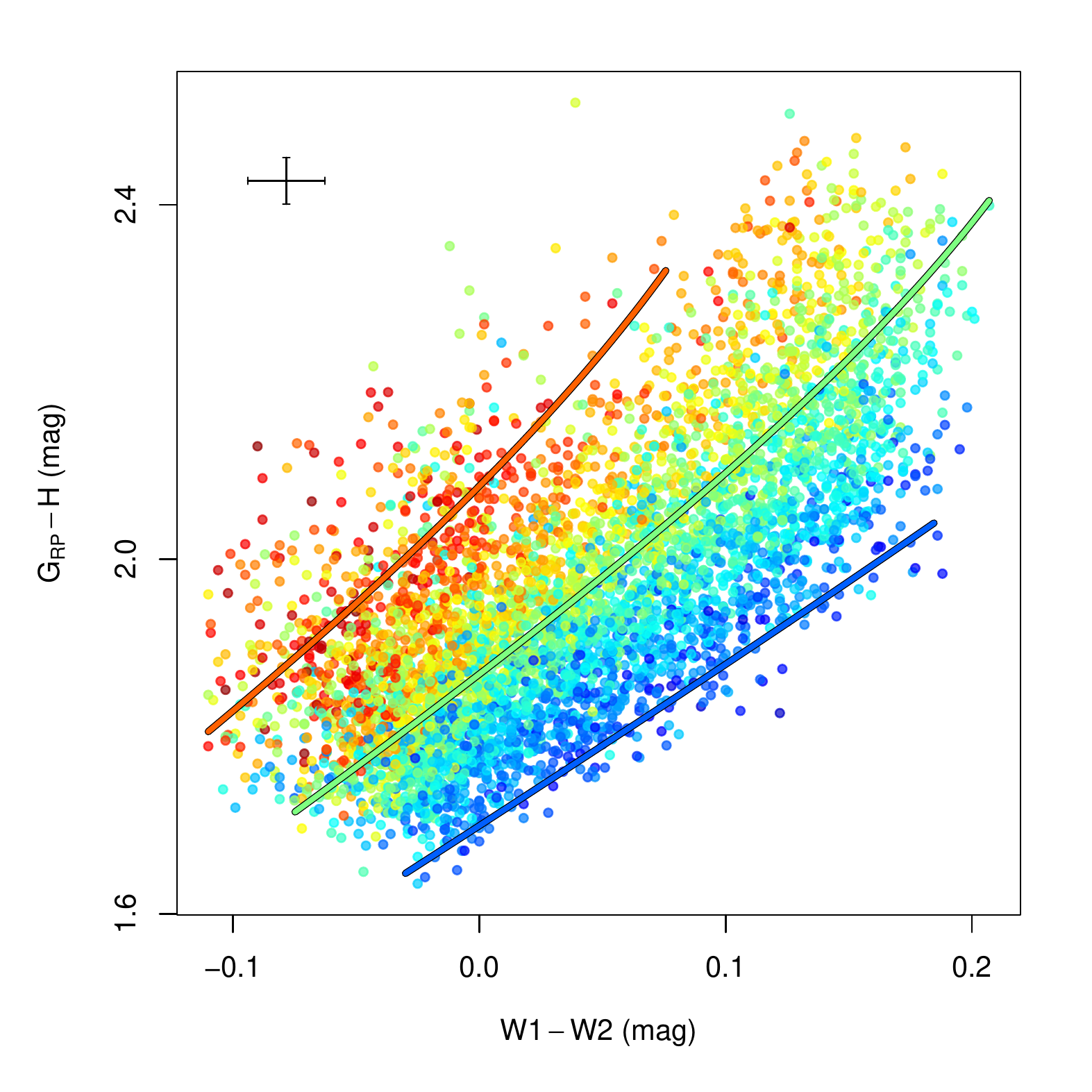} &
         \includegraphics[width=0.28\textwidth]{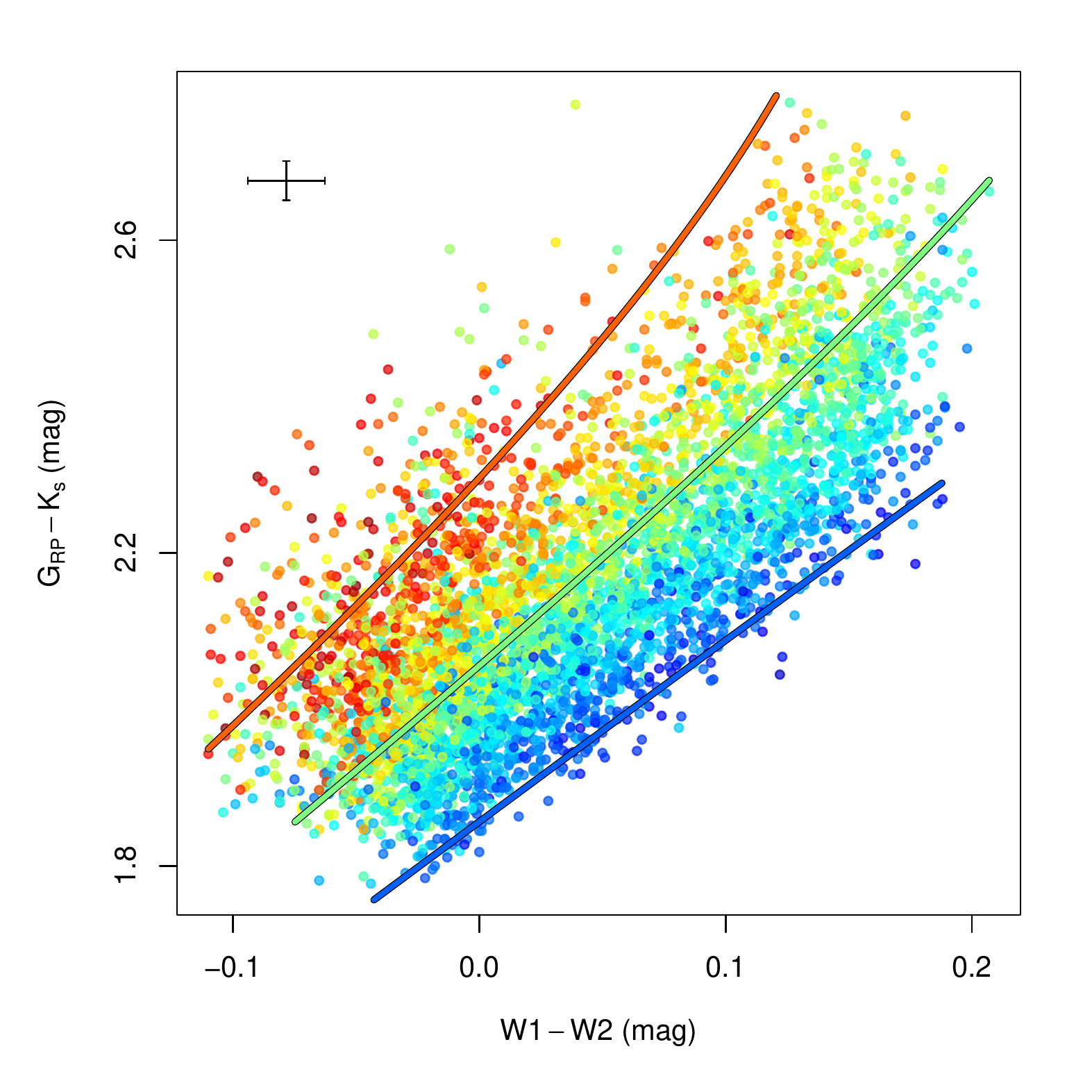} \\
         \includegraphics[width=0.28\textwidth]{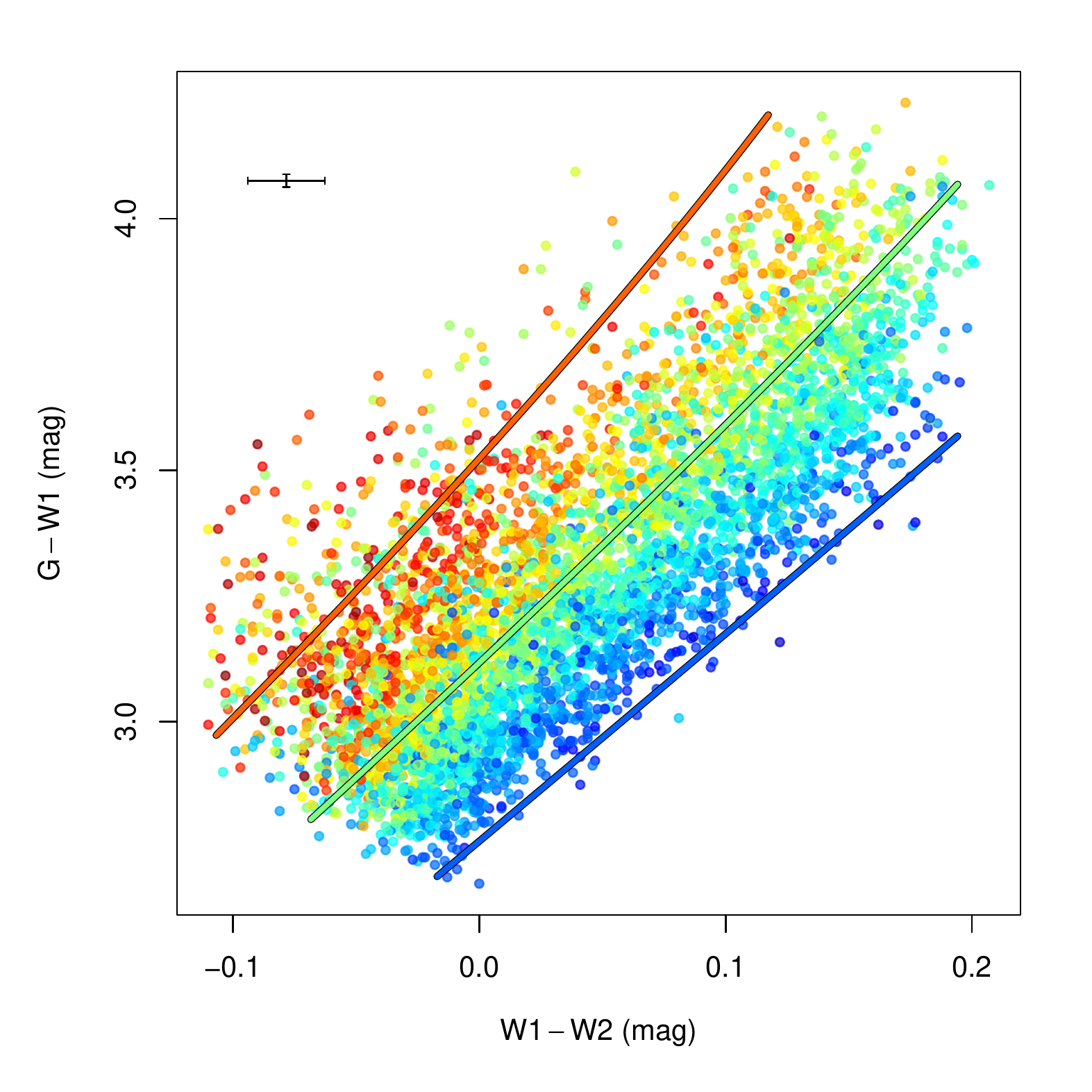} &
         \includegraphics[width=0.28\textwidth]{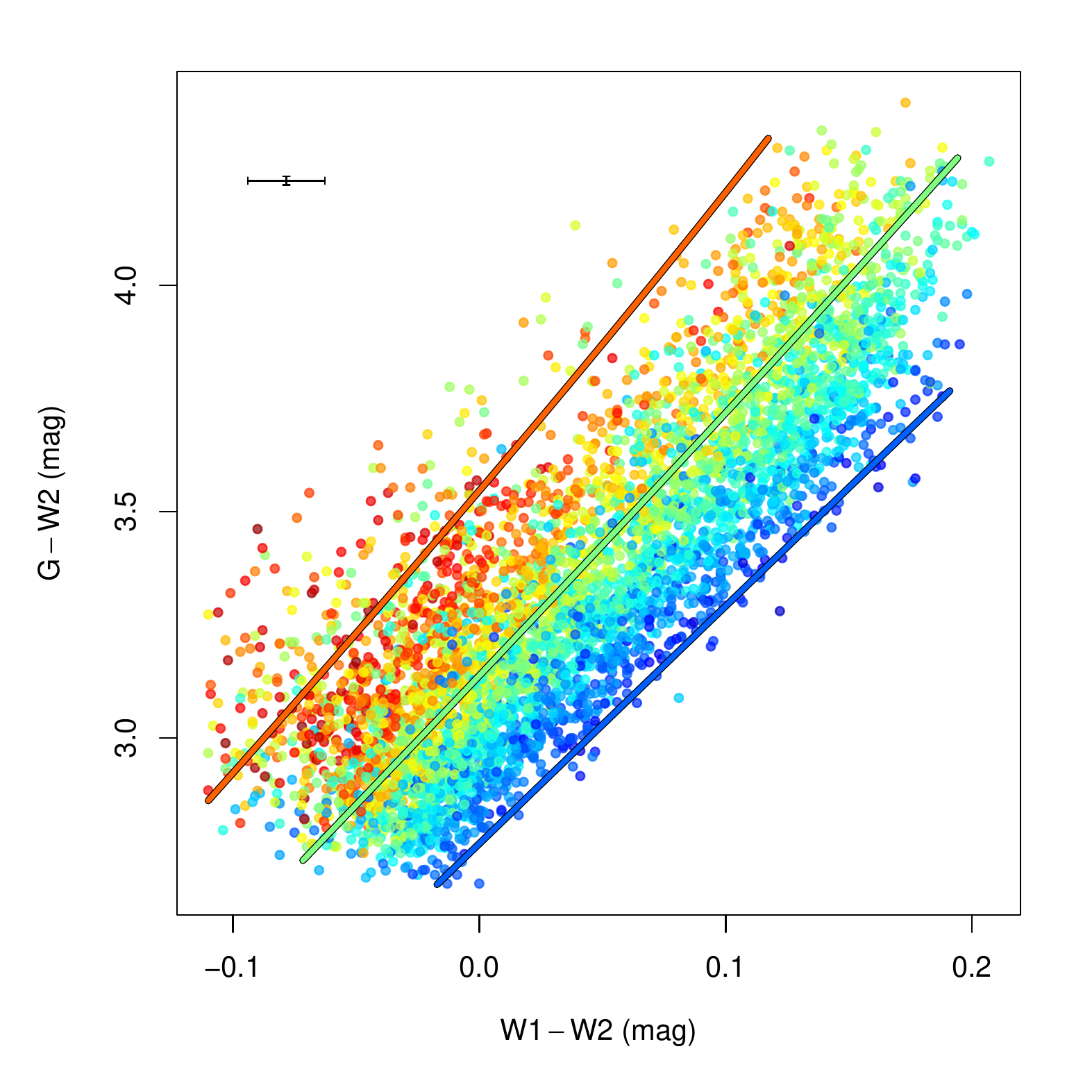} &
         \includegraphics[width=0.28\textwidth]{color_color_W1W2_BPRP_new} \\
         \includegraphics[width=0.28\textwidth]{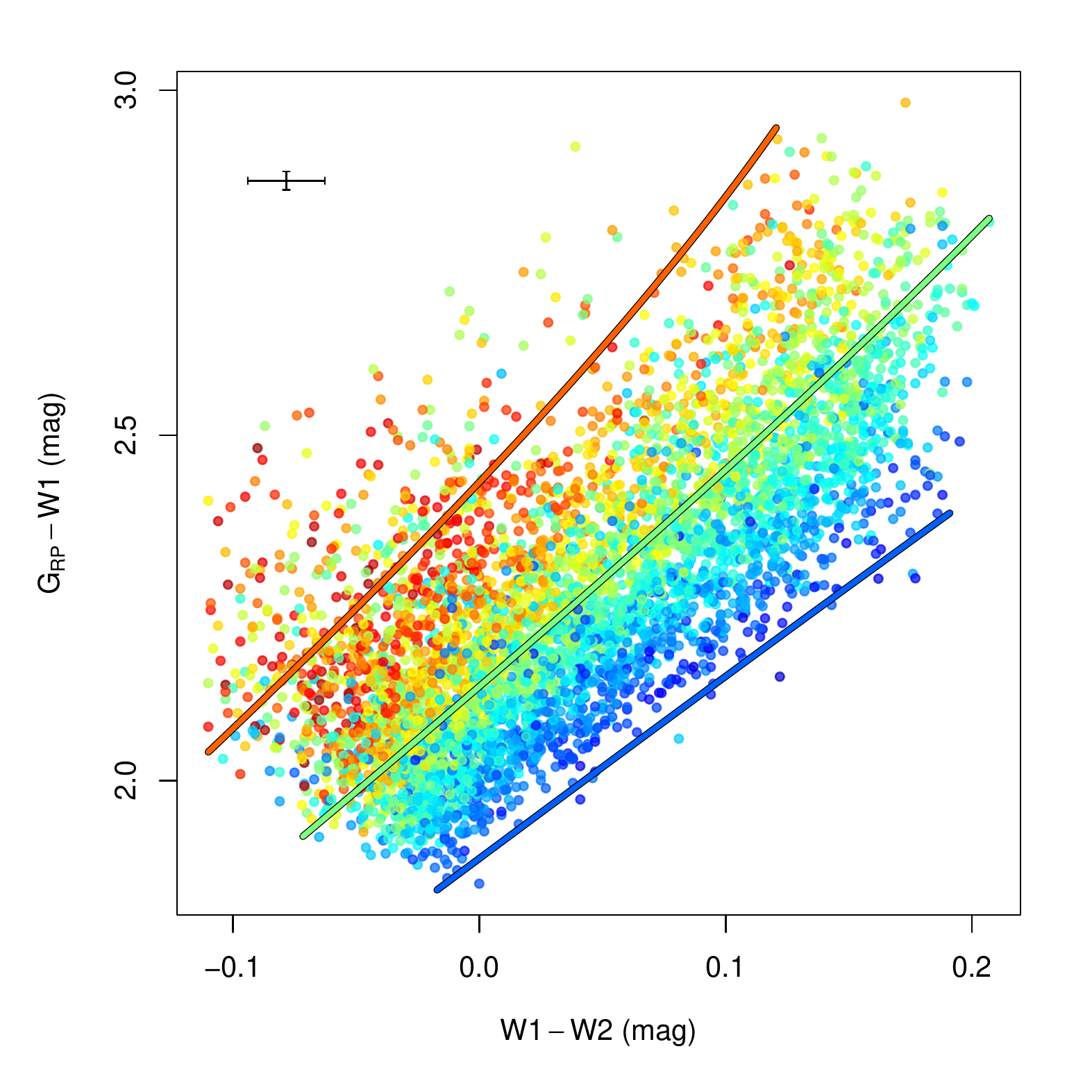} &
         \includegraphics[width=0.28\textwidth]{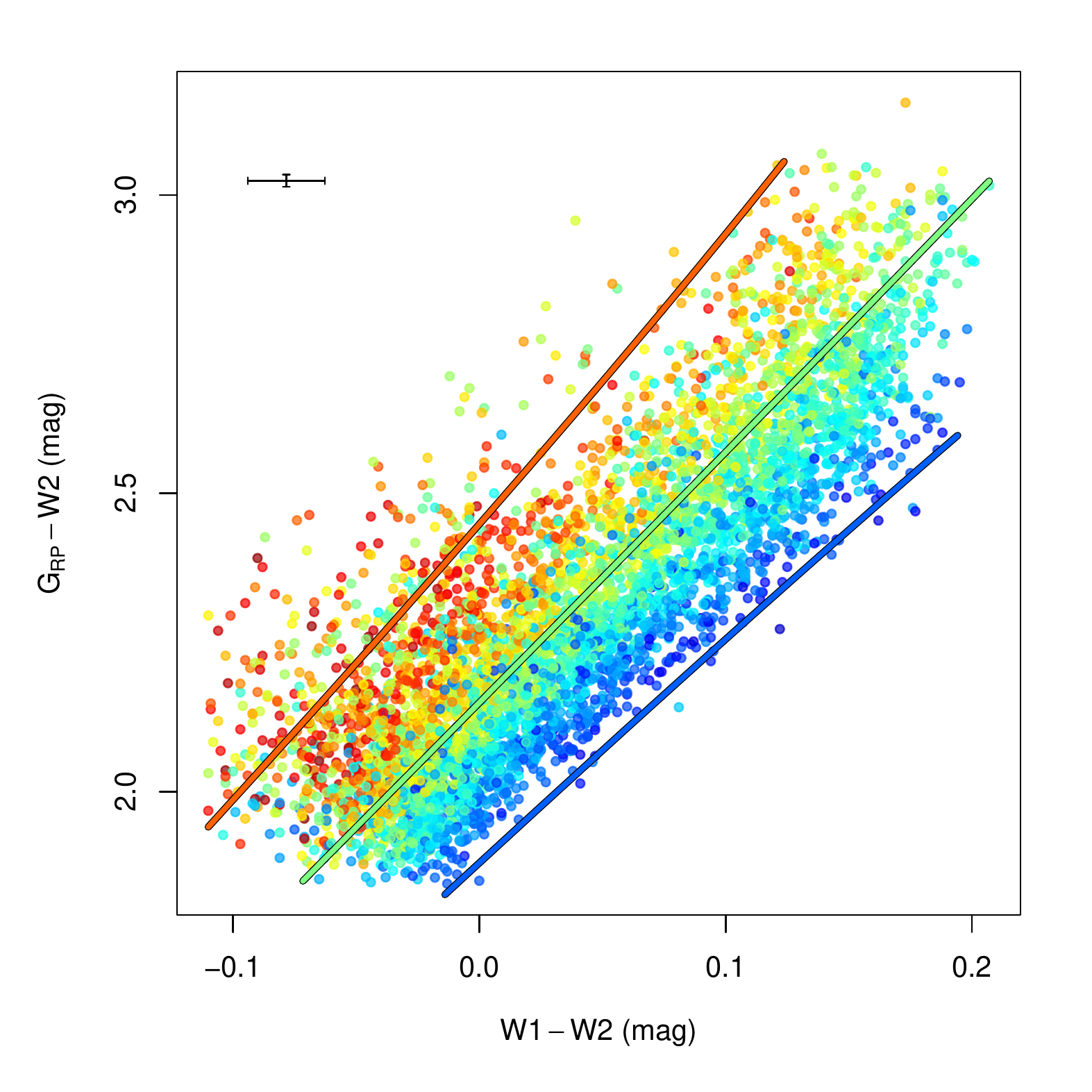} 
    \end{tabular}
    \includegraphics[width=0.25\textwidth]{metallicity_code.png}
    \caption{Color--color diagrams of the stars from \citetalias{Birky2020ApJ...892...31B}, color-coded by [Fe/H]$_\text{spec}$, with the respective calibrations given by this work (red: $\text{[Fe/H]} = +0.3$, green: $\text{[Fe/H]} = 0.0$, blue: $\text{[Fe/H]} = -0.3$). The mean uncertainties of the color indices involved are also shown in black in the bottom right corner of the panels.}
    \label{color_color_diagrams_appendix}
\end{figure*}

\renewcommand\thefigure{A.2}
\begin{figure*}
    \centering
    \begin{tabular}{ccc}
         \includegraphics[width=0.28\textwidth]{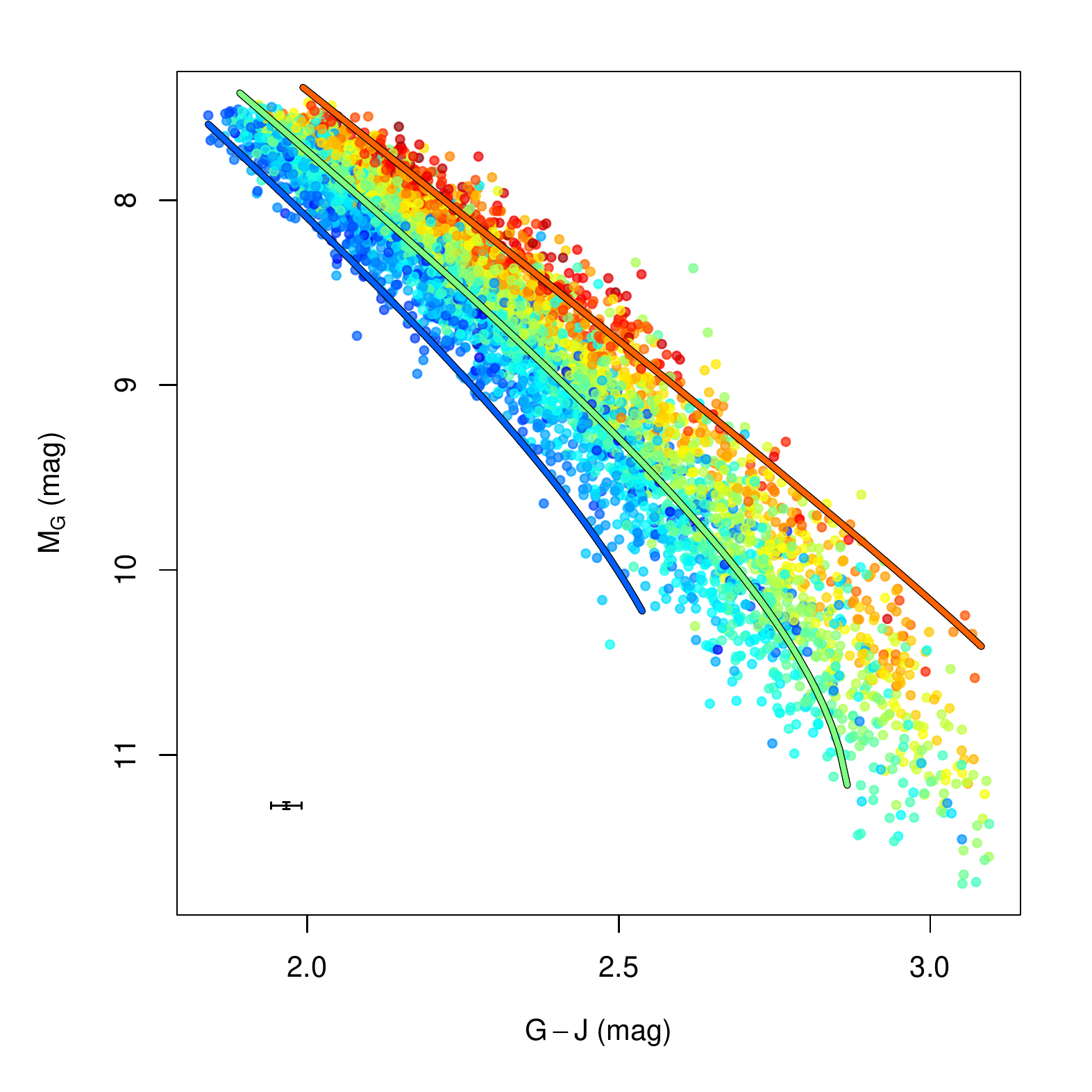} &
         \includegraphics[width=0.28\textwidth]{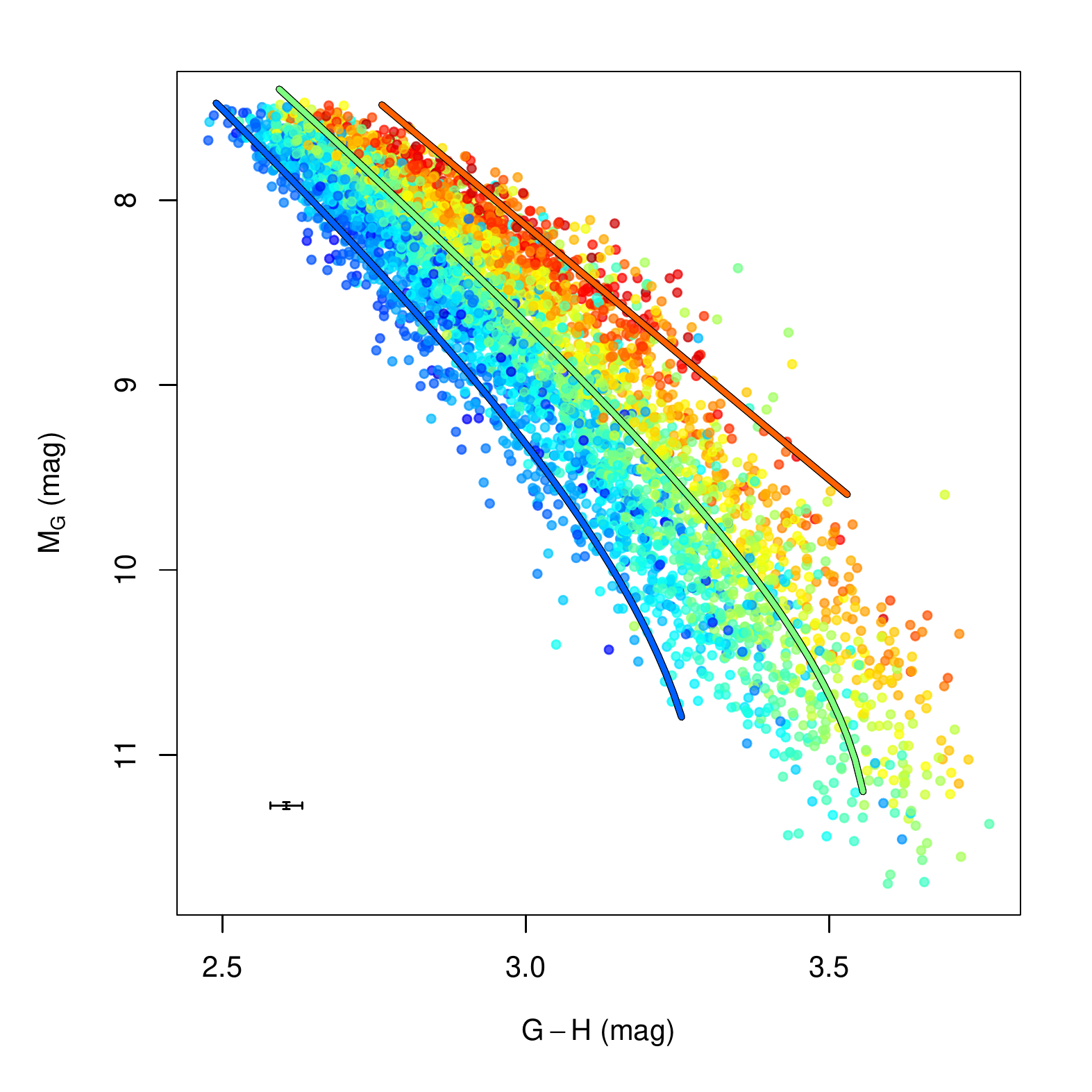} &
         \includegraphics[width=0.28\textwidth]{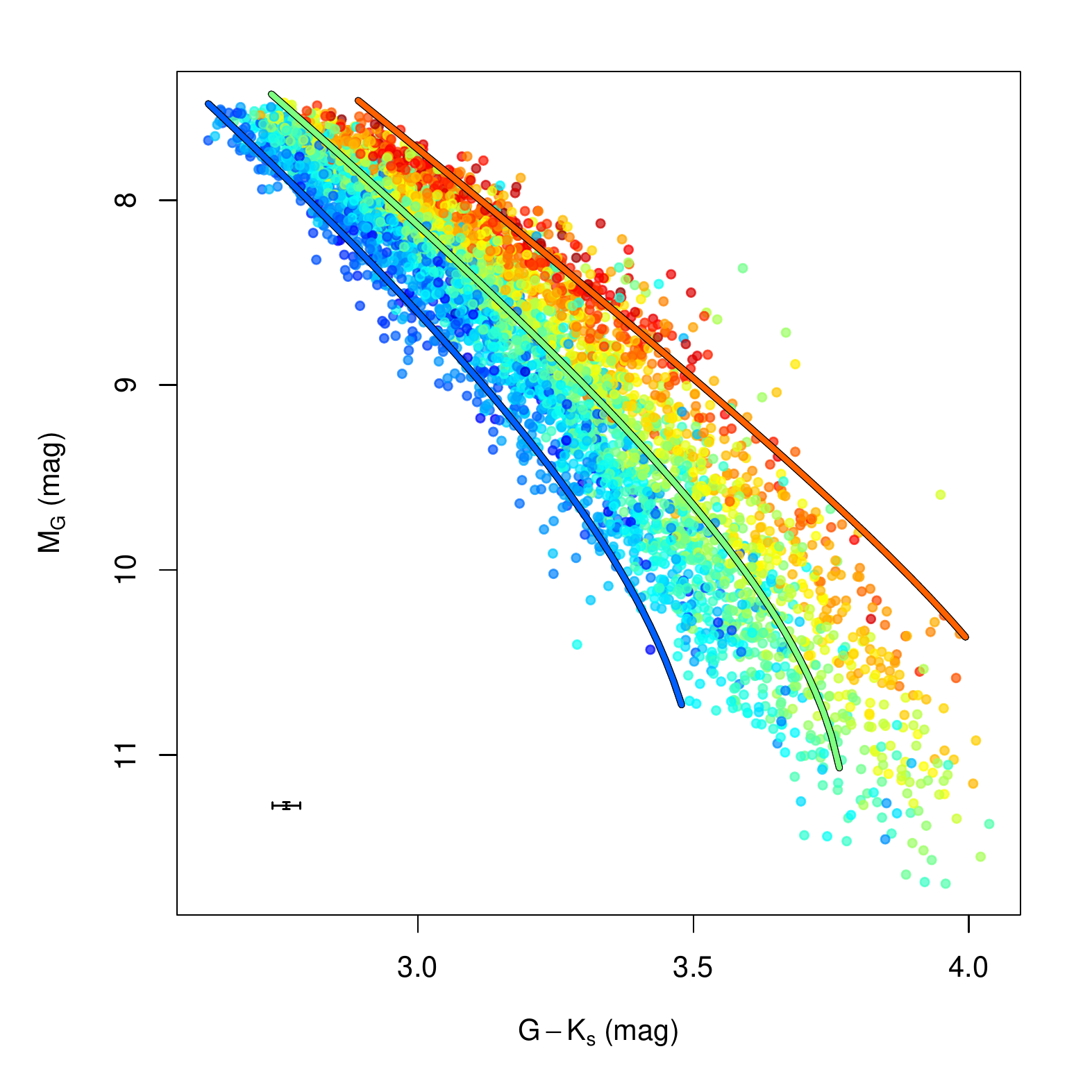} \\
         \includegraphics[width=0.28\textwidth]{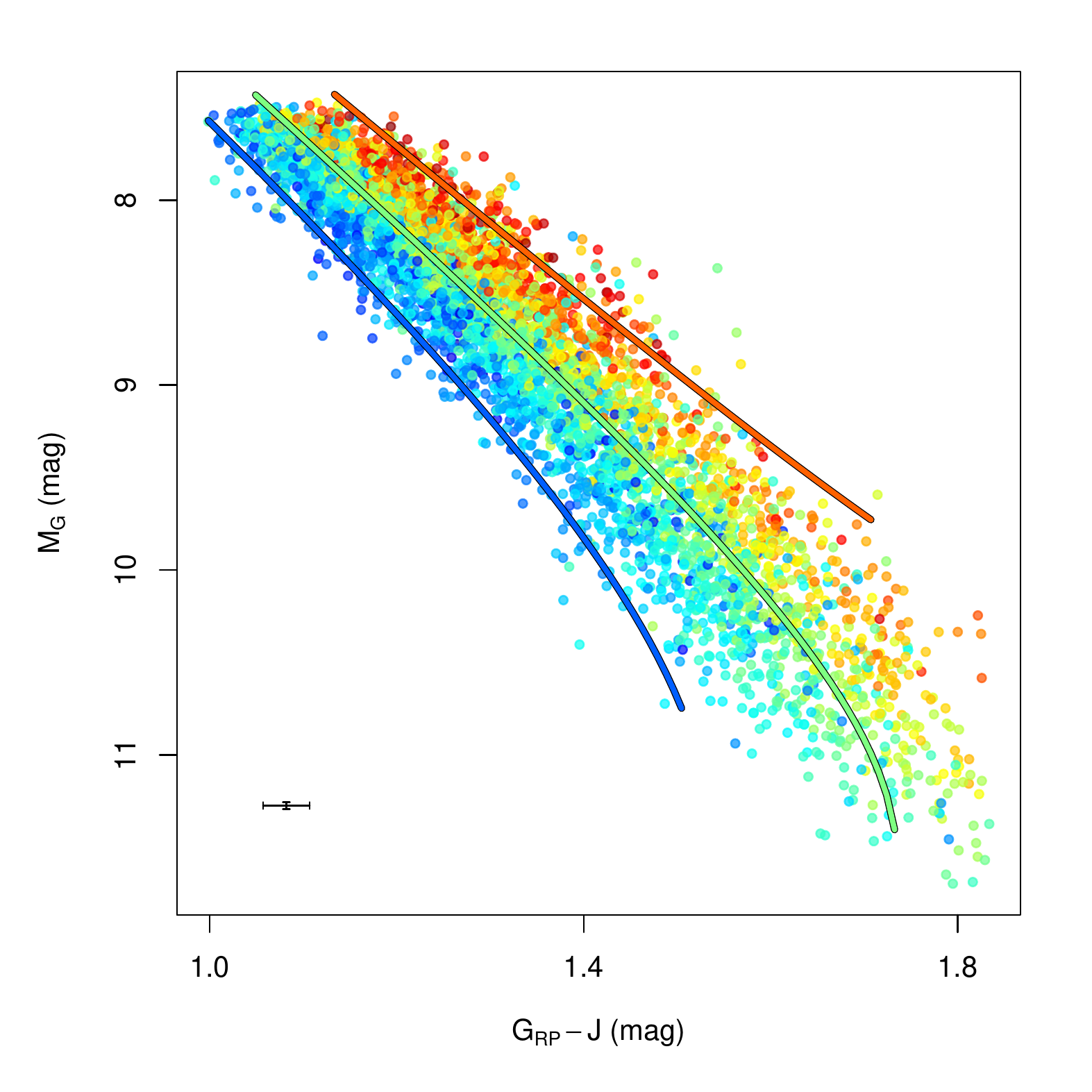} &
         \includegraphics[width=0.28\textwidth]{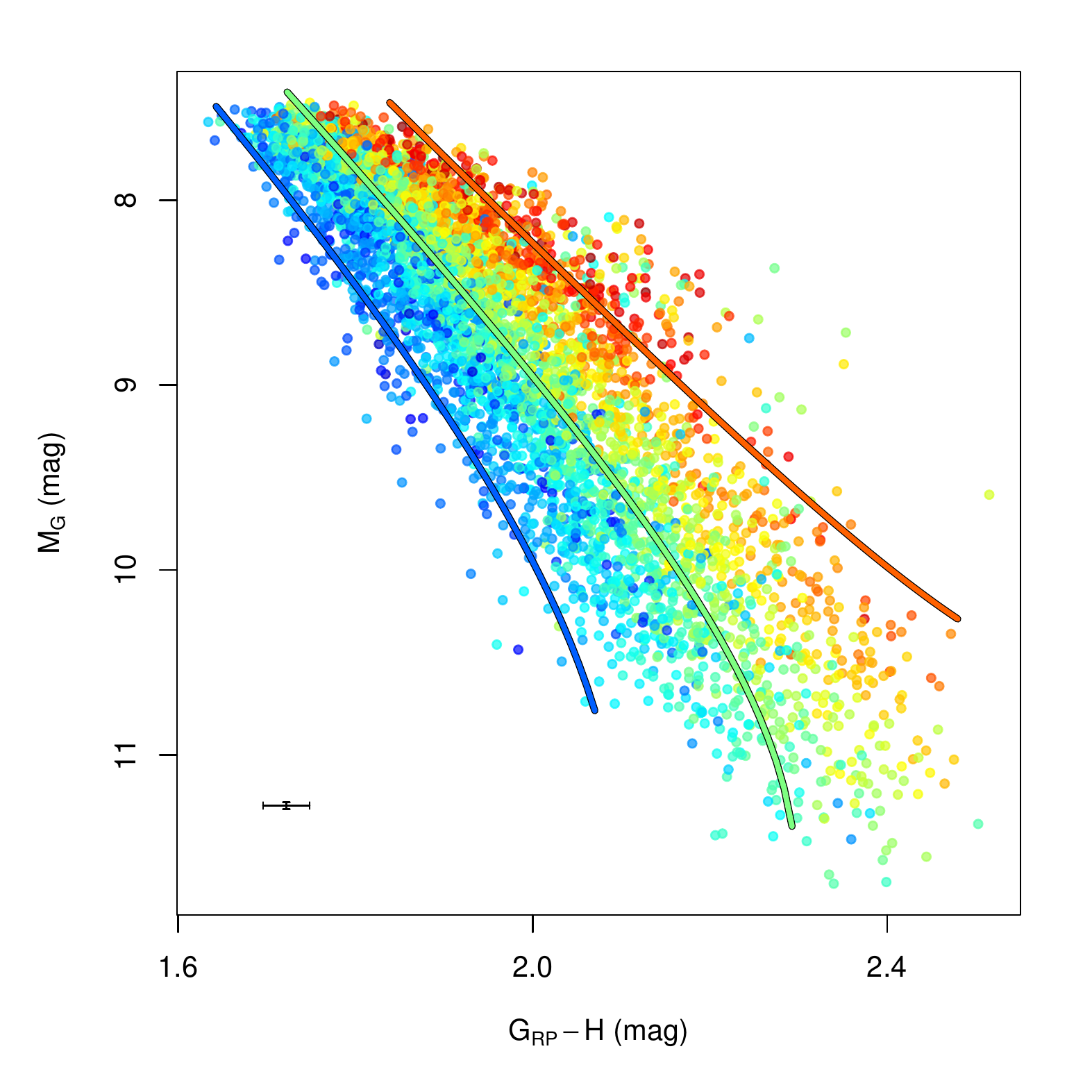} &
         \includegraphics[width=0.28\textwidth]{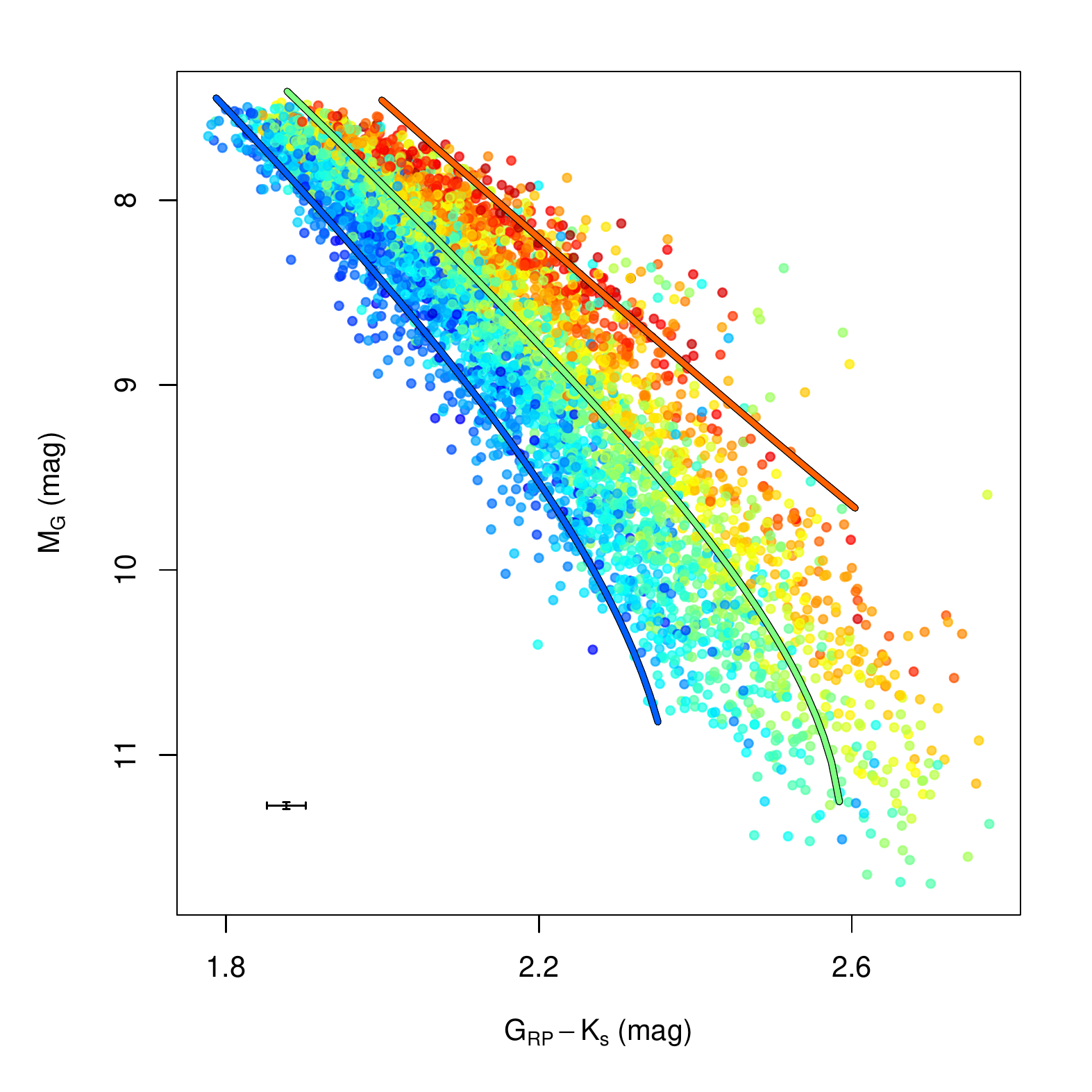} \\
         \includegraphics[width=0.28\textwidth]{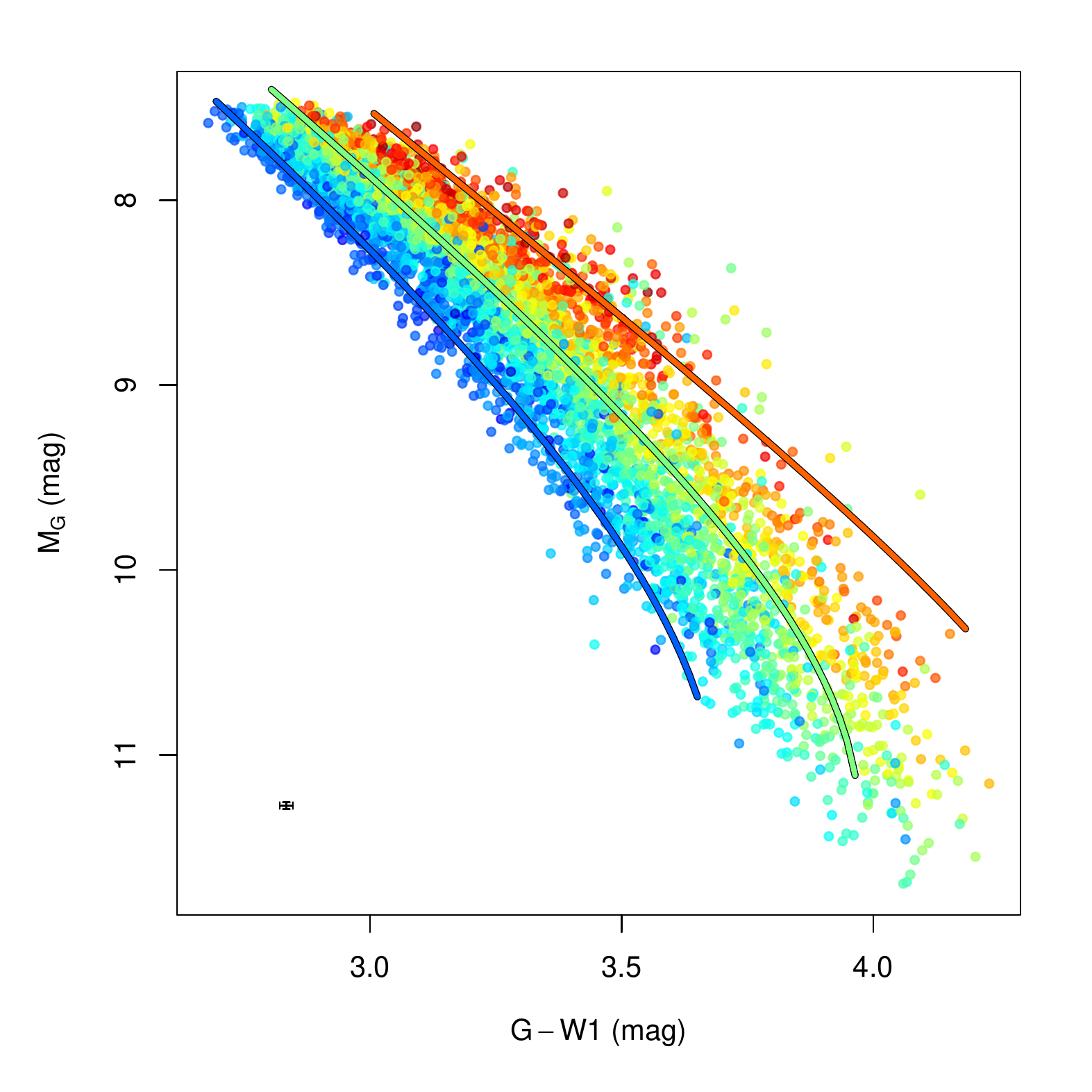} &
         \includegraphics[width=0.28\textwidth]{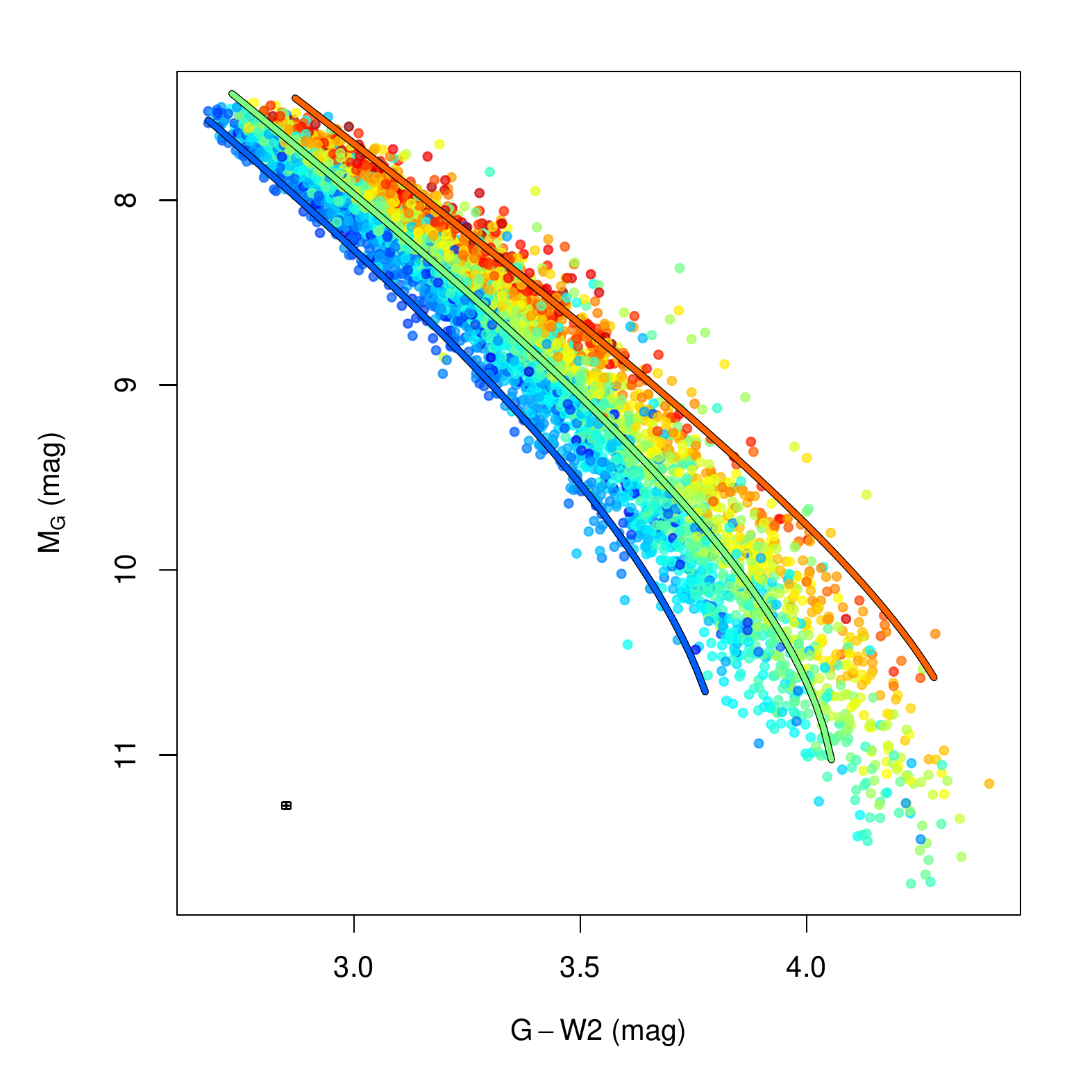} &
         \includegraphics[width=0.28\textwidth]{color_magnitude_BPRP_G_new} \\
         \includegraphics[width=0.28\textwidth]{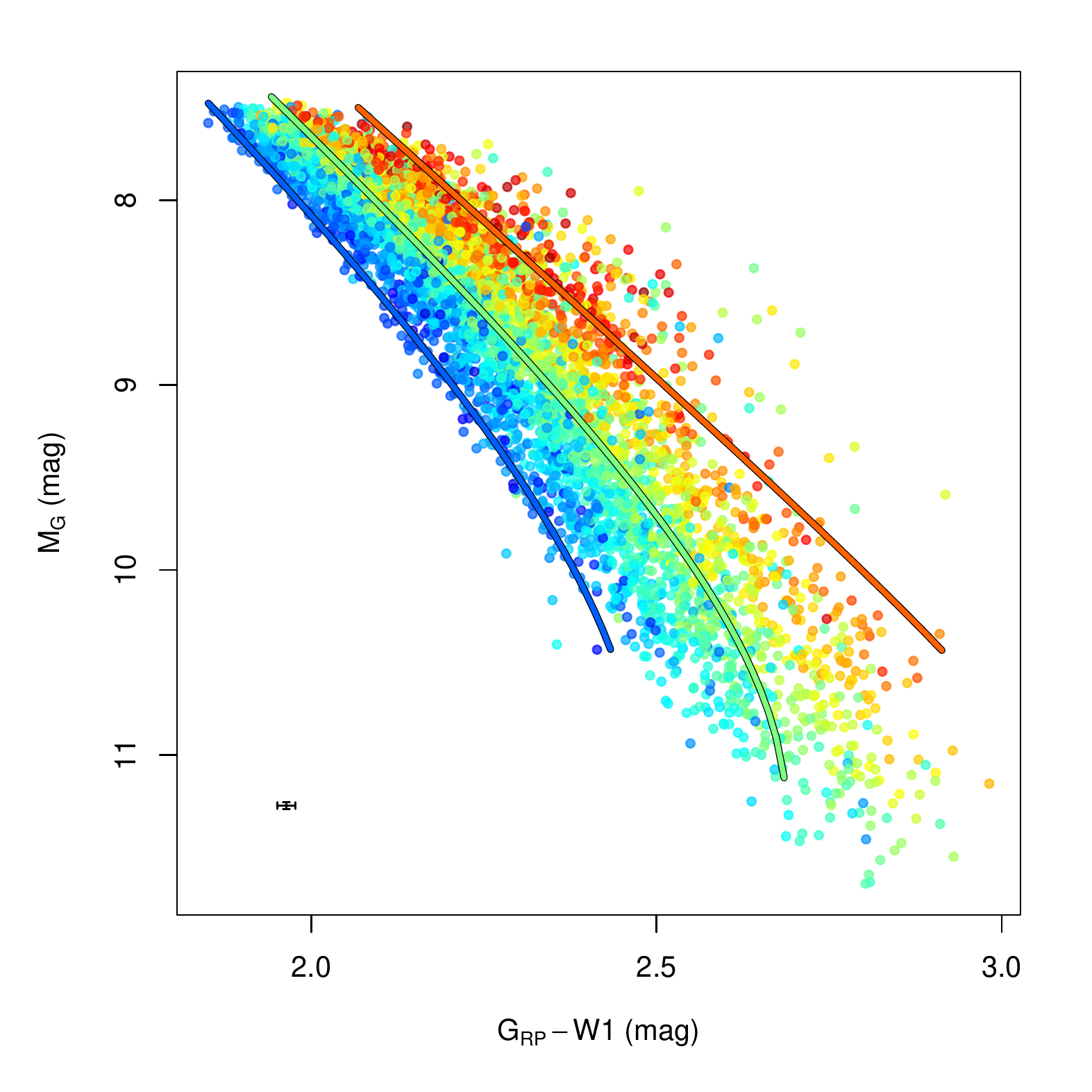} &
         \includegraphics[width=0.28\textwidth]{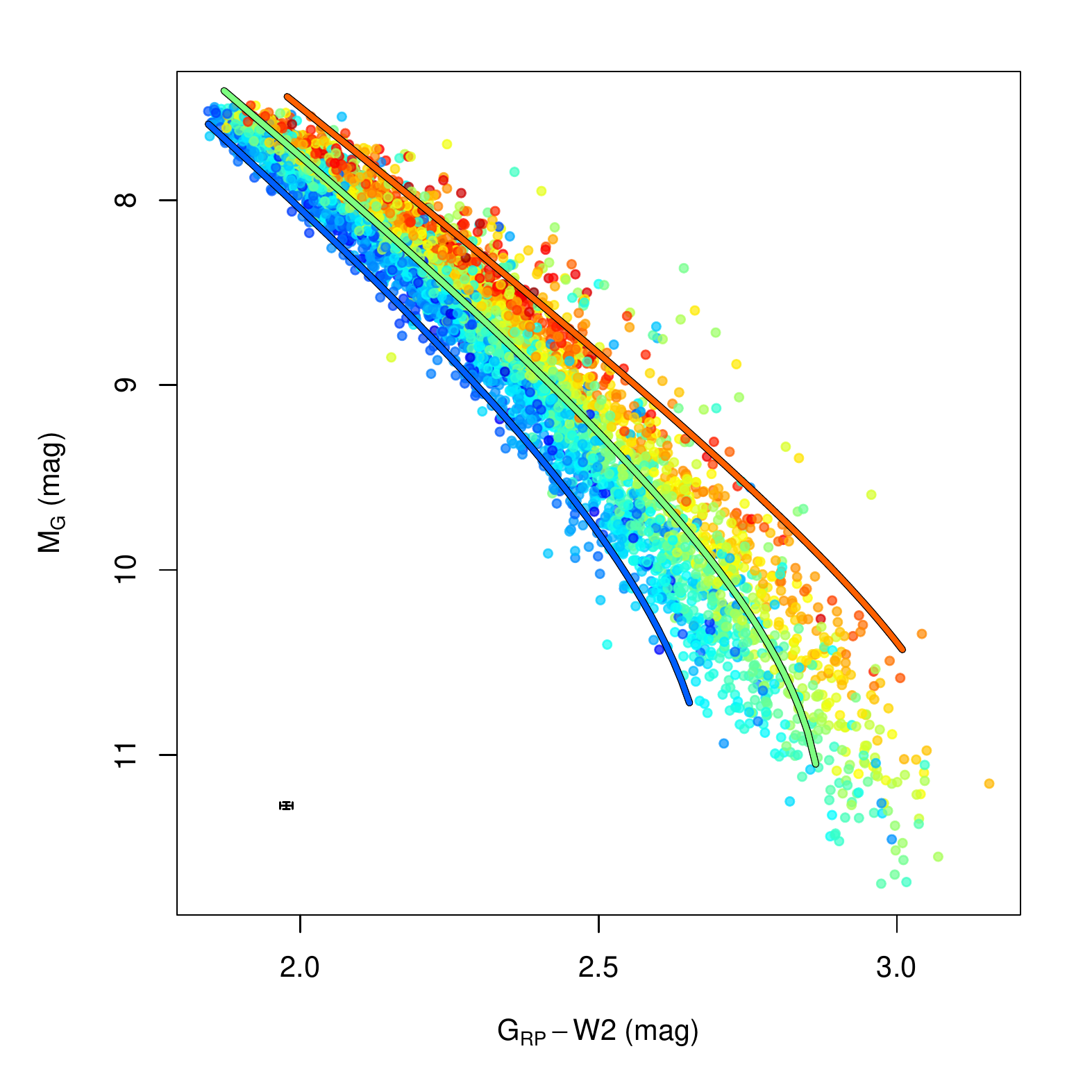} 
    \end{tabular}
    \includegraphics[width=0.25\textwidth]{metallicity_code.png}
    \caption{Same as Fig.~\ref{color_magnitude_diagrams_appendix_G} but for color--magnitude diagrams with $M_G$}
    \label{color_magnitude_diagrams_appendix_G}
\end{figure*}

\renewcommand\thefigure{A.3}
\begin{figure*}
    \centering
    \begin{tabular}{ccc}
         \includegraphics[width=0.28\textwidth]{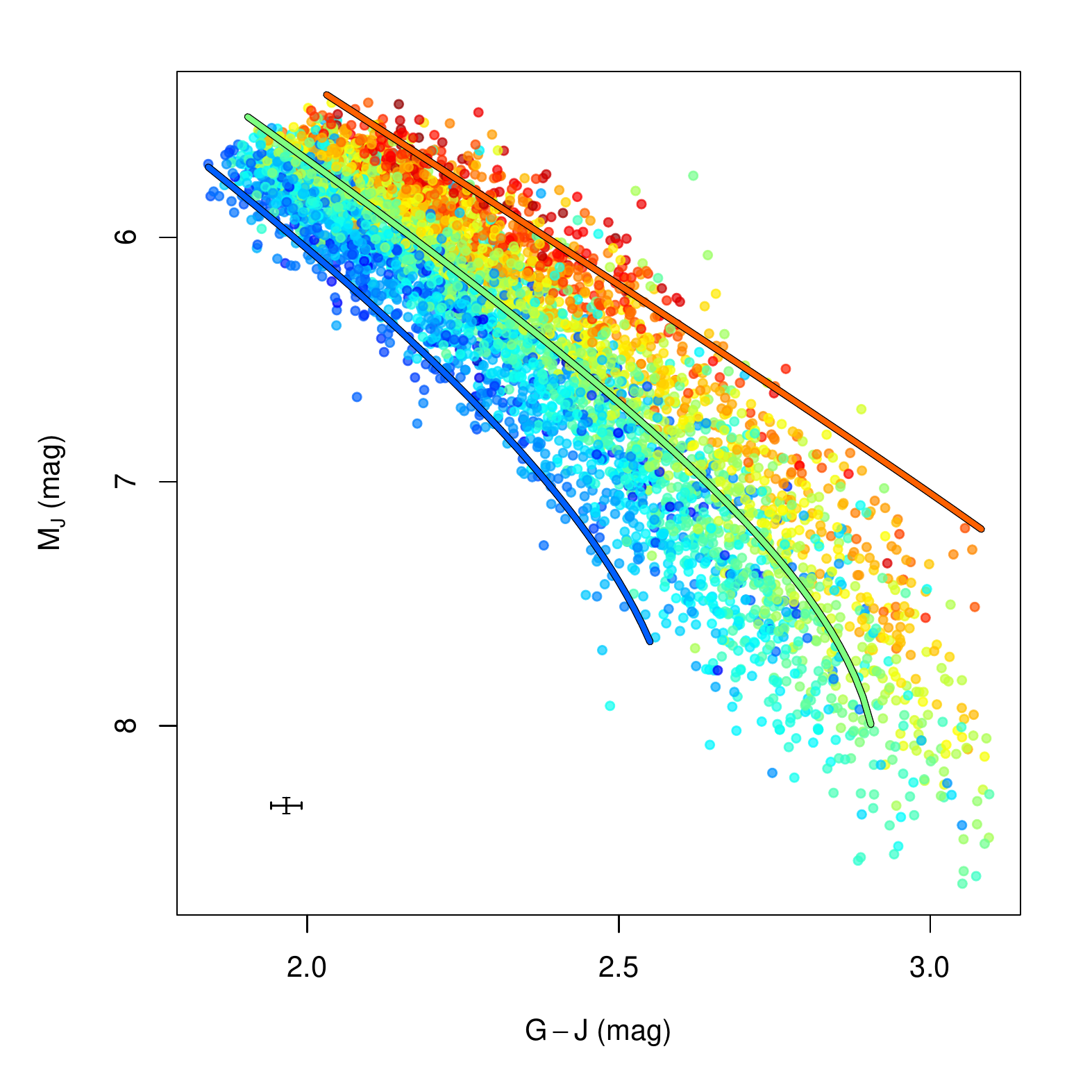} &
         \includegraphics[width=0.28\textwidth]{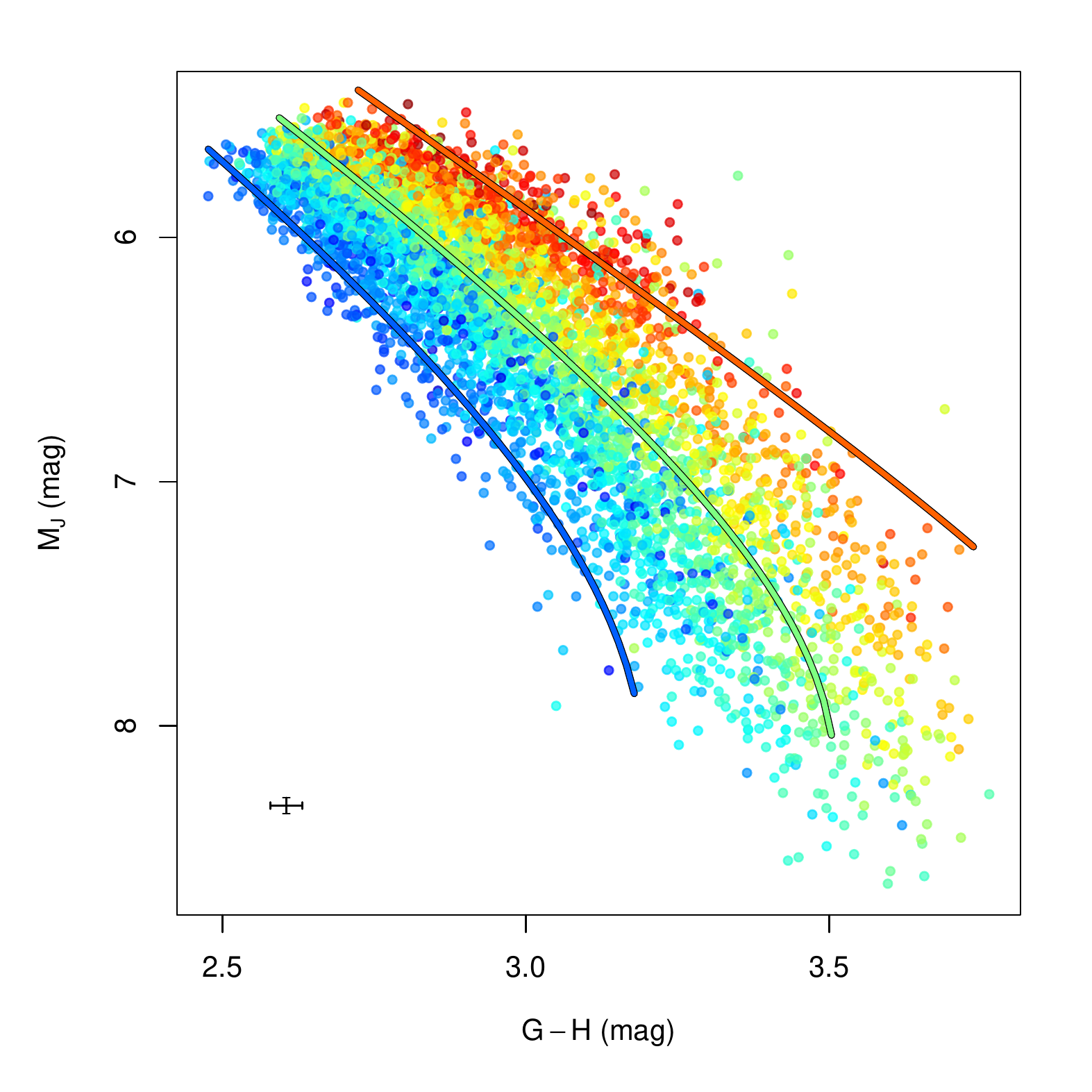} &
         \includegraphics[width=0.28\textwidth]{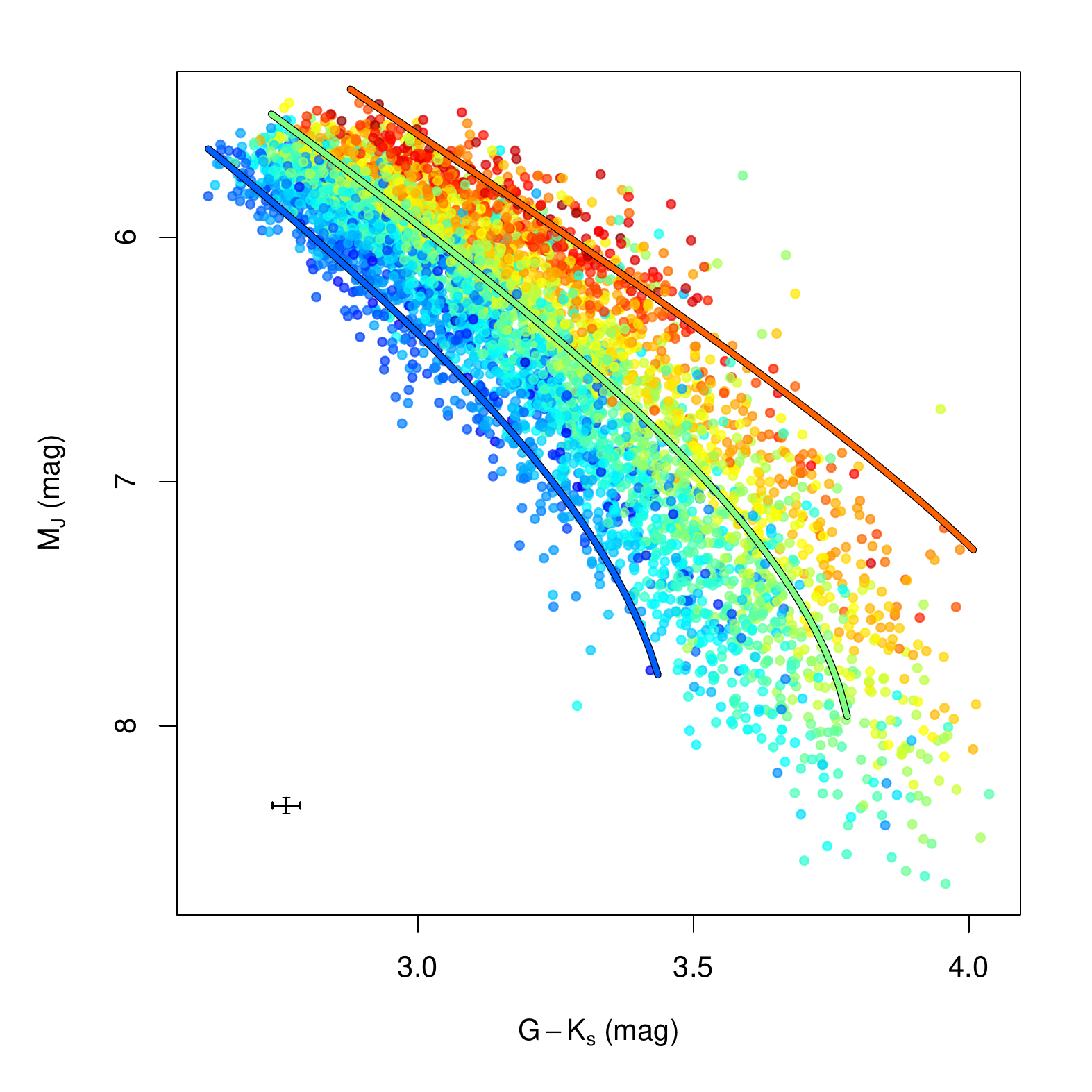} \\
         \includegraphics[width=0.28\textwidth]{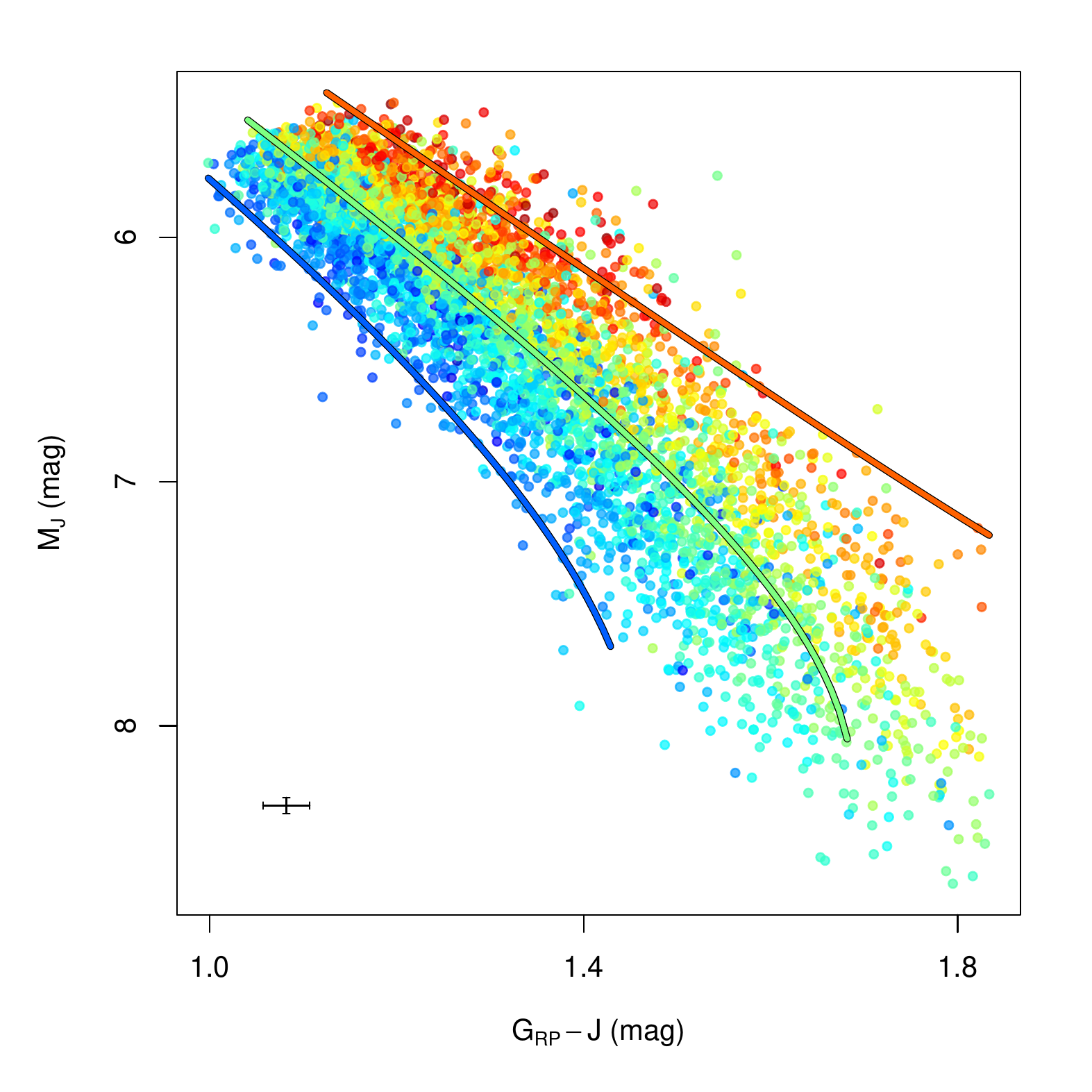} &
         \includegraphics[width=0.28\textwidth]{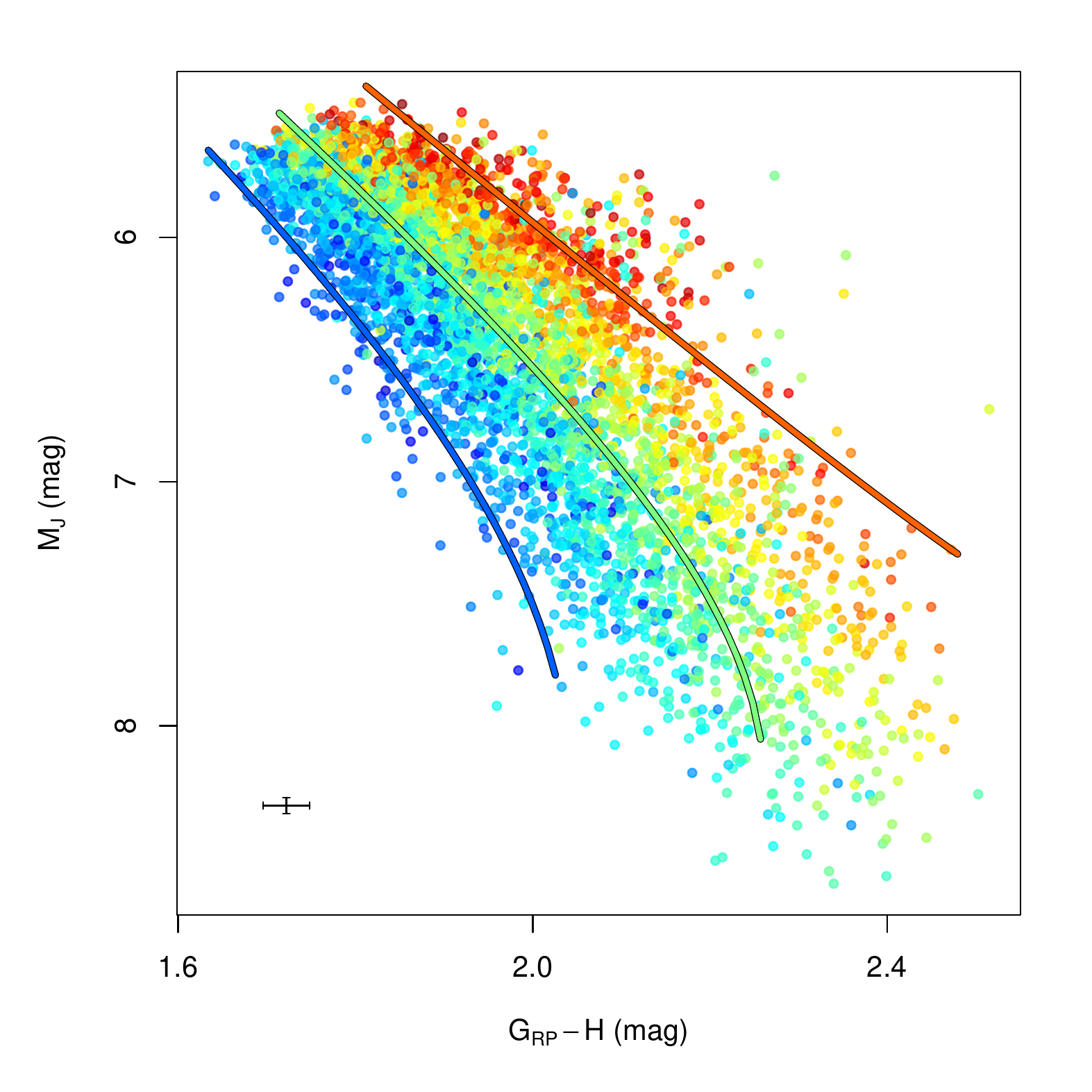} &
         \includegraphics[width=0.28\textwidth]{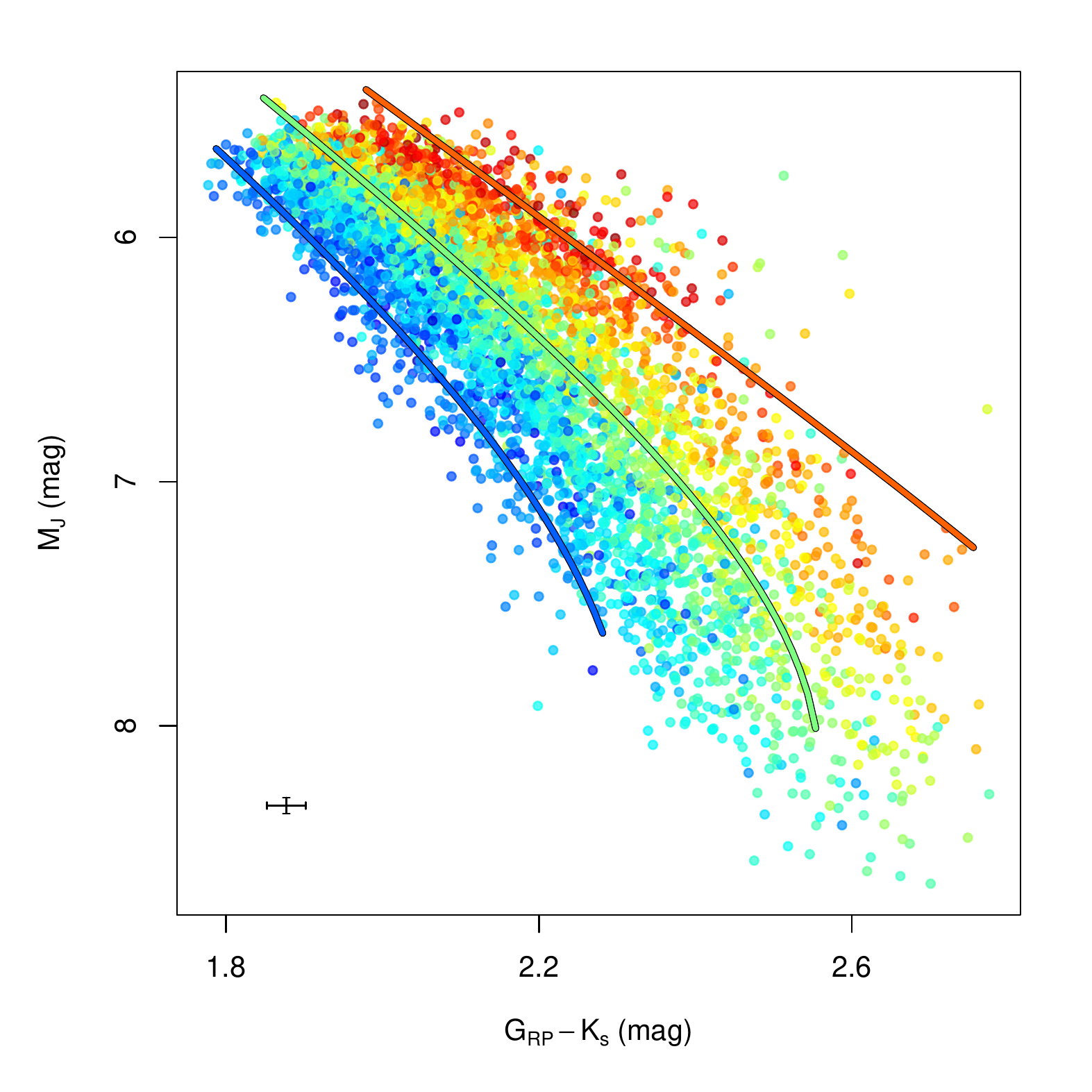} \\
         \includegraphics[width=0.28\textwidth]{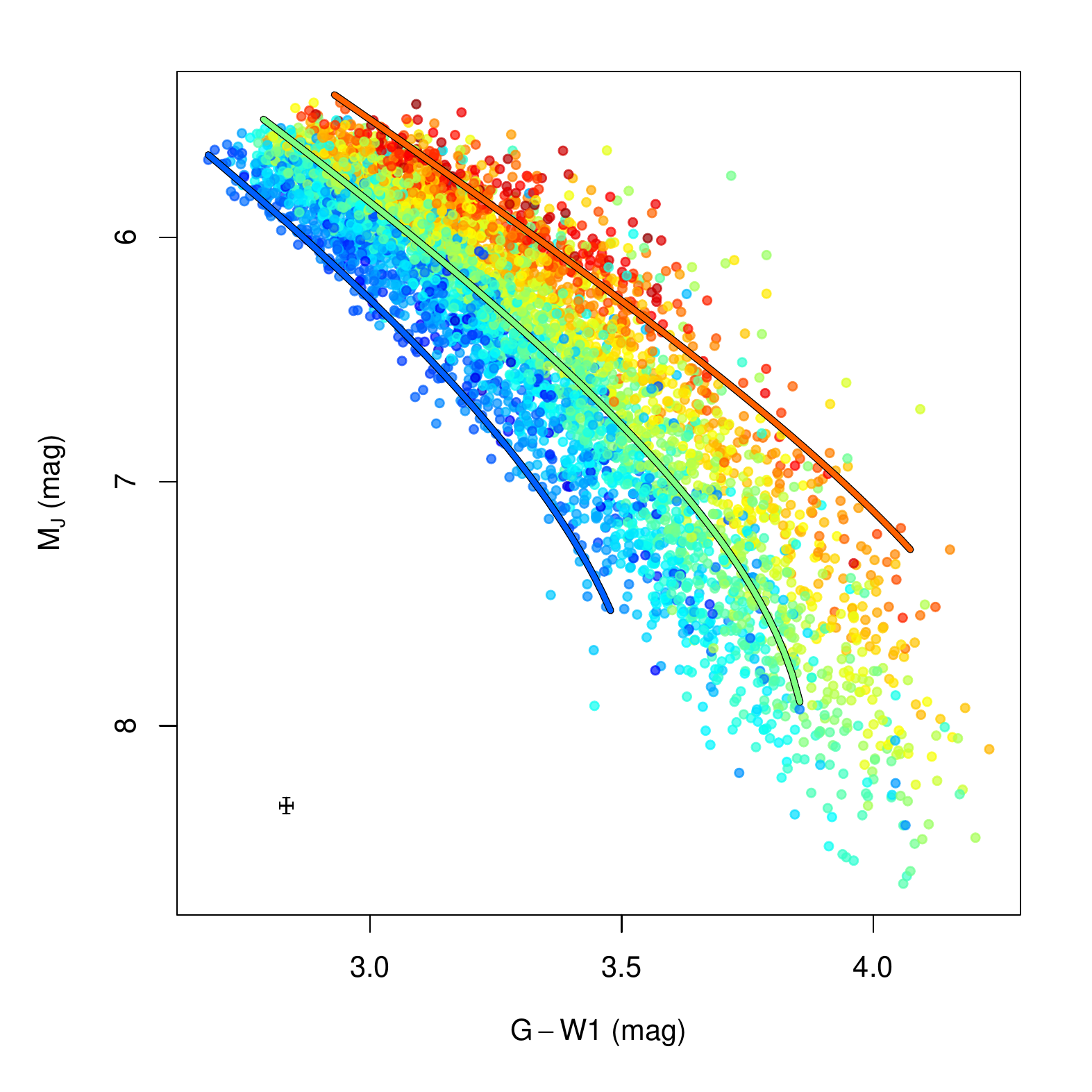} &
         \includegraphics[width=0.28\textwidth]{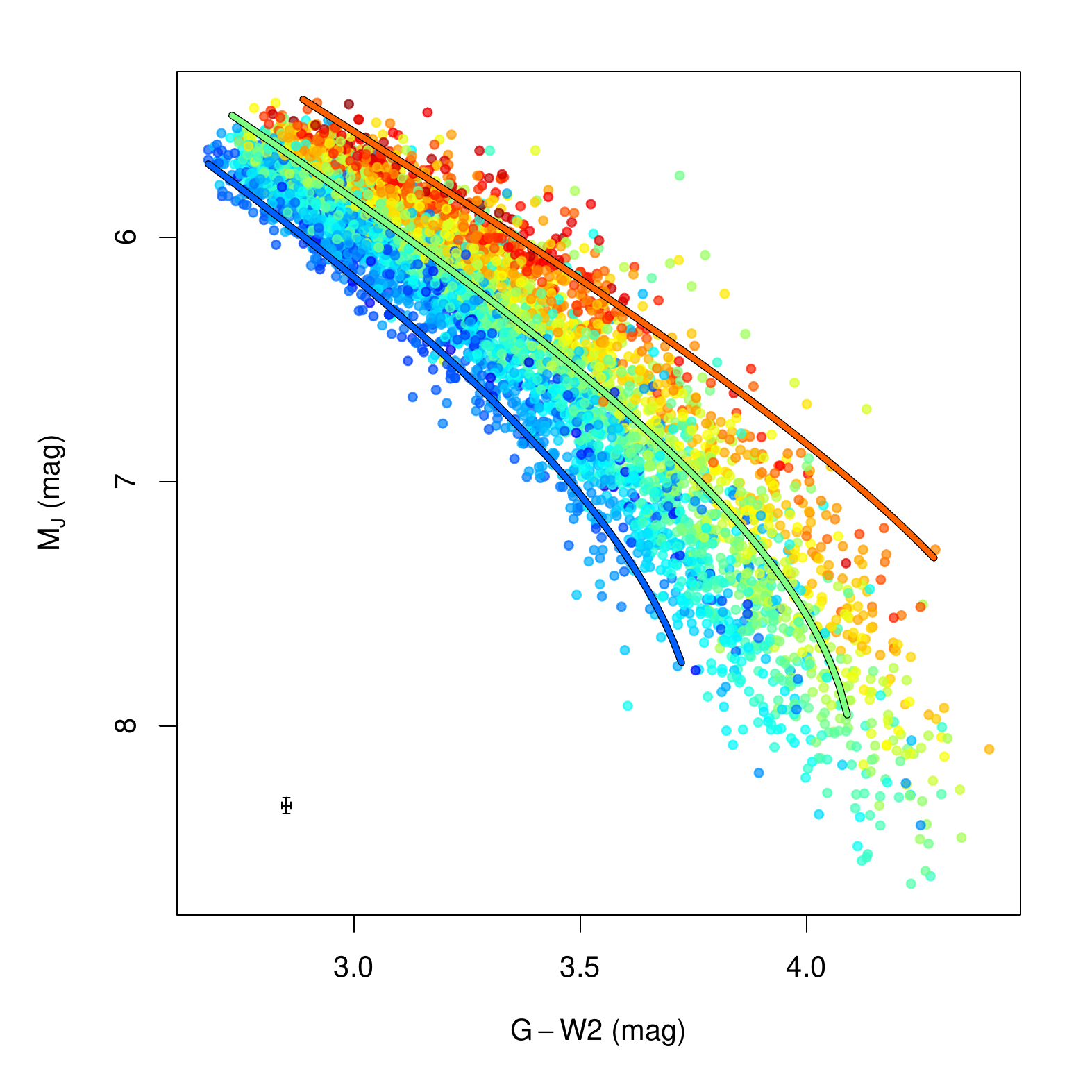} &
         \includegraphics[width=0.28\textwidth]{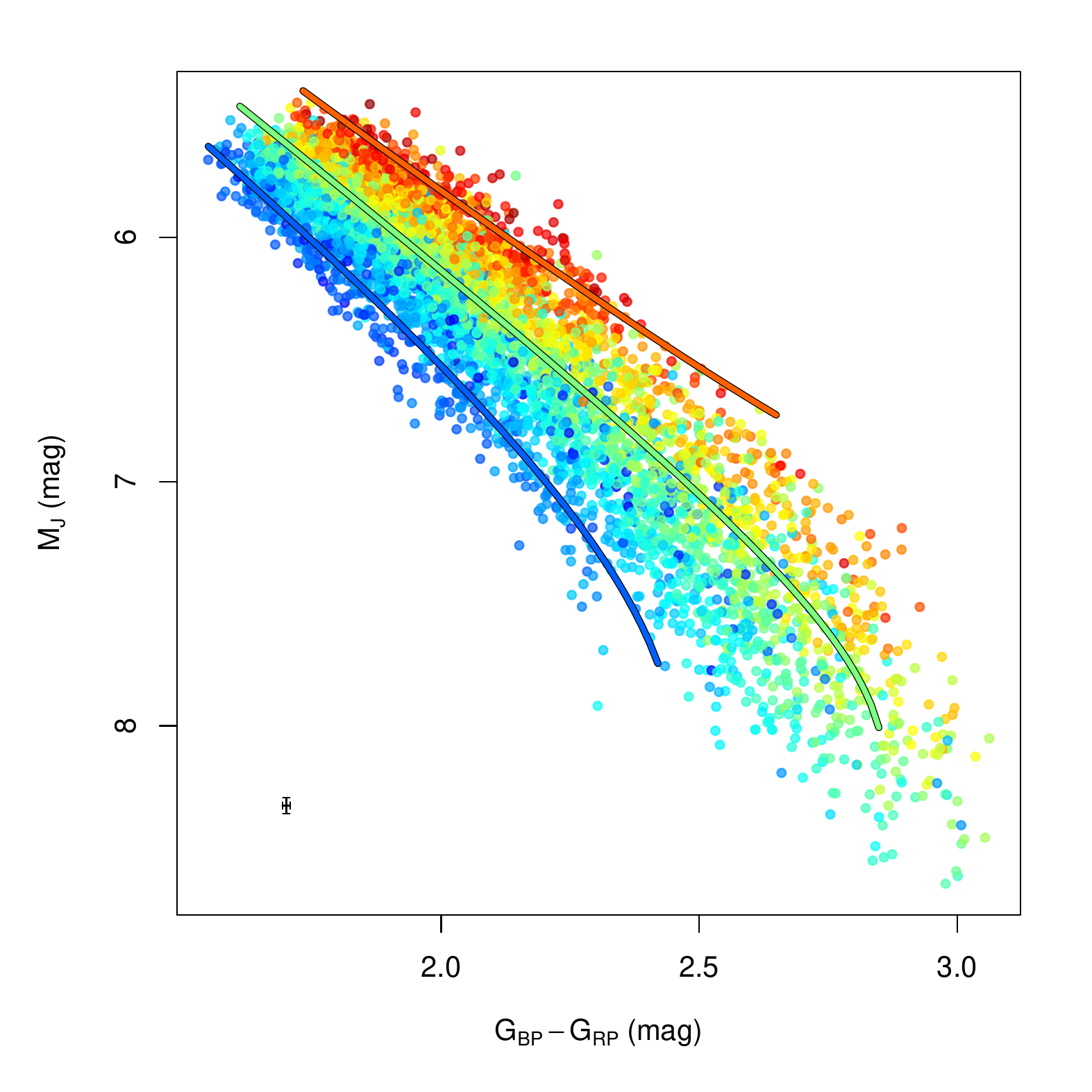} \\
         \includegraphics[width=0.28\textwidth]{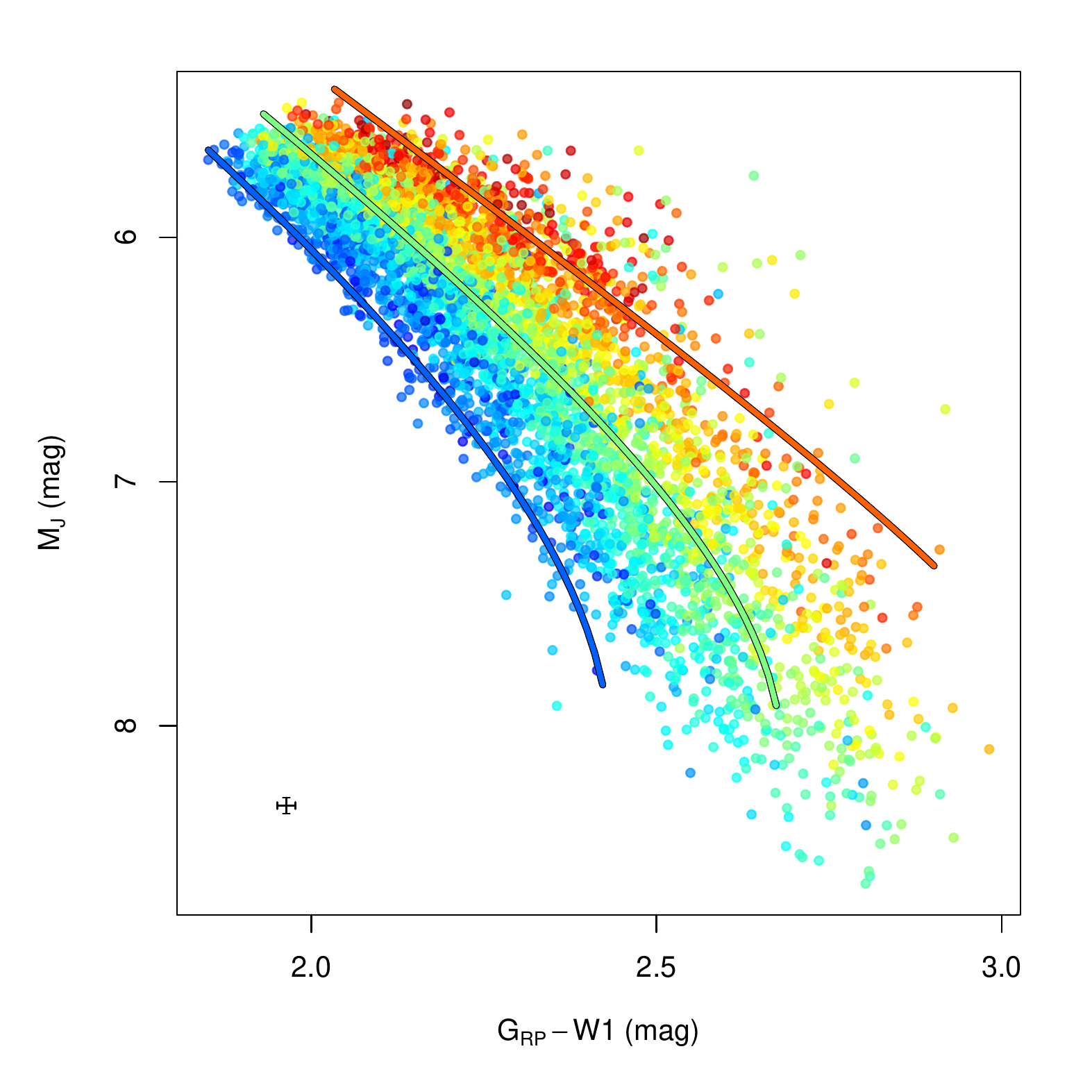} &
         \includegraphics[width=0.28\textwidth]{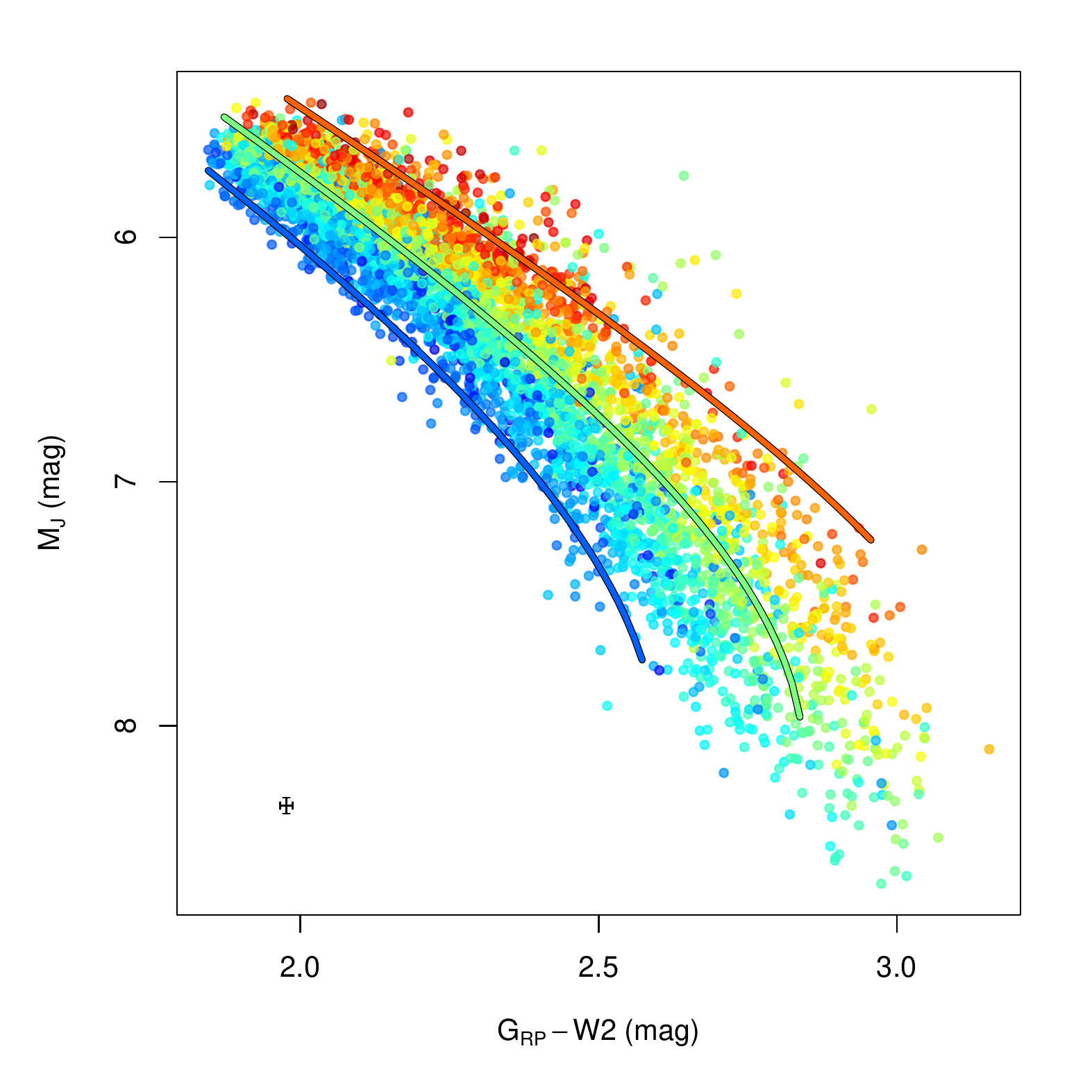} 
    \end{tabular}
    \includegraphics[width=0.25\textwidth]{metallicity_code.png}
    \caption{Same as Fig.~\ref{color_magnitude_diagrams_appendix_J} but for color--magnitude diagrams with $M_J$}
    \label{color_magnitude_diagrams_appendix_J}
\end{figure*}

\renewcommand\thefigure{A.4}
\begin{figure*}
    \centering
    \begin{tabular}{ccc}
         \includegraphics[width=0.28\textwidth]{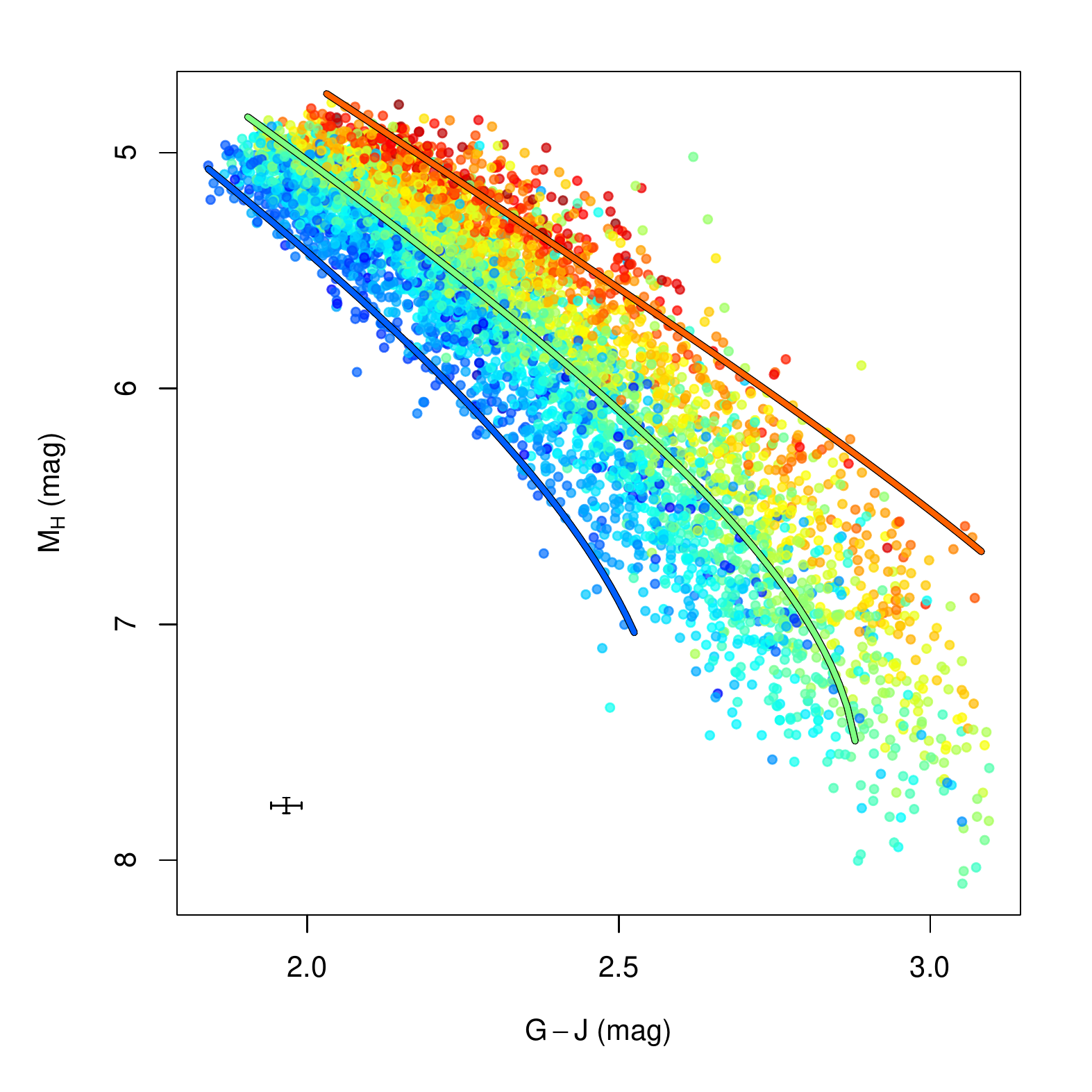} &
         \includegraphics[width=0.28\textwidth]{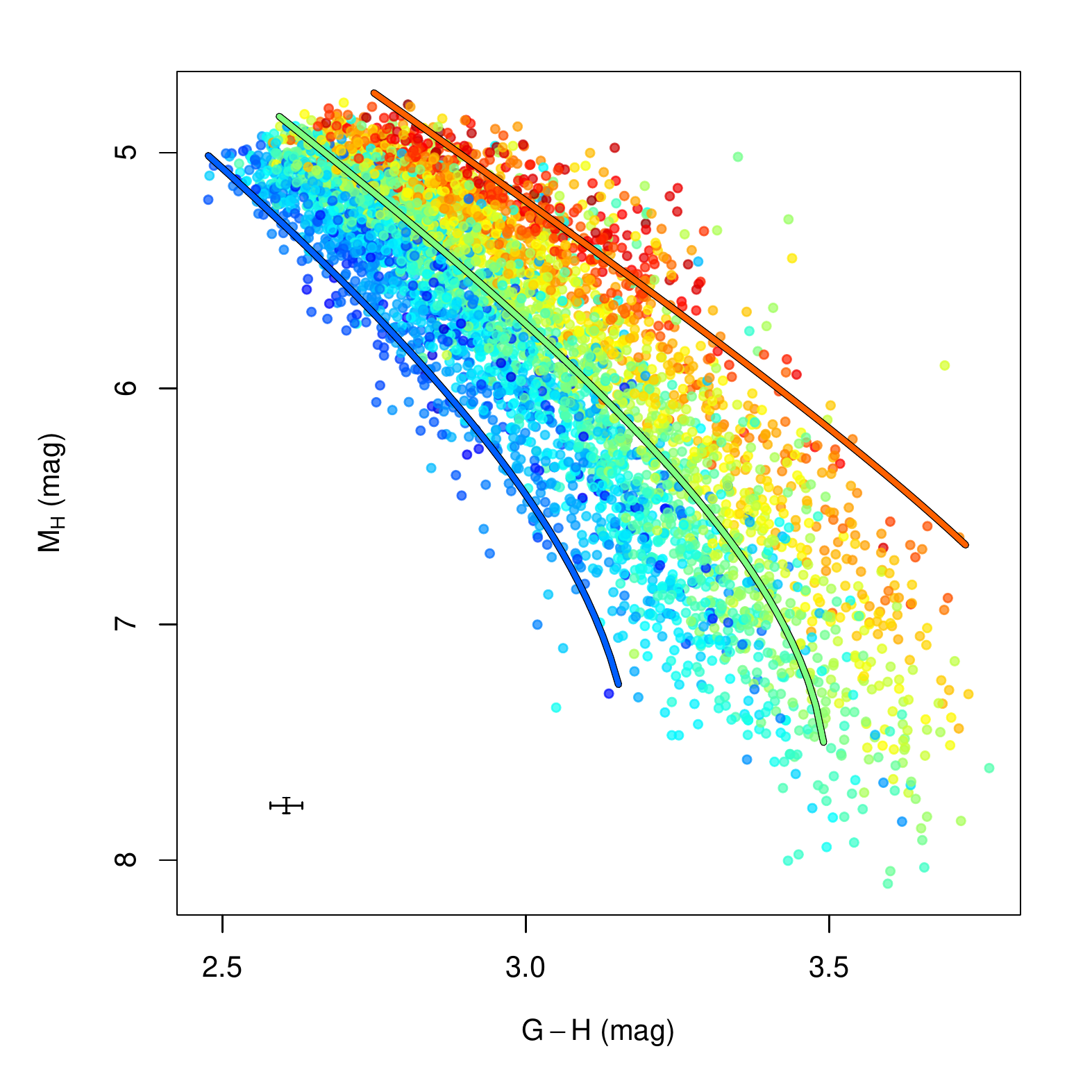} &
         \includegraphics[width=0.28\textwidth]{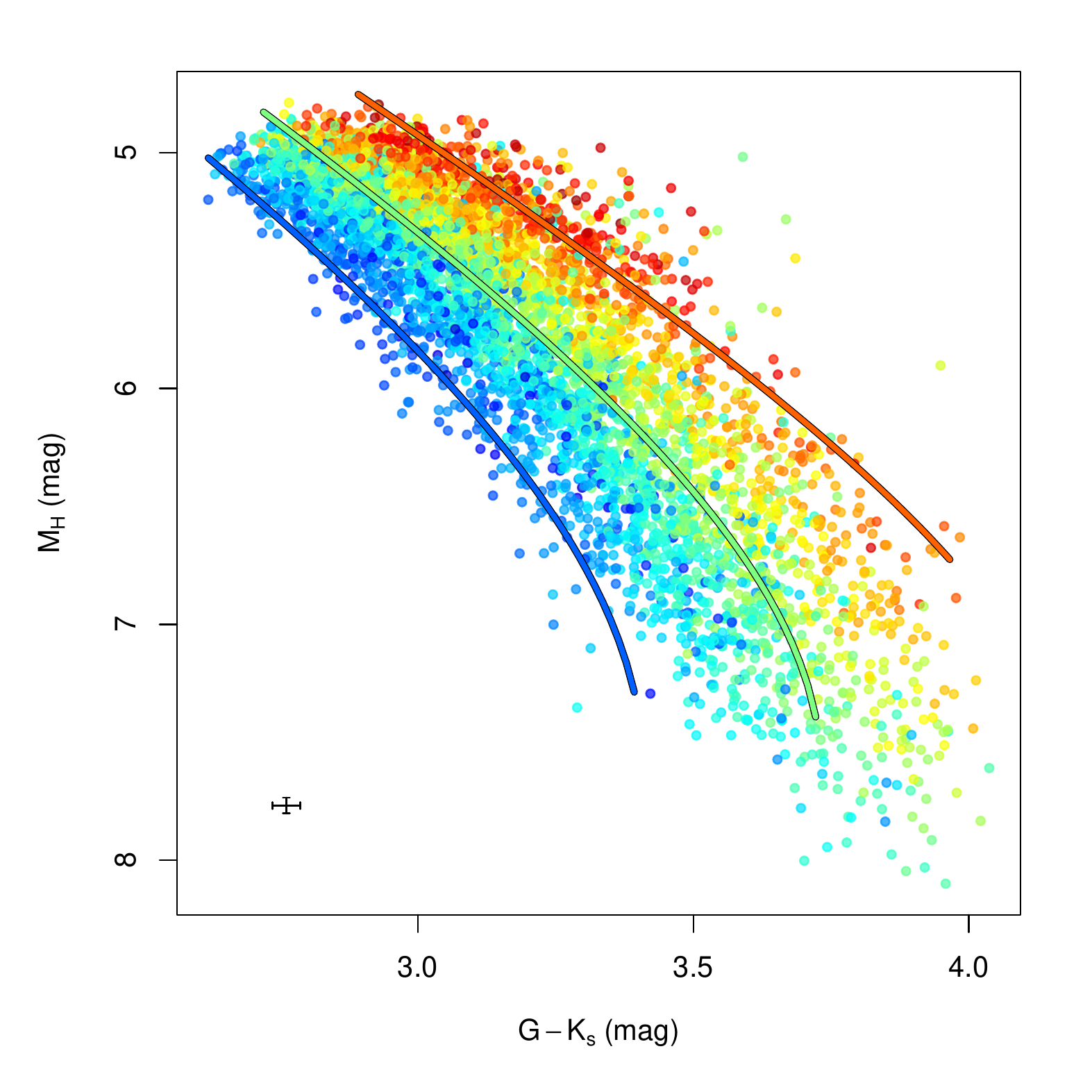} \\
         \includegraphics[width=0.28\textwidth]{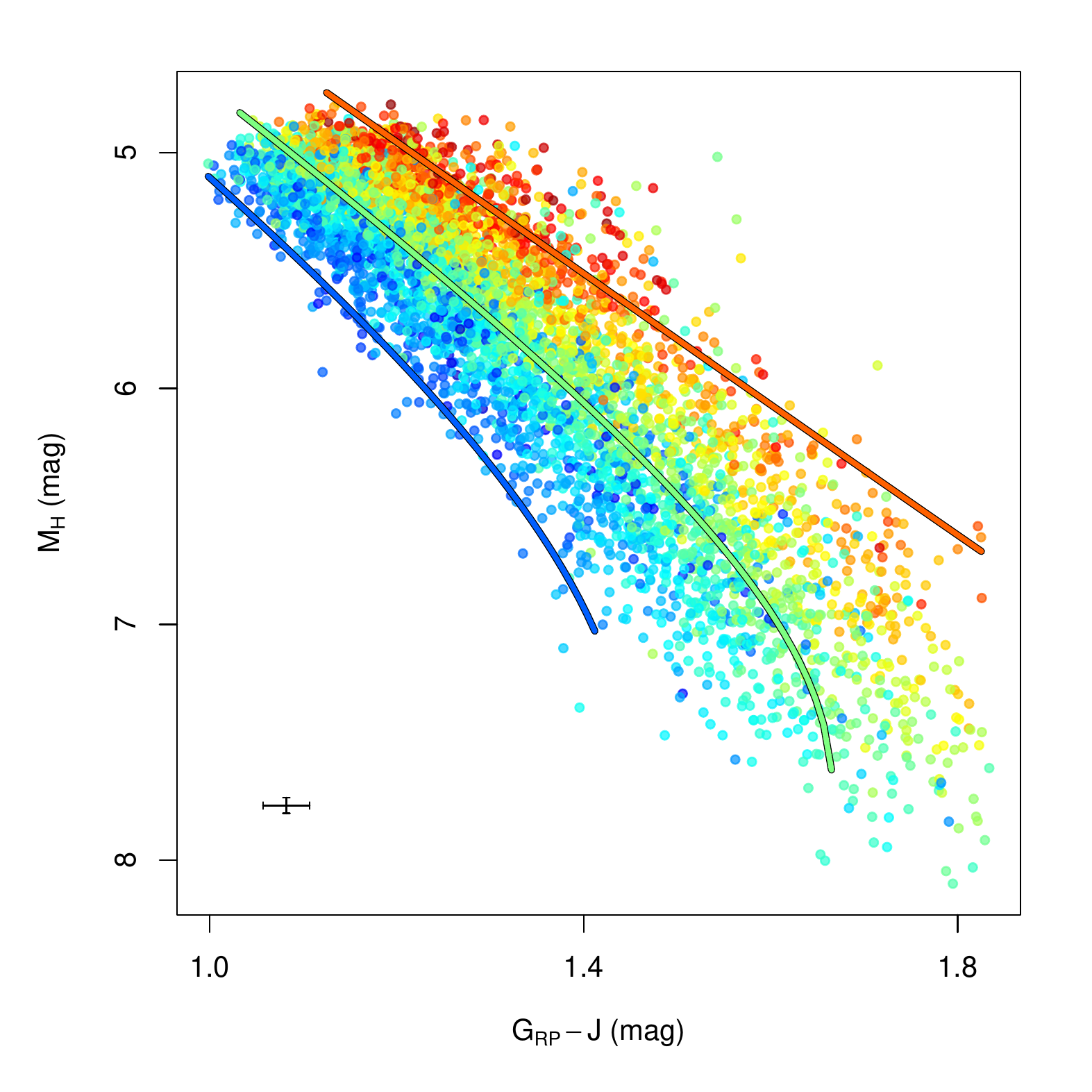} &
         \includegraphics[width=0.28\textwidth]{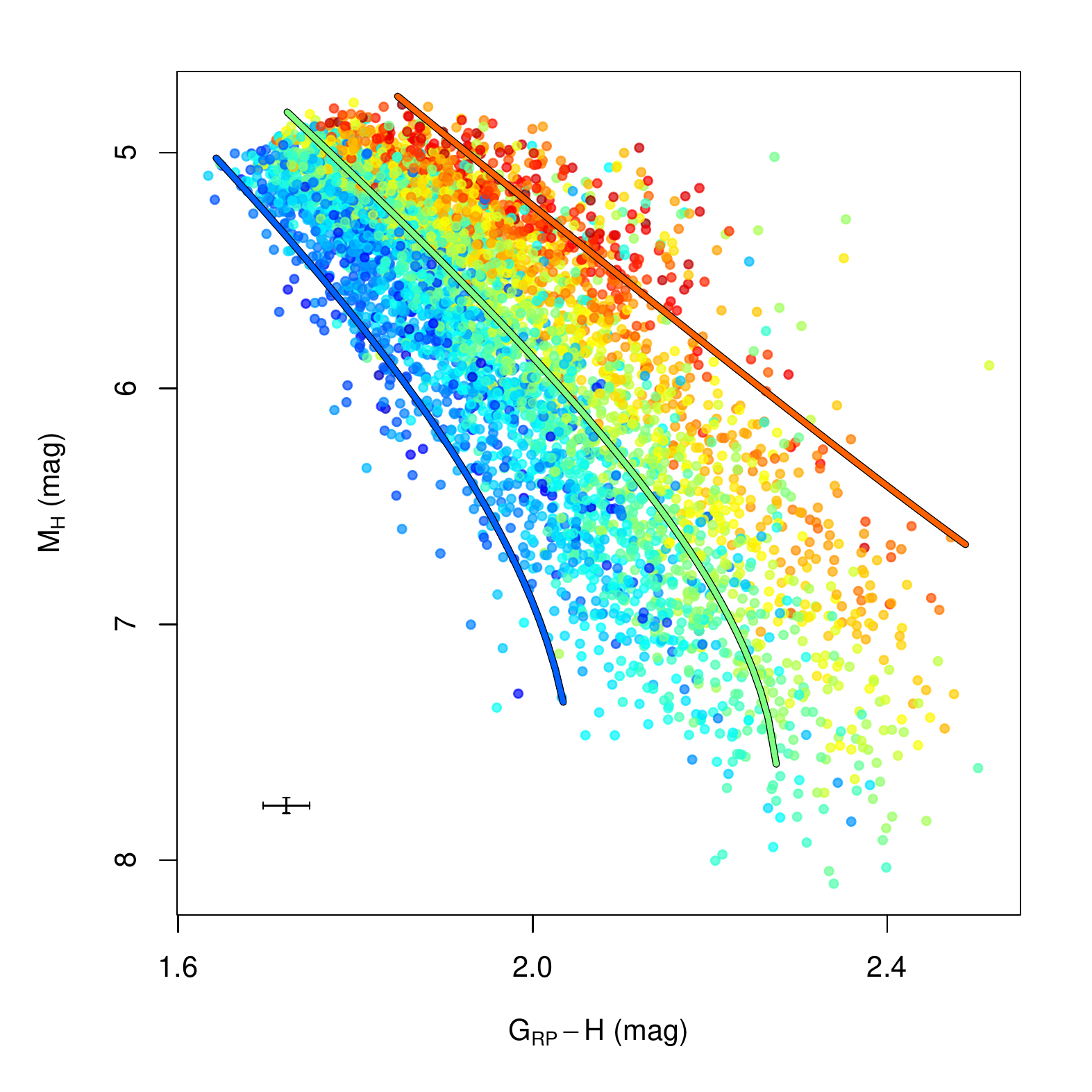} &
         \includegraphics[width=0.28\textwidth]{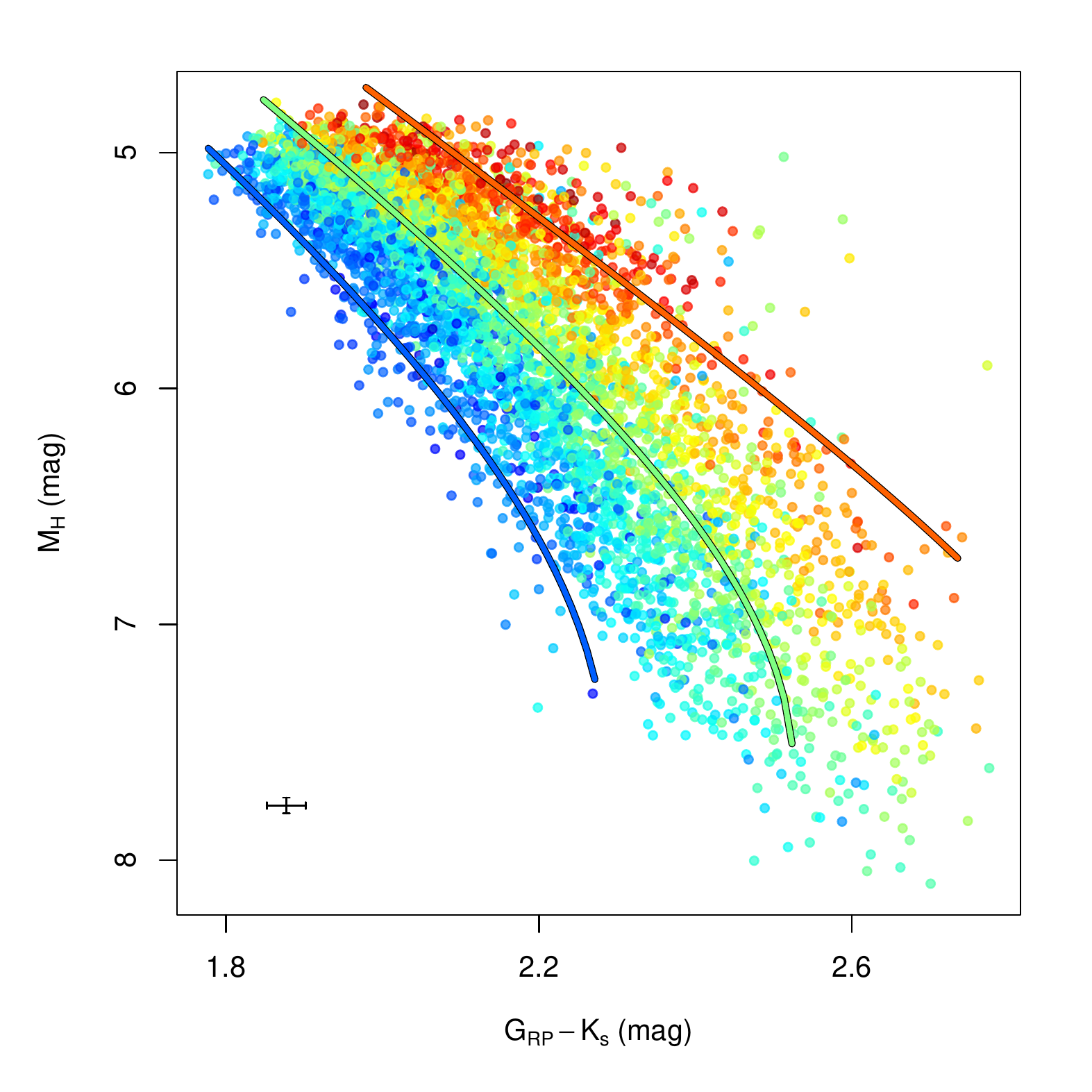} \\
         \includegraphics[width=0.28\textwidth]{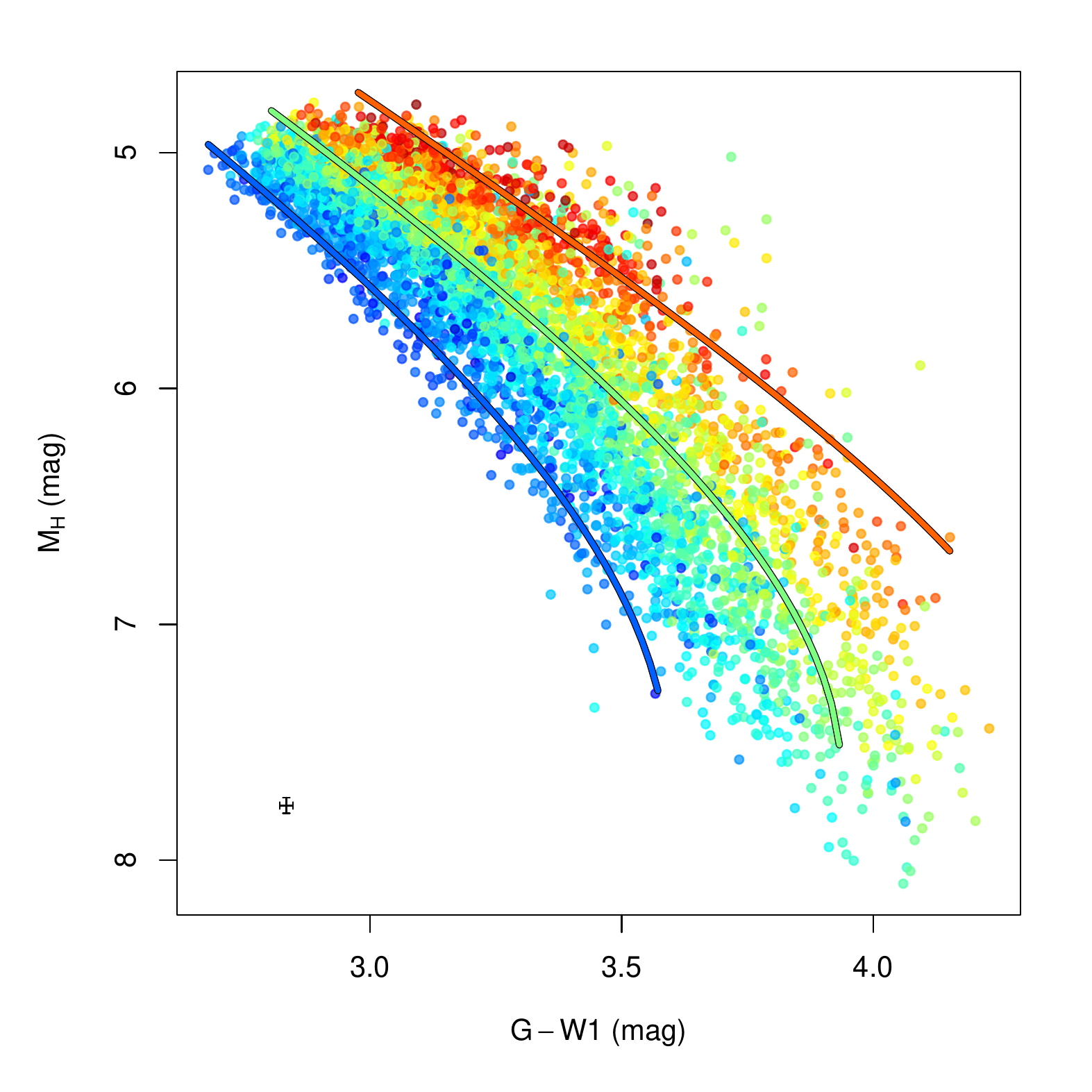} &
         \includegraphics[width=0.28\textwidth]{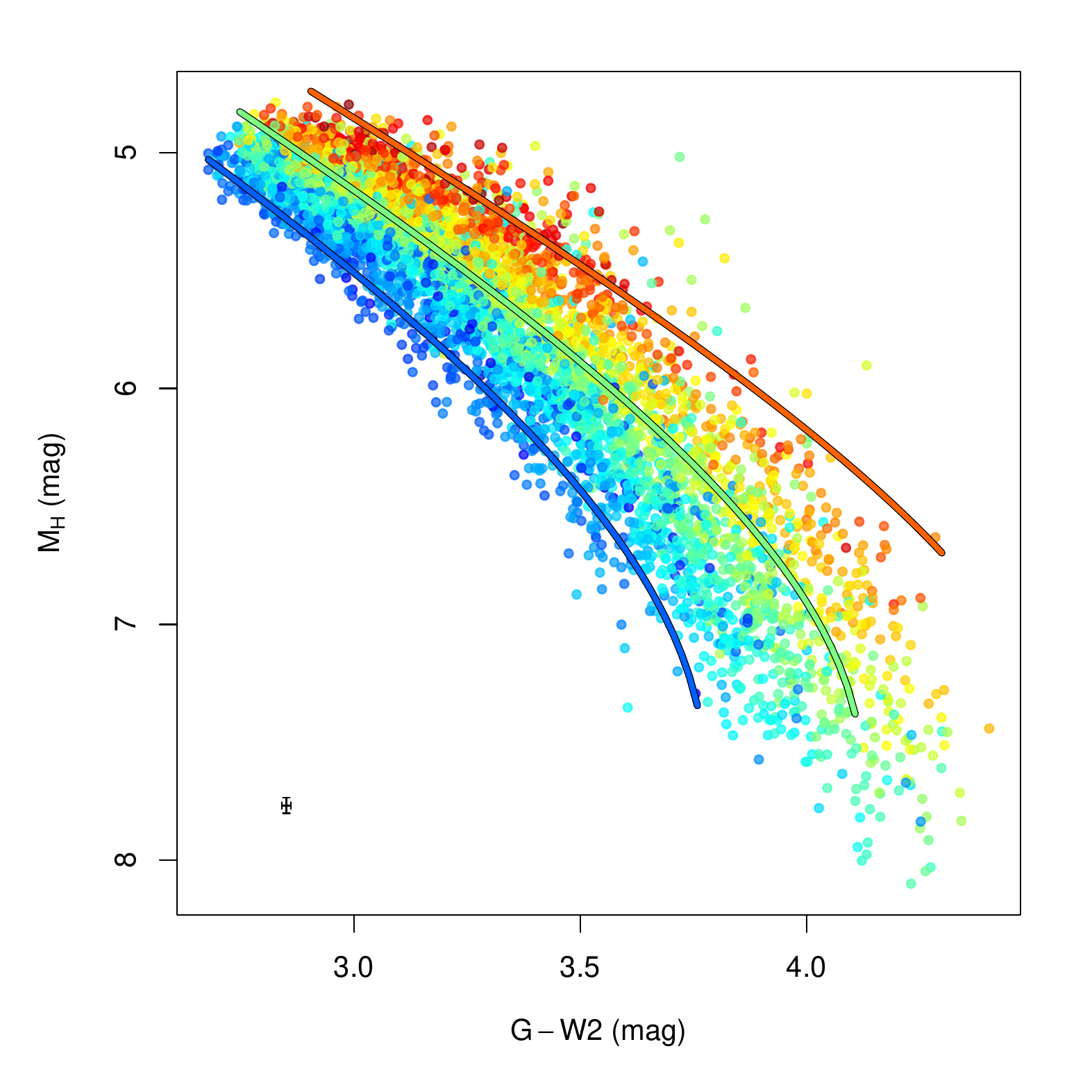} &
         \includegraphics[width=0.28\textwidth]{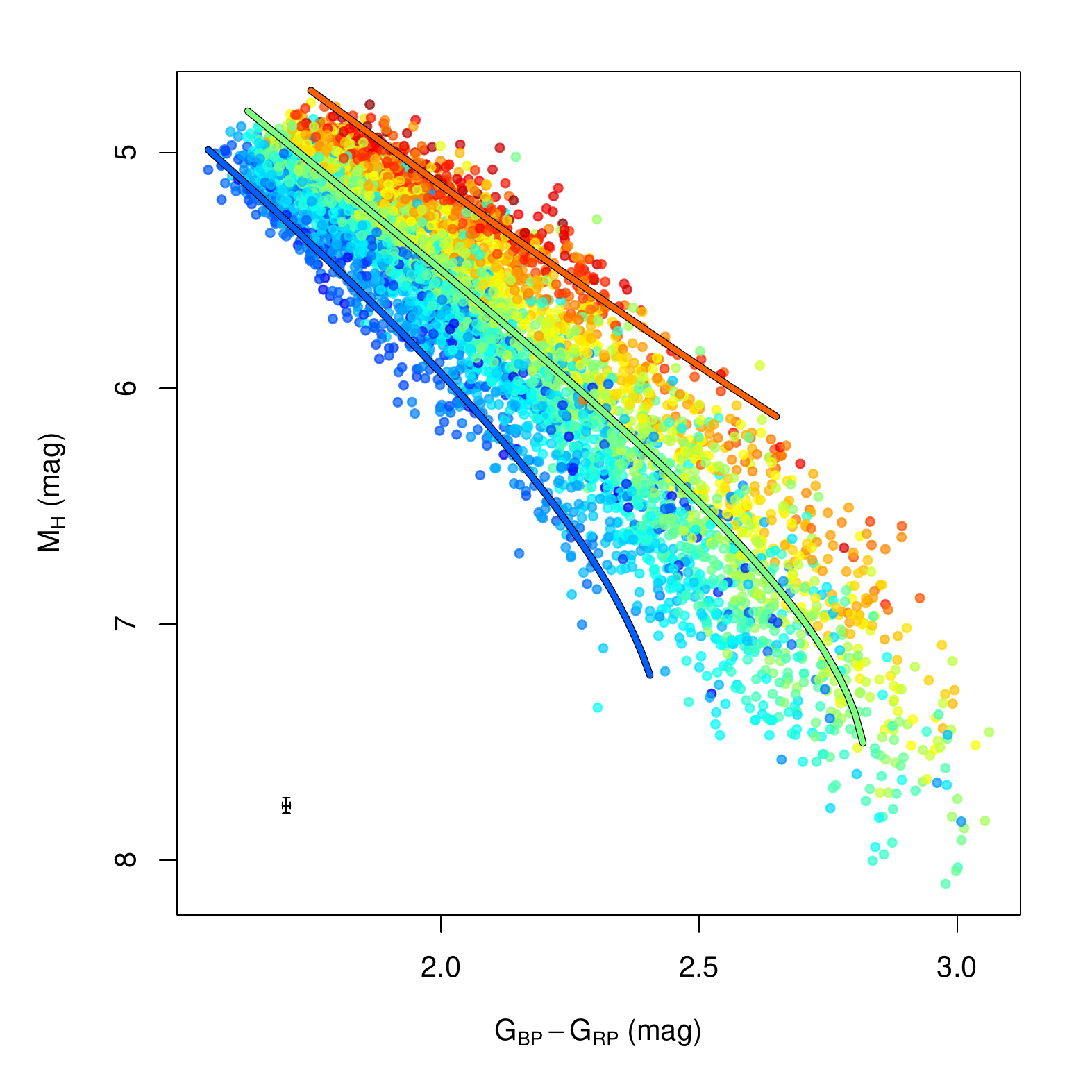} \\
         \includegraphics[width=0.28\textwidth]{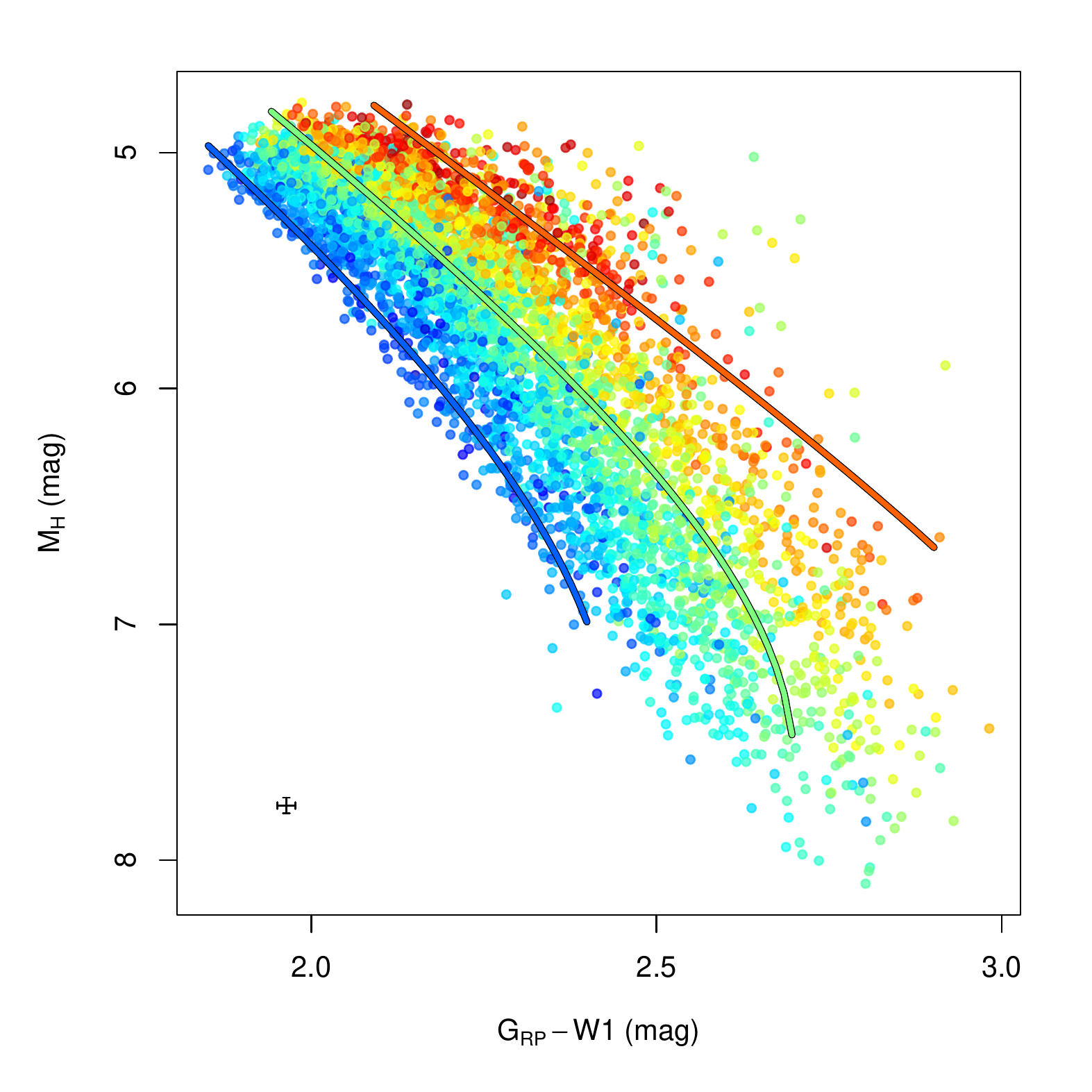} &
         \includegraphics[width=0.28\textwidth]{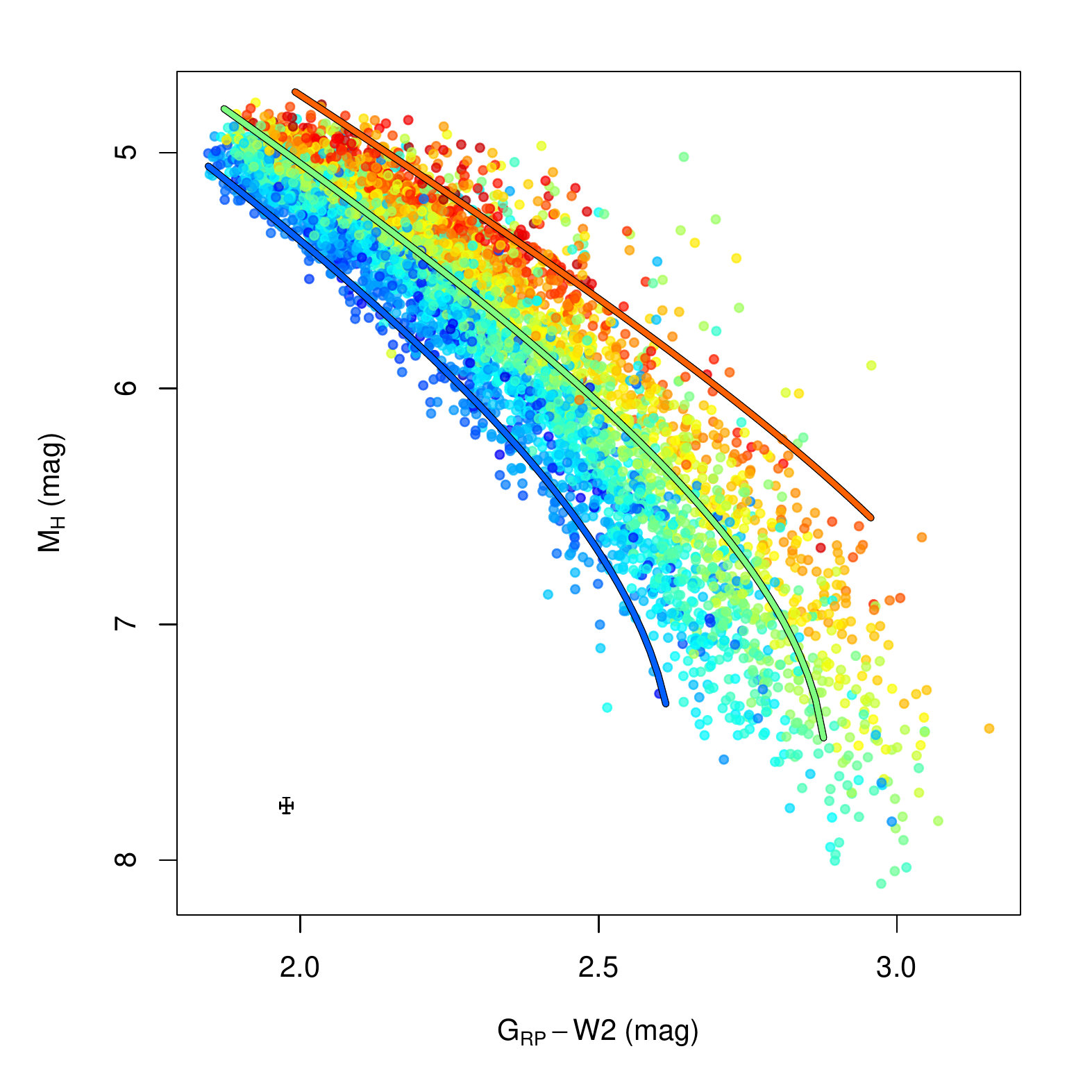} 
    \end{tabular}
    \includegraphics[width=0.25\textwidth]{metallicity_code.png}
    \caption{Same as Fig.~\ref{color_magnitude_diagrams_appendix_H} but for color--magnitude diagrams with $M_H$}
    \label{color_magnitude_diagrams_appendix_H}
\end{figure*}

\renewcommand\thefigure{A.5}
\begin{figure*}
    \centering
    \begin{tabular}{ccc}
         \includegraphics[width=0.28\textwidth]{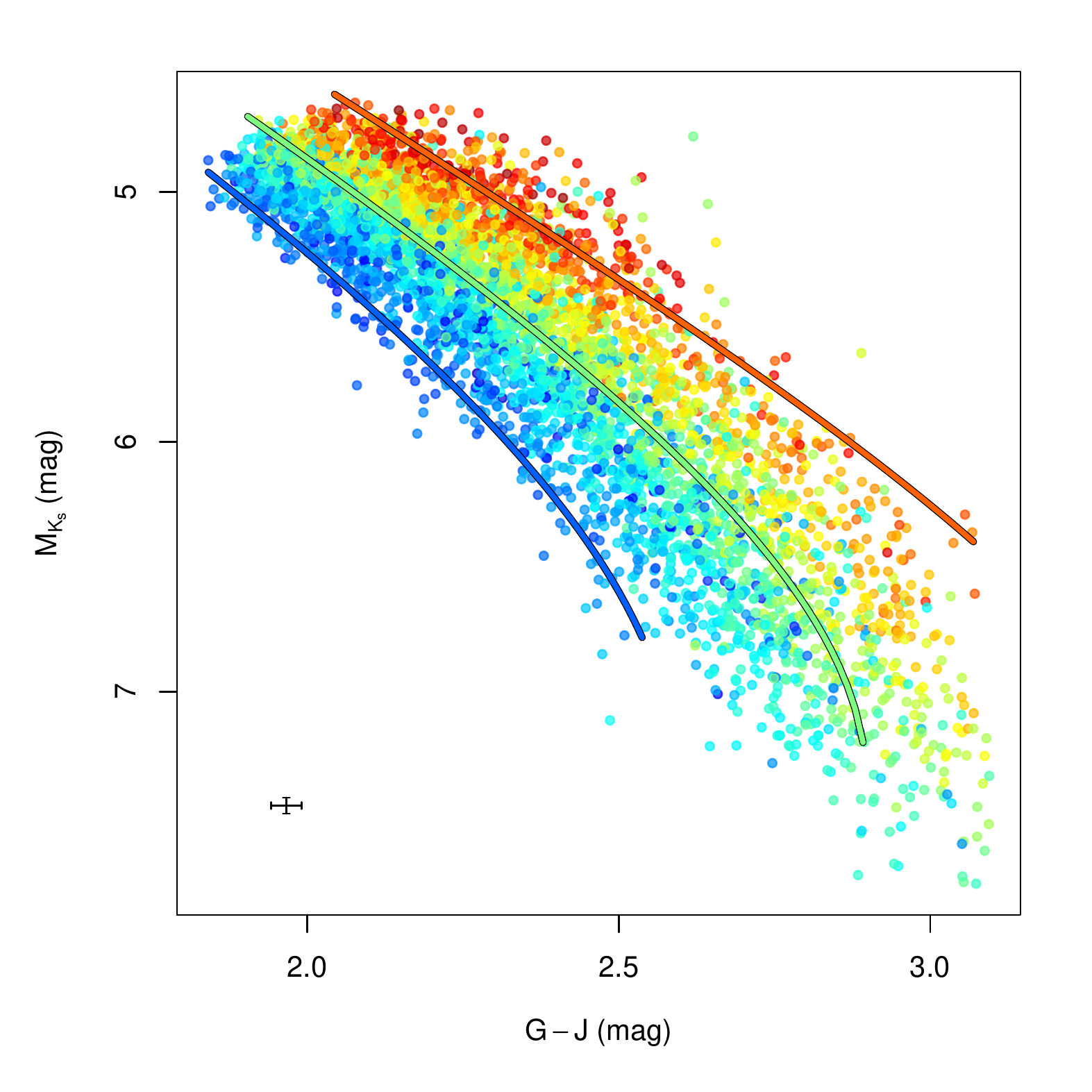} &
         \includegraphics[width=0.28\textwidth]{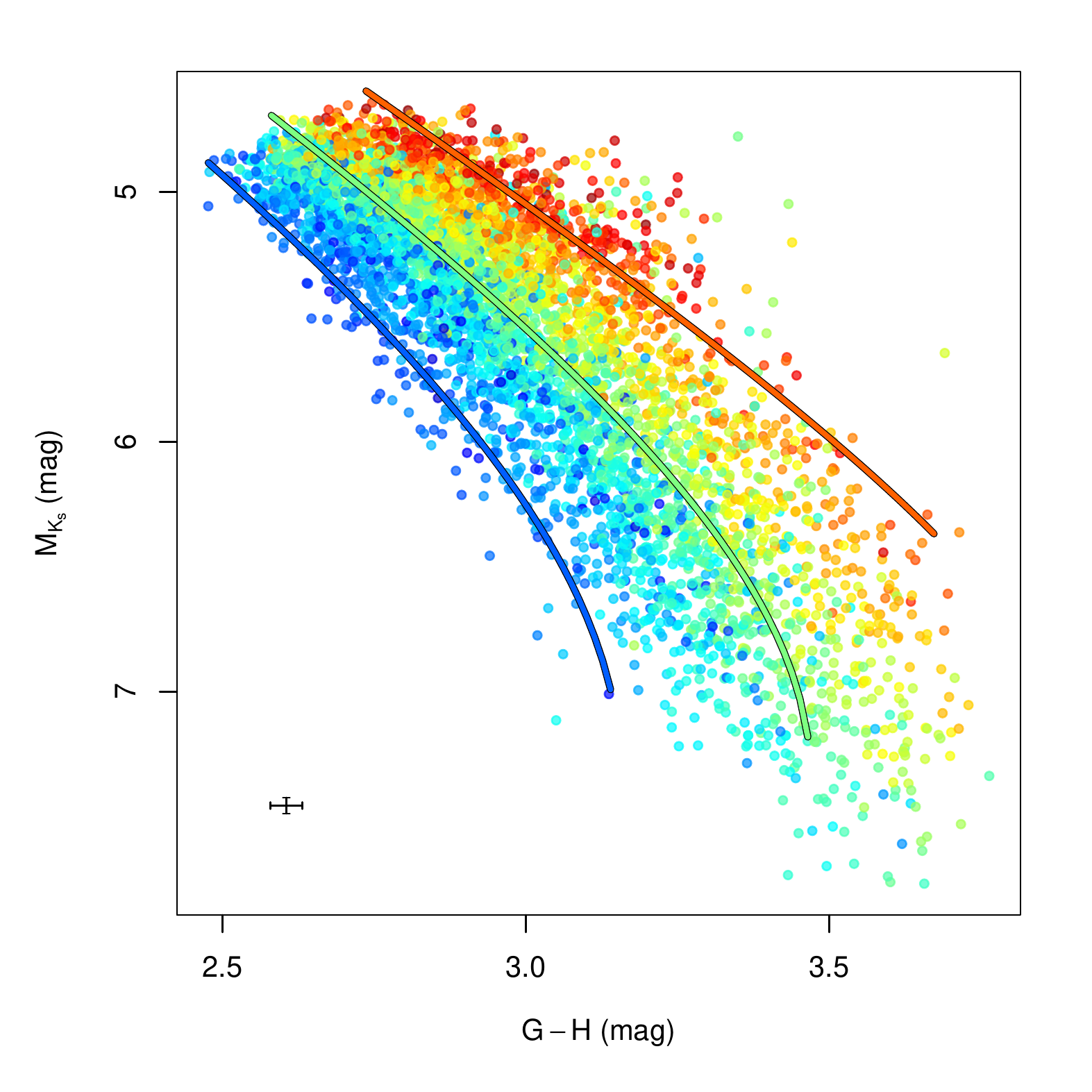} &
         \includegraphics[width=0.28\textwidth]{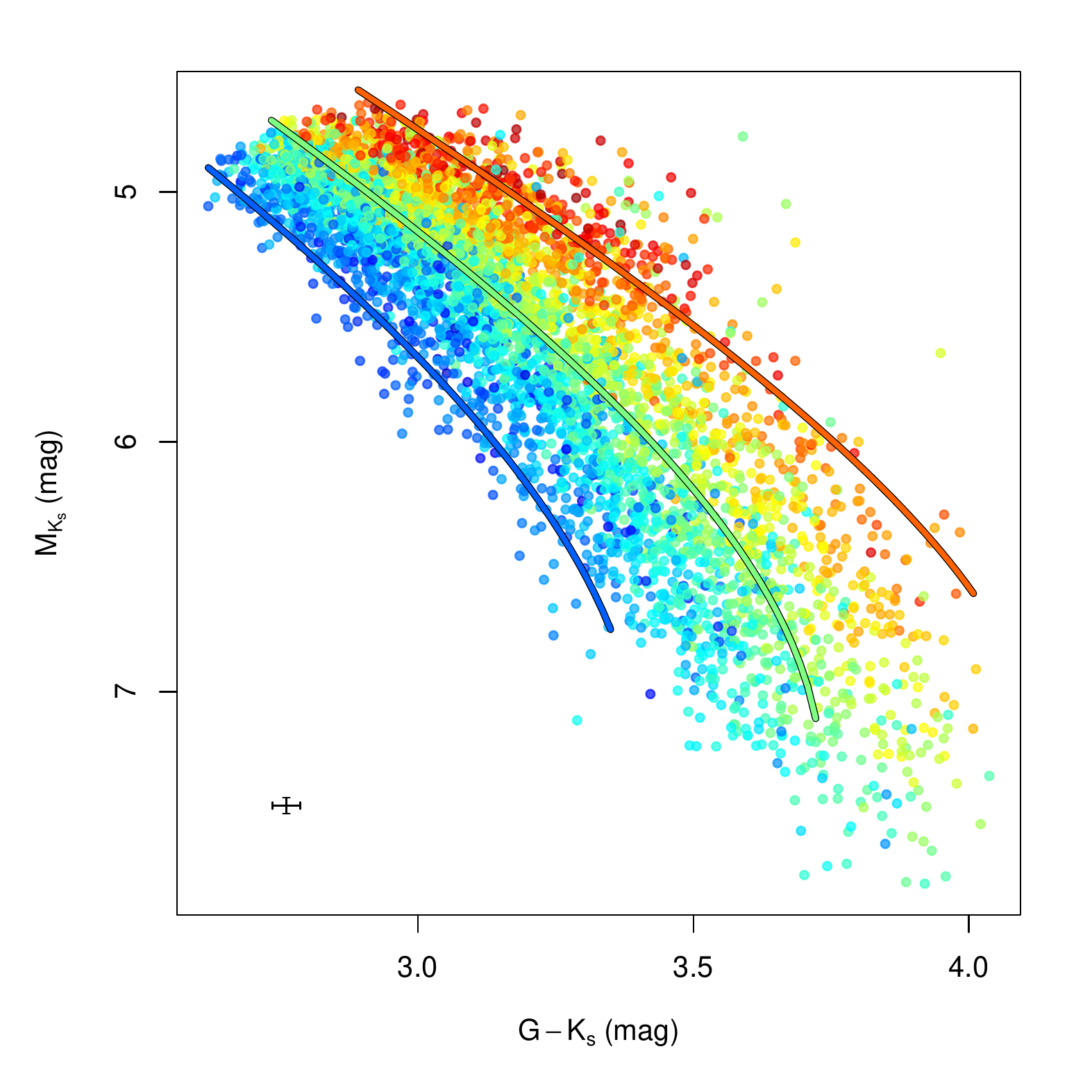} \\
         \includegraphics[width=0.28\textwidth]{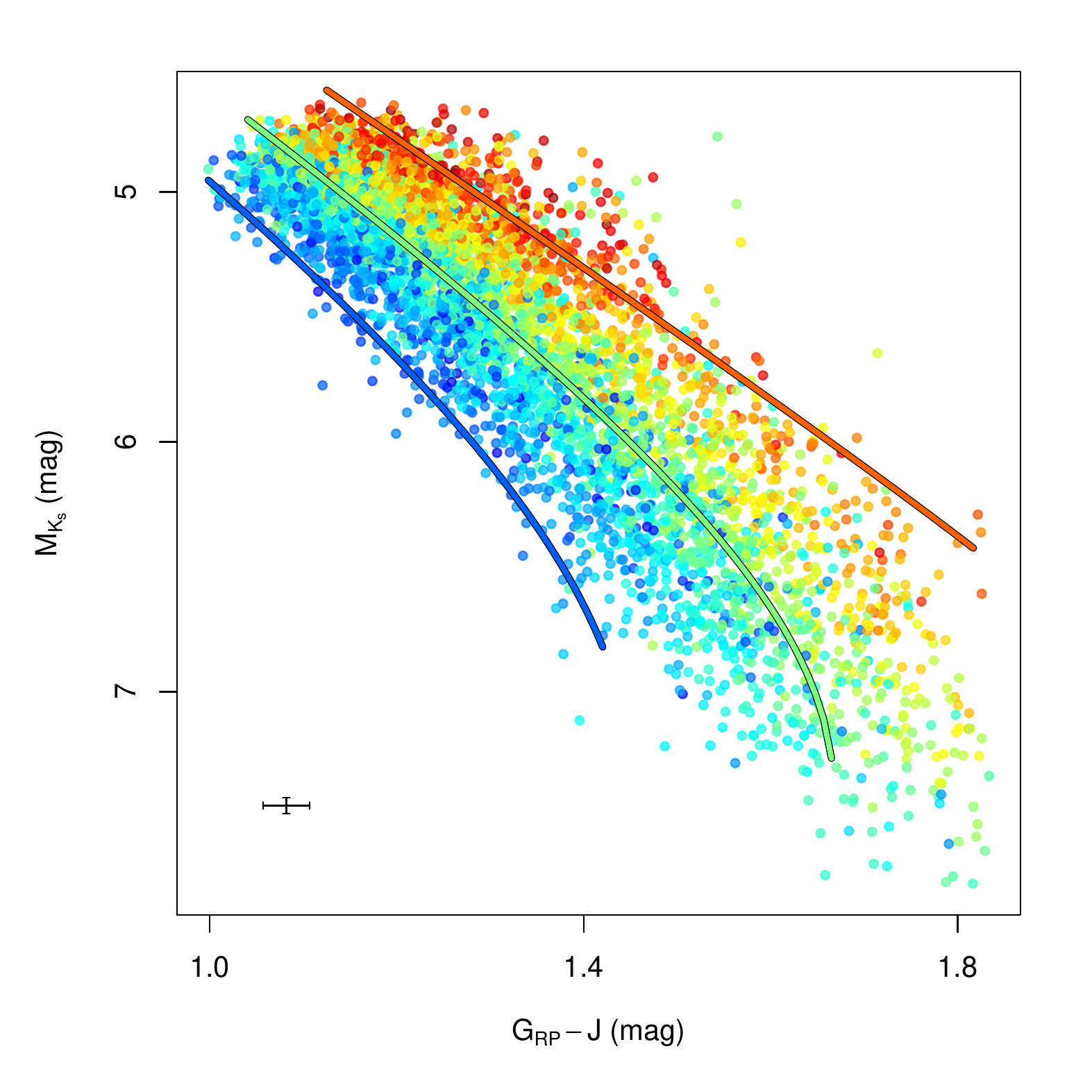} &
         \includegraphics[width=0.28\textwidth]{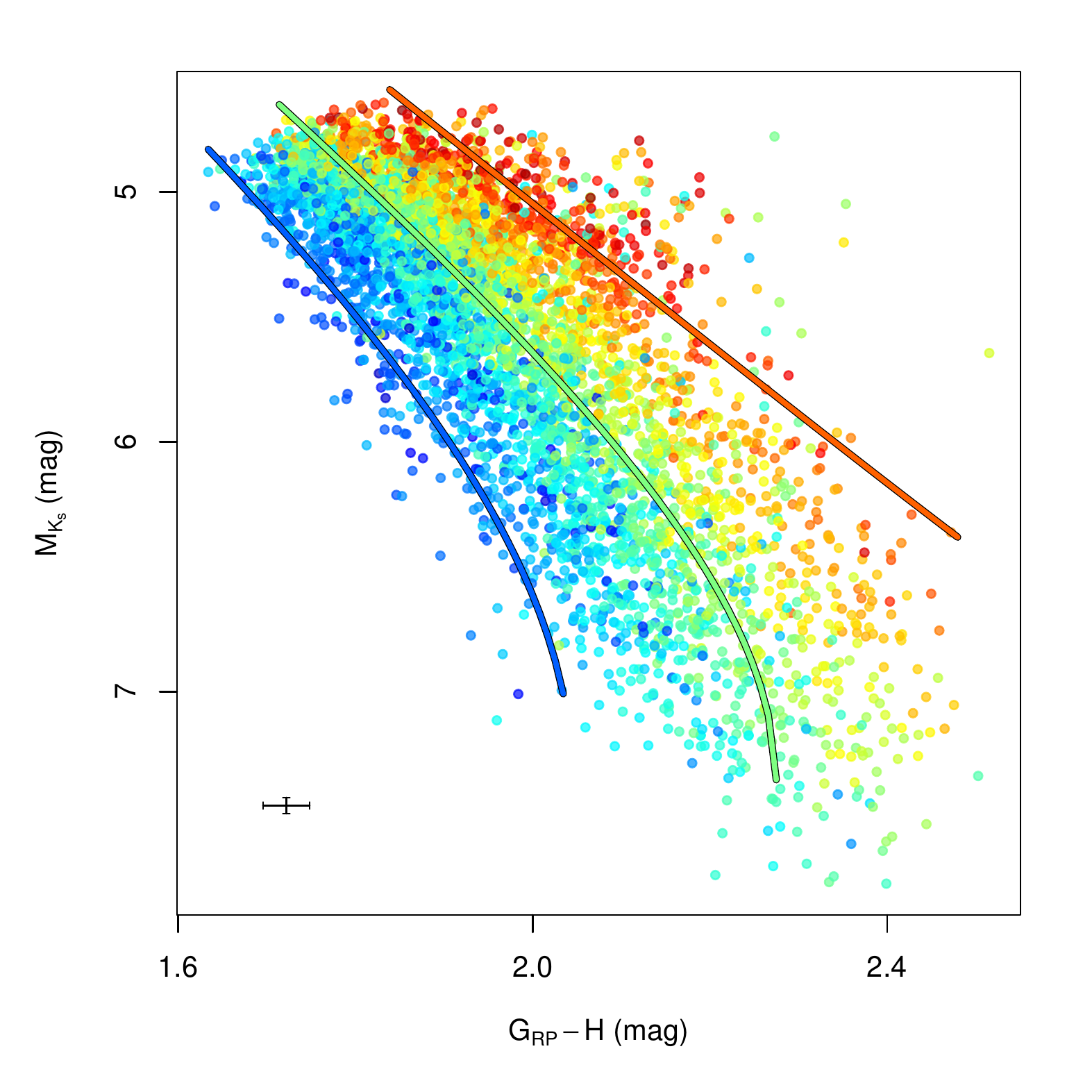} &
         \includegraphics[width=0.28\textwidth]{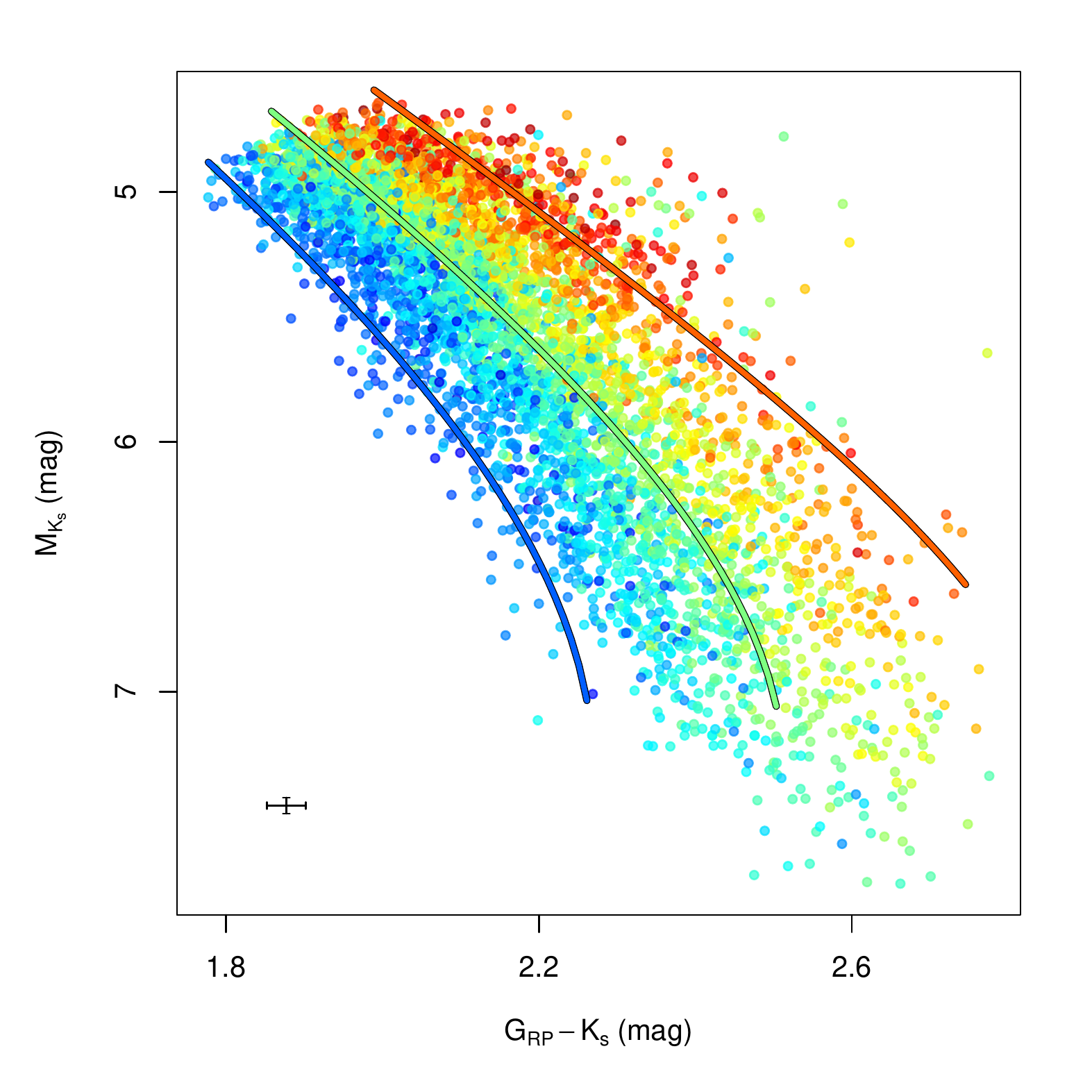} \\
         \includegraphics[width=0.28\textwidth]{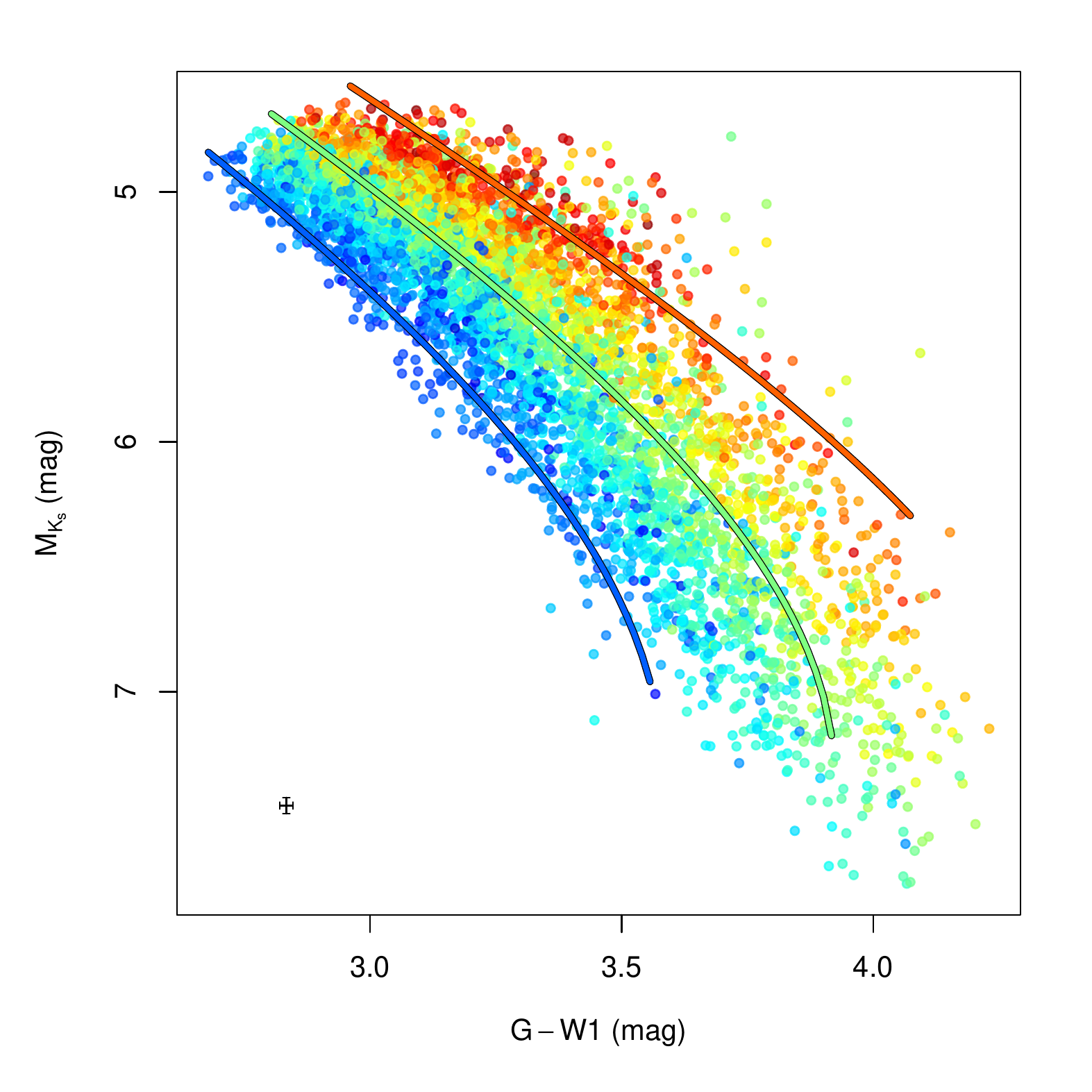} &
         \includegraphics[width=0.28\textwidth]{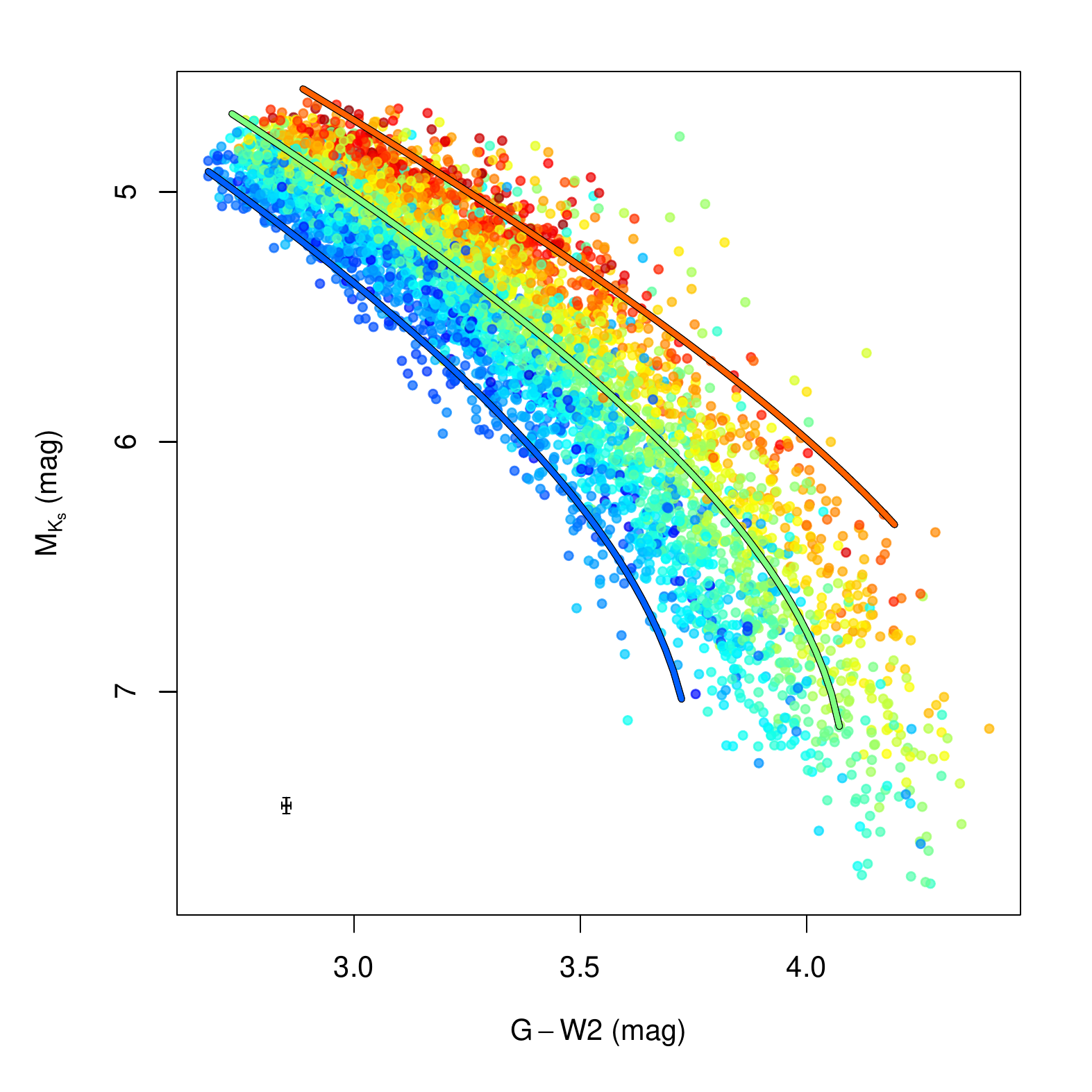} &
         \includegraphics[width=0.28\textwidth]{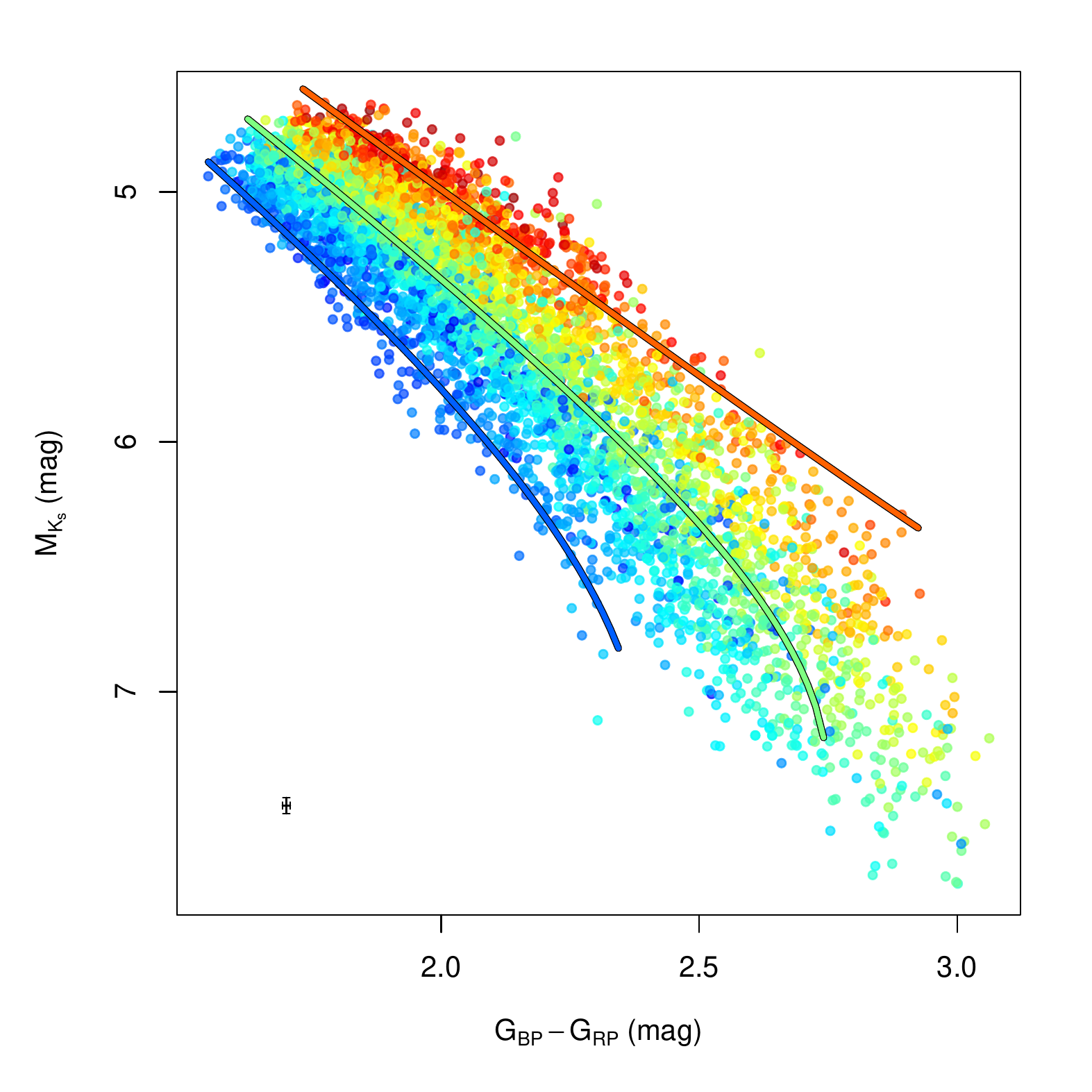} \\
         \includegraphics[width=0.28\textwidth]{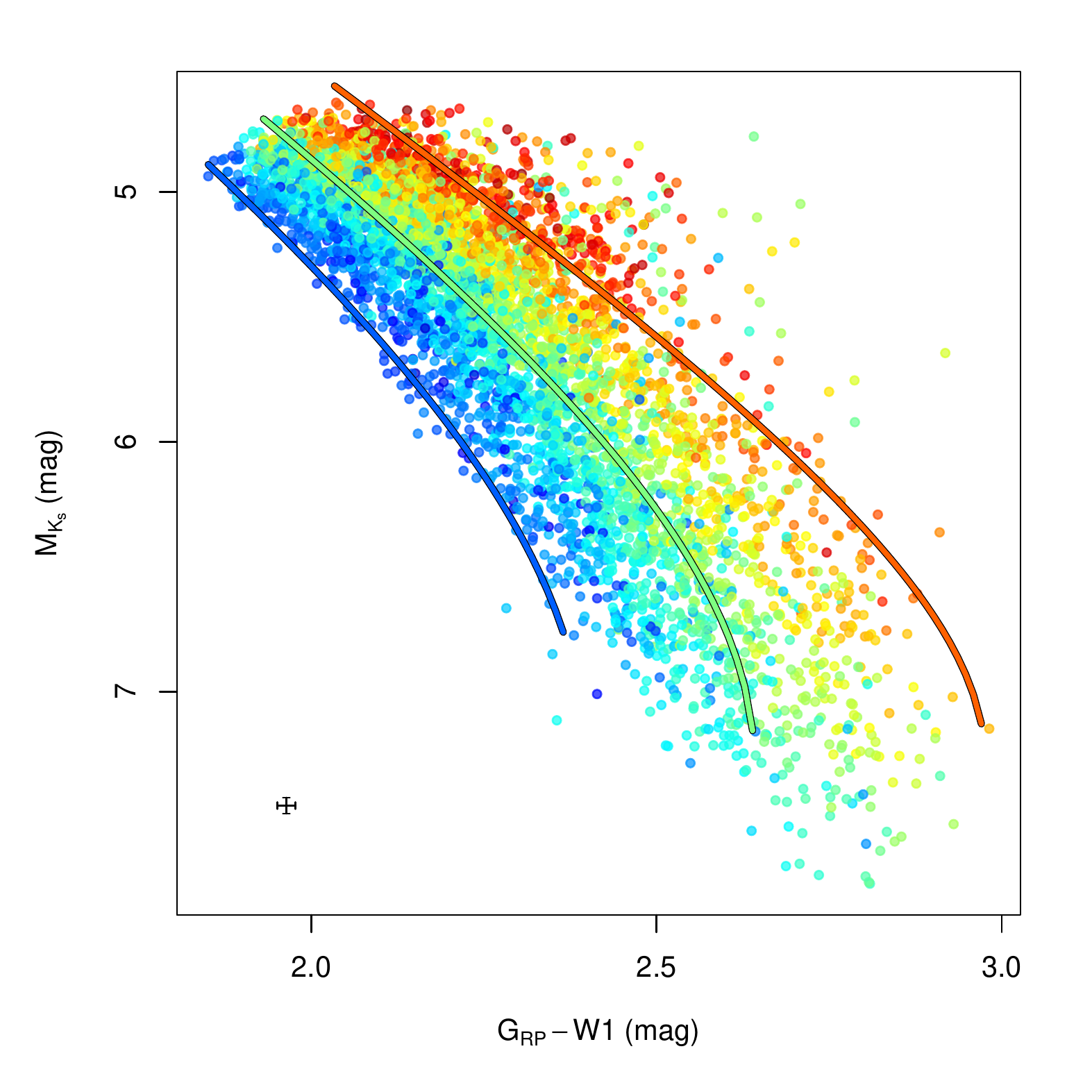} &
         \includegraphics[width=0.28\textwidth]{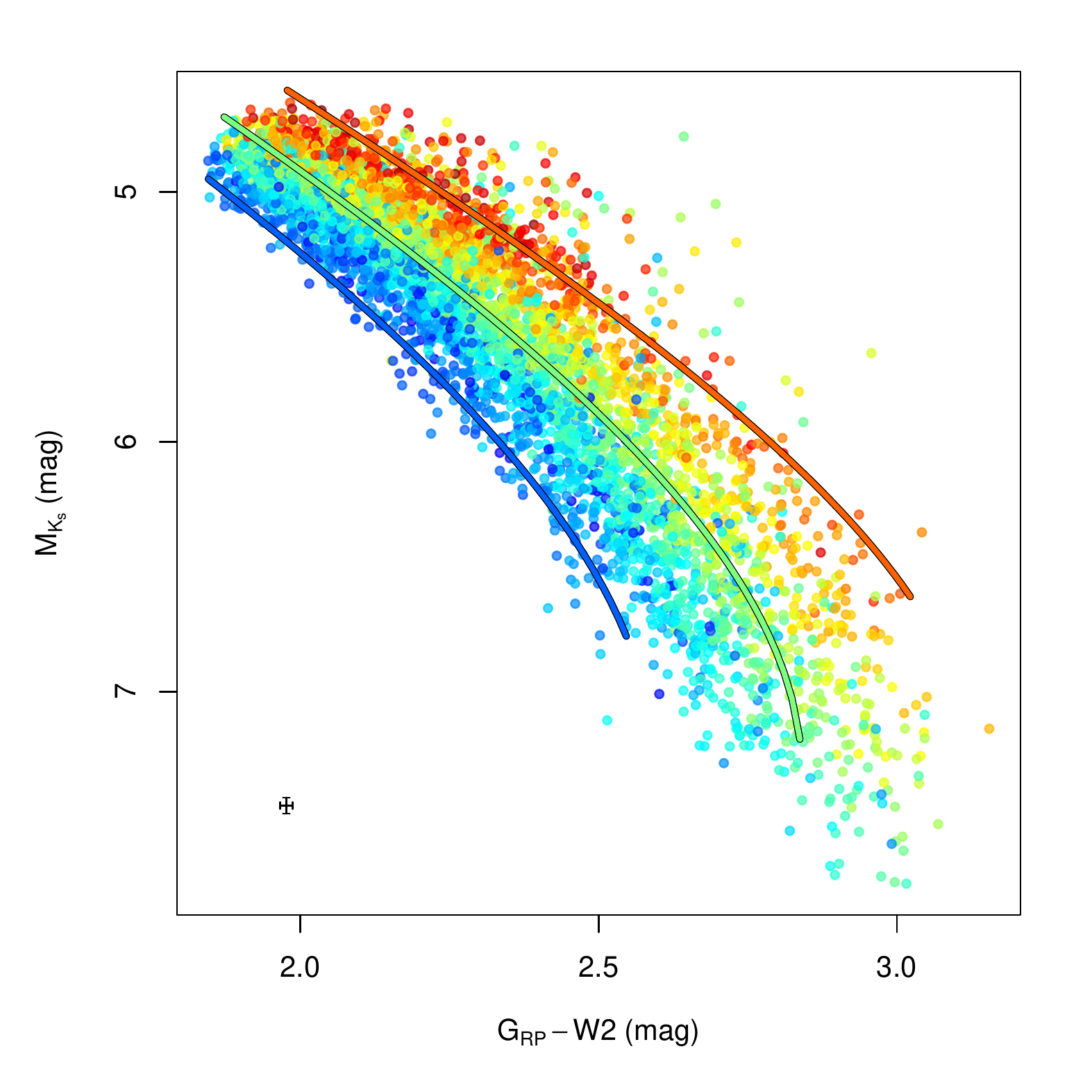} 
    \end{tabular}
    \includegraphics[width=0.25\textwidth]{metallicity_code.png}
    \caption{Same as Fig.~\ref{color_magnitude_diagrams_appendix_K} but for color--magnitude diagrams with $M_{K_s}$}
    \label{color_magnitude_diagrams_appendix_K}
\end{figure*}

\renewcommand\thefigure{A.6}
\begin{figure*}
    \centering
    \includegraphics[width=0.9\textwidth]{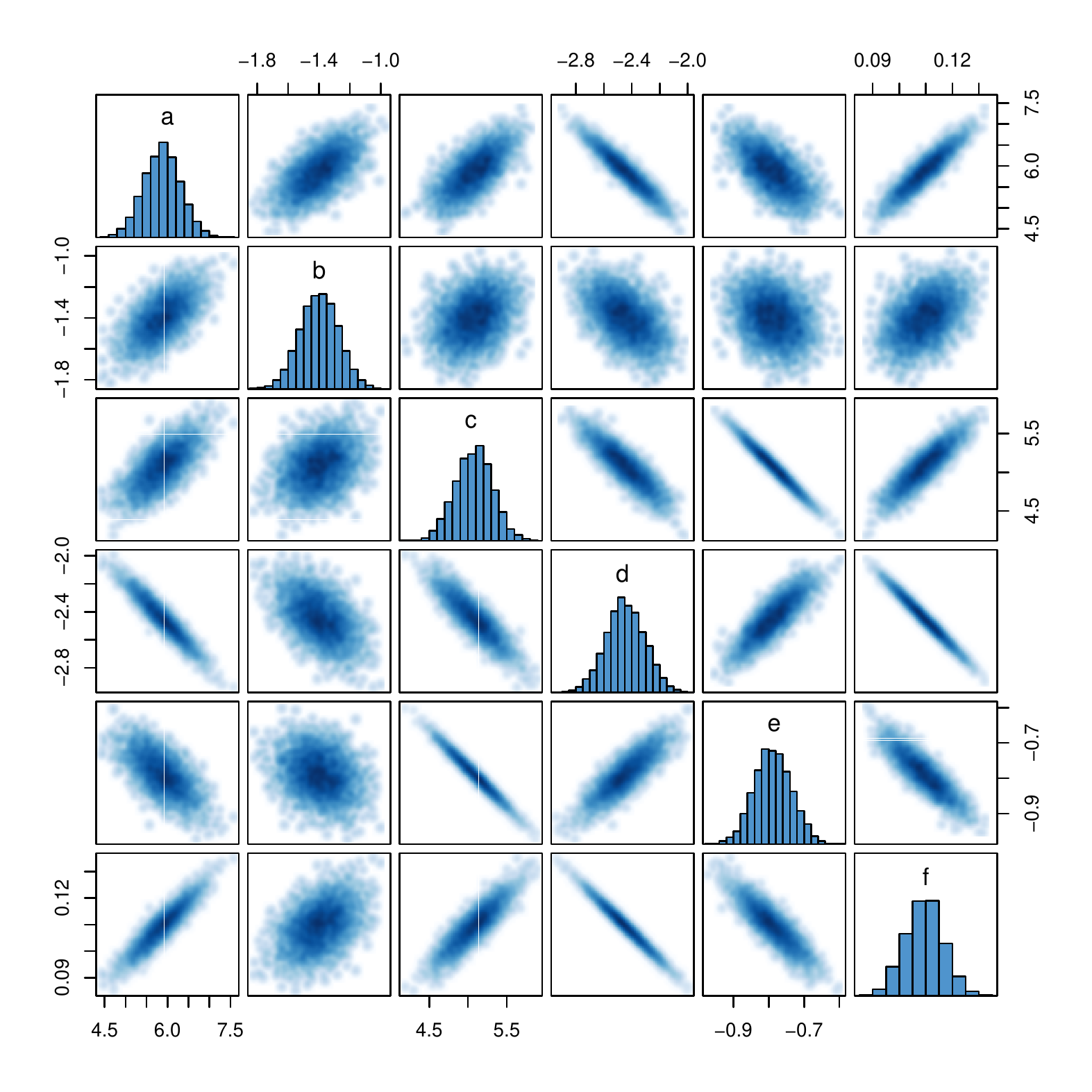}
    \caption{Pairs plot of the coefficients corresponding to our best model: $W1-W2$ vs. $G_\text{BP}-G_\text{RP}$ vs. $M_G$.}
    \label{pairsplot}
\end{figure*}

\renewcommand\thefigure{A.7}
\begin{figure*}
    \centering
    \includegraphics[width=0.9\textwidth]{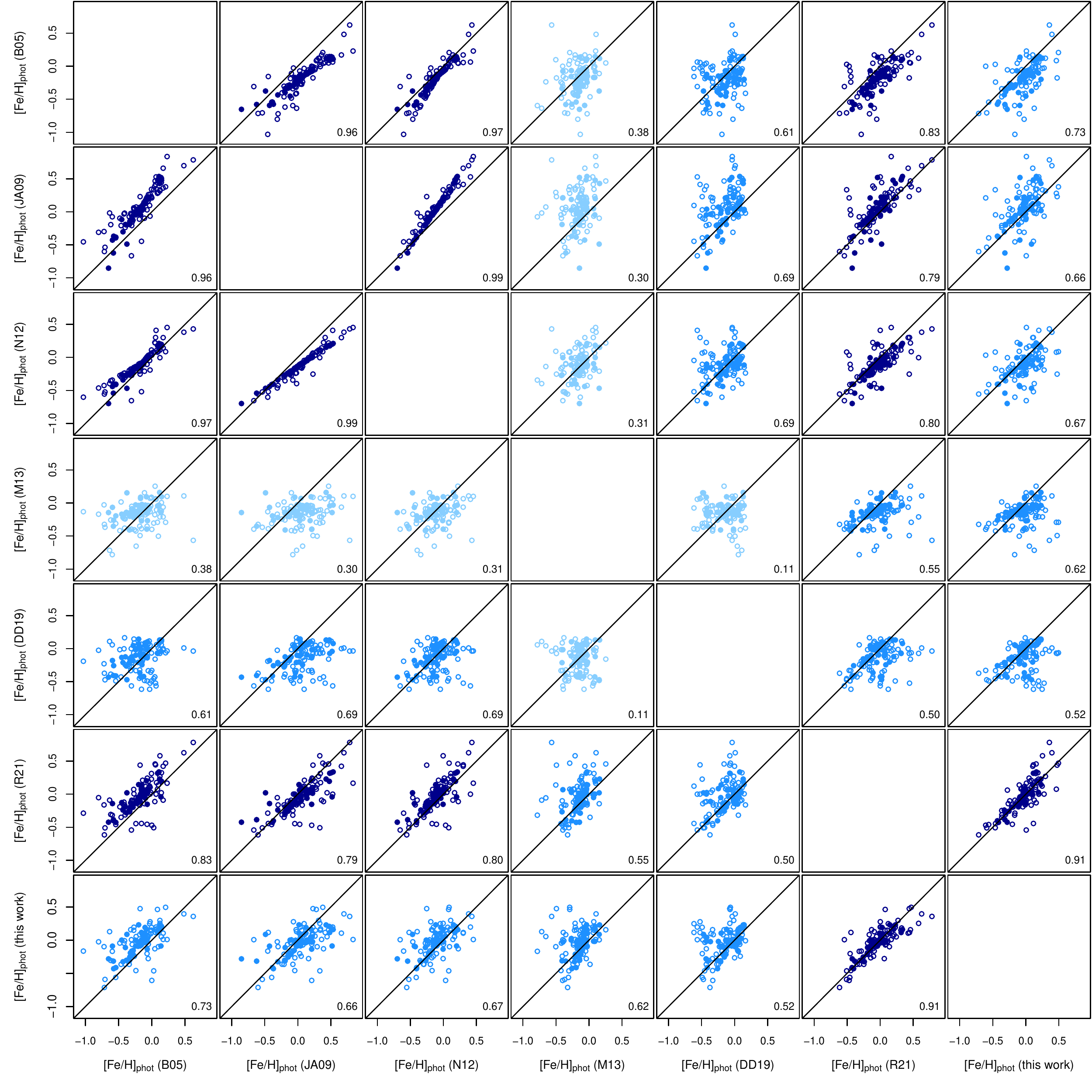}
    \caption{Comparison between the different photometric estimations for the metallicity of the M-dwarf companions. The panels are color-coded by the correlation coefficient $r$ given in the bottom right corner (dark blue: $r\geq0.75$; blue: $0.50\leq r<0.75$; light blue: $r<0.50$).}
    \label{photometric_comparisons}
\end{figure*}

\section{Long table}

\renewcommand{\thetable}{B.1}
\begin{table*}
    \centering
    \footnotesize
    \caption{Estimated metallicities of M-dwarf secondaries in FGK+M systems.} 
    \label{binaries}
    \begin{tabular}{ll cc c rr}
\hline
\hline
\noalign{\smallskip}
WDS & Name & $\alpha$ & $\delta$ & Sp. & [Fe/H]$_\text{spec}$ & [Fe/H]$_\text{phot}$ \\
 & & ~ & ~ & type & Montes+18 & This work\\
 \hline
 \noalign{\smallskip}

  J00467-0426 & LP 646-9 & 00:46:43.4 & $-$04:24:46 & M4.0\,V & $-0.10\pm 0.07$ & $0.09\pm 0.08$\\
  J01450-0104 & LP 588-44 & 01:44:57.0 & $-$01:03:04 & M2.0 & $-0.37\pm 0.03$ & $-0.33\pm 0.08$\\
  J02556+2652 & HD 18143 C & 02:55:35.8 & $+$26:52:21 & M4.0\,V & $0.18\pm 0.05$ & $0.16\pm 0.08$\\
  J03078+2533 & HD 19381B & 03:07:58.3 & $+$25:32:02 & M3.5\,V & $0.11\pm 0.02$ & $0.11\pm 0.08$\\
  J03203+0902 & HD 20727B & 03:20:42.5 & $+$09:02:10 & M4.0\,V & $-0.24\pm 0.02$ & $-0.03\pm 0.08$\\
  J03356+4253 & Wolf 191 & 03:35:28.5 & $+$42:53:35 & M0.5\,V & $-0.37\pm 0.03$ & $-0.43\pm 0.08$\\
  J03520+3947 & TYC 2868-639-1 & 03:51:58.1 & $+$39:46:57 & M0.0\,V & $-0.13\pm 0.05$ & $-0.03\pm 0.08$\\
  J03556+5214 & LSPM J0355+5214 & 03:55:36.9 & $+$52:14:29 & M2.5\,V & $-0.32\pm 0.02$ & $-0.32\pm 0.08$\\
  J05003+2508 & HD 31867B & 05:00:19.5 & $+$25:07:51 & M1.0\,V & $-0.01\pm 0.02$ & $-0.02\pm 0.08$\\
  J05264+0351 & 2MASS J05262029+0351111 & 05:26:20.3 & $+$03:51:11 & M1.5\,V & $0.01\pm 0.03$ & $-0.09\pm 0.08$\\
  J05466+0110 & HD 38529 B & 05:46:19.4 & $+$01:12:47 & M2.5\,V & $0.32\pm 0.02$ & $0.23\pm 0.08$\\
  J06319+0039 & G 106-52 & 06:31:23.7 & $+$00:36:45 & M1.5\,V & $-0.29\pm 0.02$ & $-0.30\pm 0.08$\\
  J06332+0528 & HD 46375 B & 06:33:12.1 & $+$05:27:53 & M2.0\,V & $0.23\pm 0.06$ & $0.23\pm 0.08$\\
  J06461+3233 & HD 263175 B & 06:46:07.5 & $+$32:33:15 & M1.0\,V & $-0.38\pm 0.03$ & $-0.41\pm 0.08$\\
  J07041+7514 & LP 16-395 & 07:04:09.5 & $+$75:14:37 & M4.0\,V & $0.18\pm 0.02$ & $0.10\pm 0.08$\\
  J08138+6306 & NLTT 19115 & 08:14:19.0 & $+$63:04:40 & M1.5\,V & $-0.09\pm 0.01$ & $-0.19\pm 0.08$\\
  J08484+2042 & 2MASS J08482492+2042188 & 08:48:24.9 & $+$20:42:18 & M1.5\,V & $-0.03\pm 0.02$ & $-0.14\pm 0.08$\\
  J08492+0329 & LEP 33 & 08:49:02.3 & $+$03:29:47 & M4 & $0.05\pm 0.02$ & $0.04\pm 0.08$\\
  J09008+2347 & HD 77052 B & 09:00:53.2 & $+$23:46:59 & M2.5\,V & $0.04\pm 0.02$ & $-0.02\pm 0.08$\\
  J09029+0600 & 2MASS J09025320+0602095 & 09:02:53.2 & $+$06:02:10 & M1.5\,V & $-0.12\pm 0.02$ & $-0.09\pm 0.08$\\
  J09058+5532 & NLTT 20915 & 09:05:51.2 & $+$55:32:18 & M3.5\,V & $0.03\pm 0.02$ & $-0.02\pm 0.08$\\
  J09152+2323 & BD+23 2063B & 09:15:10.1 & $+$23:21:33 & M0.0\,V & $0.21\pm 0.02$ & $0.19\pm 0.08$\\
  J11403+0931 & LP 493-31 & 11:40:20.8 & $+$09:30:45 & M1.5\,V & $-0.28\pm 0.03$ & $-0.25\pm 0.08$\\
  J11455+4740 & G 122-46 & 11:47:21.7 & $+$47:45:57 & M2.5 & $-0.45\pm 0.02$ & $-0.31\pm 0.08$\\
  J11523+0957 & LP 493-64 & 11:52:17.9 & $+$10:00:39 & M4.0\,V & $0.22\pm 0.06$ & $0.16\pm 0.08$\\
  J12051+1933 & BD+20 2678B & 12:05:11.9 & $+$19:31:41 & M2 & $-0.07\pm 0.03$ & $0.02\pm 0.08$\\
  J12372+3545 & 2MASS J12371547+3549176 & 12:37:15.5 & $+$35:49:18 & M1.5\,V & $-0.05\pm 0.02$ & $-0.03\pm 0.08$\\
  J13315-0800 & HD 117579B & 13:31:29.8 & $-$07:59:59 & M0.0\,V & $-0.18\pm 0.02$ & $-0.09\pm 0.08$\\
  J14050+0157 & NLTT 36190 & 14:04:55.8 & $+$01:57:23 & M2 & $-0.01\pm 0.02$ & $0.00\pm 0.08$\\
  J14336+0920 & HD 127871B & 14:33:39.9 & $+$09:20:10 & M3.5\,V & $-0.11\pm 0.03$ & $-0.09\pm 0.08$\\
  J14415+1336 & HD 129290B & 14:41:30.3 & $+$13:37:36 & M1.0\,V & $-0.12\pm 0.02$ & $-0.08\pm 0.08$\\
  J15123+3939 & LP 222-50 & 15:11:51.5 & $+$39:33:02 & M2.5\,V & $-0.21\pm 0.03$ & $-0.21\pm 0.08$\\
  J15353+6005 & LP 99-392 & 15:35:25.7 & $+$60:05:08 & M3.5\,V & $-0.13\pm 0.06$ & $-0.02\pm 0.08$\\
  J16239+0315 & G 17-23 & 16:33:02.8 & $+$03:11:37 & M3.0\,V & $0.01\pm 0.04$ & $0.00\pm 0.08$\\
  J17428+1646 & 2MASS J17425203+1643476 & 17:42:52.0 & $+$16:43:48 & M1.5\,V & $-0.23\pm 0.03$ & $-0.21\pm 0.08$\\
  J17477+2748 & G 182-27 & 17:47:44.3 & $+$27:47:07 & M1.5\,V & $-0.06\pm 0.02$ & $0.00\pm 0.08$\\
  J18006+6833 & LDS 1460B & 18:00:37.0 & $+$68:32:54 & K7\,V & $-0.02\pm 0.02$ & $0.00\pm 0.12$\\
  J18006+2934 & HD 164595B & 18:00:45.4 & $+$29:33:57 & M2.0\,V & $-0.08\pm 0.01$ & $-0.09\pm 0.08$\\
  J18090+2409 & 2MASS J18090192+2409041 & 18:09:01.9 & $+$24:09:04 & M1.0\,V & $-0.13\pm 0.02$ & $-0.11\pm 0.08$\\
  J18292+1142 & 2MASS 18291369+1141271 & 18:29:13.7 & $+$11:41:27 & K5.0\,V & $0.28\pm 0.02$ & $0.21\pm 0.08$\\
  J19321-1116 & HD 183870B & 19:32:08.1 & $-$11:19:57 & M3.5\,V & $-0.10\pm 0.06$ & $-0.02\pm 0.08$\\
  J21324-2058 & LP 873-74 & 21:32:21.0 & $-$20:58:10 & M0.5\,V & $-0.19\pm 0.03$ & $-0.15\pm 0.08$\\
  J21575+2856 & 2MASS J21572970+2854494 & 21:57:29.7 & $+$28:54:50 & M1.5\,V & $0.15\pm 0.02$ & $0.10\pm 0.08$\\
  J22090-1754 & LP 819-37 & 22:08:54.2 & $-$17:47:52 & M2.5\,V & $-0.42\pm 0.02$ & $-0.28\pm 0.08$\\
  J22159+5440 & HD 211472B & 22:16:02.6 & $+$54:40:00 & M4.0\,V & $-0.05\pm 0.02$ & $0.00\pm 0.08$\\
  J22311+4509 & HD 213519B & 22:31:06.5 & $+$45:09:44 & M3 & $0.00\pm 0.01$ & $-0.06\pm 0.08$\\ 
  
\noalign{\smallskip}
\hline
         \end{tabular}
\end{table*}




\end{document}